\documentclass[pdflatex,sn-mathphys-num]{sn-jnl}

\usepackage{rotating}
\usepackage{graphicx}%
\usepackage{multirow}%
\usepackage{amsmath,amssymb,amsfonts}%
\usepackage{amsthm}%
\usepackage{lineno}
\usepackage{mathrsfs}%
\usepackage[title]{appendix}%
\usepackage{xcolor}%
\usepackage{bbm}%
\usepackage{textcomp}%
\usepackage{subcaption}
\usepackage{manyfoot}%
\usepackage{booktabs}%
\usepackage{algorithm}%
\usepackage{algorithmicx}
\usepackage{algpseudocode}%
\usepackage{listings}
\usepackage{caption}  
\usepackage{longtable}
\usepackage{float}

\usepackage{printlen} 
\theoremstyle{thmstyleone}%

\theoremstyle{thmstyletwo}%
\theoremstyle{thmstylethree}%
\renewcommand{\appendixname}{}
\begin{document}

\title[Article Title]{An extension to reversible jump Markov chain Monte Carlo for change point problems with heterogeneous temporal dynamics}

\author*[1]{\fnm{Emily} \sur{Gribbin}}\email{egribbin02@qub.ac.uk}

\author[2]{\fnm{Benjamin} \sur{Davis}}

\author[2]{\fnm{Daniel} \sur{Rolfe}}

\author*[1]{\fnm{Hannah} \sur{Mitchell}}\email{h.mitchell@qub.ac.uk}

\affil*[1]{\orgdiv{Mathematical Sciences Research Centre, School of Mathematics and Physics}, \orgname{Queen's University Belfast} \orgaddress{\street{University Road}, \city{Belfast}, \postcode{BT7 1NN}, \state{Northern Ireland}, \country{United Kingdom}}}

\affil[2]{\orgdiv{OCTOPUS Group, Central Laser Facility, Research Complex at Harwell}, \orgname{Science and Technology Facilities Council}, \orgaddress{\street{Appleton Laboratory}, \city{Oxfordshire}, \postcode{OX11 0FA}, \state{England}, \country{United Kingdom}}}

\abstract{Detecting brief changes in time-series data remains a major challenge in fields where short-lived states carry important biological, physical or operational meaning. In single-molecule localisation microscopy, this problem is particularly acute as fluorescent molecules used to tag protein oligomers display heterogenous photophysical behaviour that can complicate photobleach step analysis; a key step in resolving nanoscale protein organisation. Existing methods to perform such analyses often require extensive filtering or prior calibration, and can fail to accurately account for blinking or reversible dark states which have the potential to contaminate downstream analysis processes. In this paper, an extension to RJMCMC is proposed for change point detection problems with heterogeneous temporal dynamics. This approach is applied to the problem of estimating per-frame active fluorophore counts from one-dimensional integrated intensity traces derived from Fluorescence Localisation Imaging with Photobleaching (FLImP), where compound change point pair moves are introduced to better account for short-lived events known as blinking and dark states. The approach is validated using both simulated and experimental data, demonstrating improved accuracy and robustness when compared with the current state-of-the-art photobleach step analysis methods and with the existing analysis approach for FLImP data. This Compound Reversible jump Markov chain Monte Carlo (CRJMCMC) algorithm requires no prior calibration from labelled data and minimal user input. It performs reliably across a wide range of fluorophore counts and signal-to-noise conditions, with signal-to-noise ratio (SNR) down to 0.001 and counts as high as nineteen fluorophores, while also effectively estimating low counts typically observed in FLImP when studying EGFR oligomerisation. Beyond single molecule imaging, this work has applications for a variety of time series change point detection problems with heterogeneous state persistence. For example, electrocorticography brain-state segmentation, fault detection in industrial process monitoring, realised volatility in financial time series, speech segmentation and environmental sensor monitoring.}


\maketitle

\section{Introduction}\label{Sec:Intro}
Identifying change points in time-series data is fundamental to understanding dynamic systems in a variety of applications from financial markets to healthcare monitoring \cite{aminikhanghahi2017survey}. Reversible jump Markov chain Monte Carlo (RJMCMC) offers a principled solution to such problems, enabling trans-dimensional estimation of the number and location of change points, and so is capable of modelling discrete events embedded in continuous, noisy data. However, when temporal dynamics are heterogeneous within the model, short-lived level changes are often missed or mistakenly considered as noise or outliers, despite the potential to mark biologically, physically, or operationally meaningful events \cite{WyseFrielRue2010,gyarmati2011modelling, lu2018intelligent,CasiniPerron2021, girgaonkar2024mitigation}.

Standard RJMCMC approaches often fail to account for short-lived events accurately for two main reasons. 1) Short-lived events are weakly penalised in reversible jump Markov chain Monte Carlo, as their brief duration contributes little to the overall likelihood, highlighting a broader challenge in capturing heterogeneous temporal dynamics. 2) Modelling these short-lived events typically requires the addition of two closely spaced change points; one to enter and one to exit the state. The first proposed change point often introduces a temporary mismatch with the observed data, reducing the model likelihood and, in the case of RJMCMC, resulting in a low acceptance probability. This makes addition of both change points unlikely unless an exhaustive search of possible configurations is performed. As a result, chains can exhibit poor convergence, with short-lived states inconsistently identified \cite{WyseFrielRue2010}. In cases where multiple change points can be added simultaneously \cite{ROTONDI2002633}, the aforementioned problems persist as acceptance is unlikely unless both change points are simultaneously correctly placed, and so samplers remain inefficient in the presence of short-lived events, as provisions are not made to explore closely-spaced changepoints.

One field where this issue is particularly pronounced is photobleach step analysis in single molecule localisation microscopy. In photobleach step analysis, fluorescently tagged protein subunits are imaged over time. The small separations between these sub-units (as low as 5nm to 50nm) lie well below the diffraction limit of conventional light microscopy ($\sim$ 250nm) and so the fluorophore point spread functions overlap appearing as a single point spread function, from which the individual fluorophores cannot be resolved. Discrete changes in one-dimensional fluorescence intensity traces are therefore used to count the number of active fluorophores per-frame. In this field, data naturally takes the form of step functions, corresponding to discrete fluorophore counts. These counts can be used to estimate the oligomeric distribution of protein populations in a sample, or, when considering the frame-wise active fluorophore count, to determine nanoscale protein separations and study protein oligomerisation. However, the reliability of each application depends directly on the accuracy of the initial fluorophore counting \cite{Hummert2021,chen2014molecular,Arant2014,Blunck2021,ChenMuller2007,leake2006stoichiometry,Singh2020,Verdaasdonk2014,ZhangGuo2014,Ulbrich2012}.

\begin{figure}[h!]
    \centering
    \includegraphics[width=\linewidth]{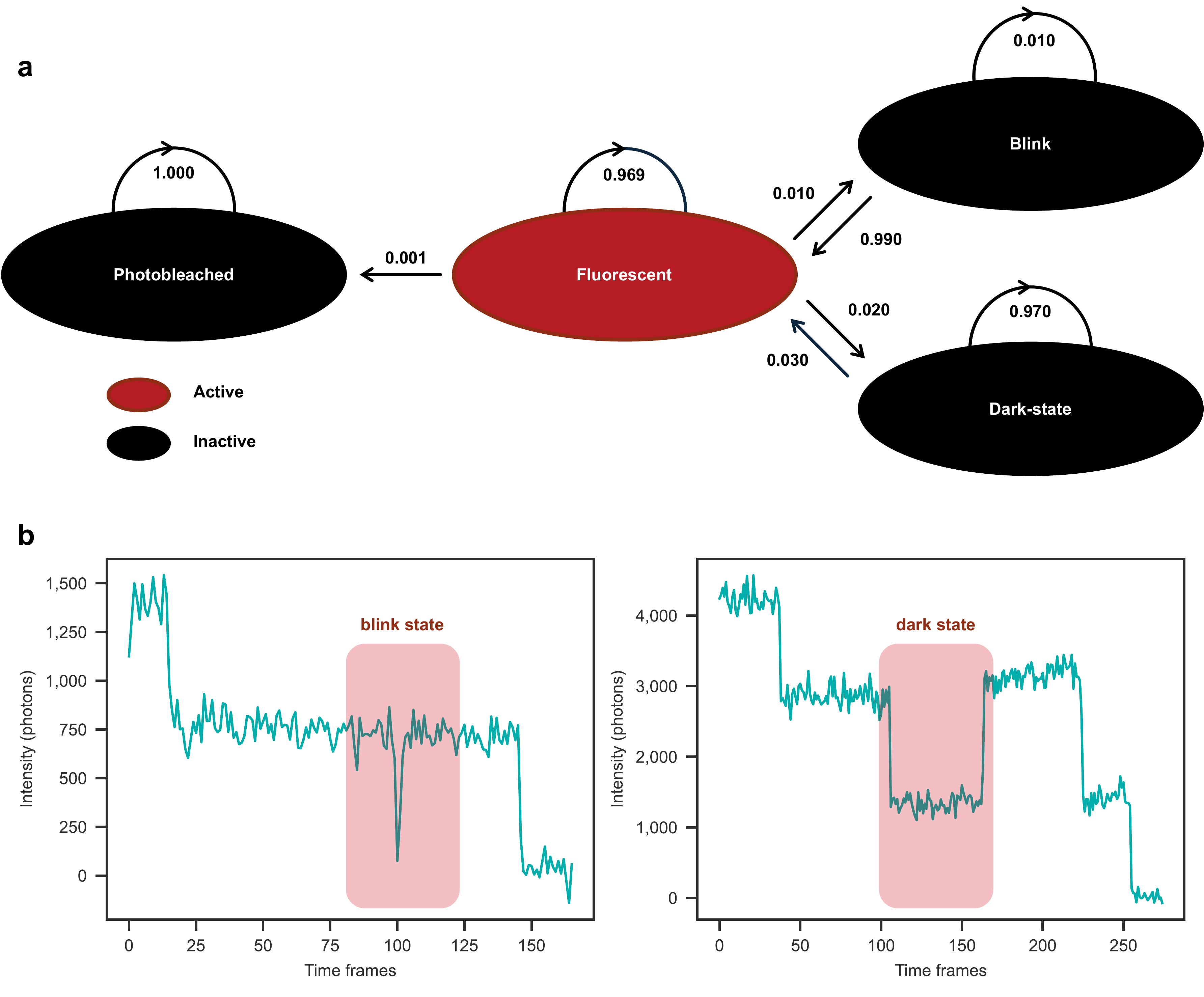}
    \caption{\textbf{Visualisation of complex photophysical behaviour of fluorophores.} (a) Markov chain describing fluorophore state transitions. Fluorophores typically exist in one of four states: bright (fluorescent), blinking (short-term dark), long-lived dark, or photobleached, and they transition between these as indicated by the arrows. Bright fluorophores emit detectable photons; blinking and dark states involve temporary loss of fluorescence, and photobleaching is a permanent transition to an off-state. Probabilities presented here have been obtained from \cite{needham2013measuring}. Note that there can be multiple dark states, but for the purposes of this study, this simplified model is implemented based on Alexa Fluor 488 fluorophores \cite{iyer2024drug}. (b) Examples of blink and dark states. Blink and dark states are reversible events which produce temporary drops in fluorescence that are visually indistinguishable from the irreversible photobleaching. Blink states are very short, typically only lasting 1-2 frames, whereas dark states are longer lived, with dwell time depending on the type of fluorophore used.}
    \label{fig:MarkovChainShortState}
\end{figure}

In photobleach step analysis, fluorophores most often reside in an active, fluorescent state and eventually move into an irreversible, inactive, photobleached state. However, fluorophores can also display complex photophysical behaviour, including short-lived ‘off’ states known as blink states and longer-lived dark states \cite{aspelmeier2015modern}, as visualised in Fig. \ref{fig:MarkovChainShortState}. These temporary changes in fluorescence can lead to incorrect estimation of per-frame active fluorophore counts, or may force partial or complete exclusion of traces, particularly when such deviations are misclassified as noise or considered analytically intractable. This issue is especially pronounced during separation localisation, where accurate per-frame fluorophore counts are critical, and the misattribution of short lived states can compromise subsequent positional measurements \citep{lee2017unraveling, rollins2015stochastic, chen2014molecular, iyer2024drug}.

Several approaches to fluorophore counting in photobleach step analysis have been proposed. However, many existing methods assume intensity is monotonically decaying and so cannot accommodate blink or dark states \citep{lee2017unraveling, garry2020bayesian}. Those that do support reversibility often suffer from high computational demands \citep{bryan2022diffraction} or limited scalability \citep{tsekouras2016novel}. Certain approaches address the unknown number of fluorophores, and consequently the unknown number of changepoints, by overestimating this number and representing the presence or absence of each with a binary indicator \cite{wallgren2019bayesian,bryan2022diffraction,MATTAMIRA20253161}. This formulation allows for changepoints to be identified but introduces inefficiencies in the estimation process. Correctly identifying related intensity parameters, namely single-fluorophore and background intensity, is also essential, yet existing methods often require prior user knowledge \citep{bryan2022diffraction}, fix intensity parameters at the beginning of analysis \citep{tsekouras2016novel}, or rely on calibration from labelled traces \citep{garry2020bayesian}, which are often not available in this field. RJMCMC has been applied to several problems in fluorescence microscopy and related imaging contexts \citep{soubies20143d, fazel2020bayesian}, however, these applications have largely focused on spatial reconstruction problems and do not address one-dimensional integrated intensity traces.

In this paper, an RJMCMC sampler for multiple change point analysis is proposed to estimate per-frame active fluorophore counts in photobleach step analysis, with a focus on short-lived blink and dark states. Compound change point moves are introduced to add or remove pairs of change points based on prior knowledge of fluorophore transitions and dwell times. These moves avoid the low-likelihood intermediate state in single change point approaches, and allow pairs of change points to be placed sufficiently close together with the guidance of an informed proposal distribution, without requiring filtering or trace exclusion. In addition, intensity parameters are initialised using both population- and trace-level information and are updated during analysis, reducing reliance on user-specified inputs and labelled calibration datasets. To demonstrate the approach, the Compound RJMCMC (CRJMCMC) method was developed for use with data derived from Fluorescence Localisation Imaging with Photobleaching (FLImP) \cite{needham2013measuring}. FLImP uses photobleach step analysis to resolve nanometre-scale separations within fluorescently labelled membrane protein oligomers, such as the Epidermal Growth Factor Receptor (EGFR), to study changes in oligomerisation following mutations in their DNA; a process linked to the development of cancers such as non-small cell lung cancer. Once per-frame active fluorophore counts have been obtained, this information can be combined with knowledge of the shape of fluorophore point spread functions to enable spatial localisation, and so accurate per-frame counts are essential \cite{iyer2024drug, needham2016egfr}. More broadly, the CRJMCMC algorithm provides a generalisable approach for detecting brief, meaningful signal changes in time-series data, with potential applications across biological, clinical, and engineering domains.

This paper is structured as follows. Section \ref{Sec:Results} validates the CRJMCMC method using both simulated and experimental FLImP data, comparing performance against leading alternatives and the existing analysis approach for FLImP data. Section \ref{Sec:Discussion} presents a discussion of the broader implications and potential extensions of the method. The methods are outlined in Section \ref{sec:Methods}; Section \ref{Sec:Multiplechange pointModel} introduces the multiple change point model developed for photobleach step analysis. Section \ref{Sec:reversible jump MCMC} outlines RJMCMC for change point detection in photobleach step analysis. Section \ref{Sec:Compoundreversible jump MCMC} incorporates the compound change point move designed to directly model short-lived dark states. Section \ref{Sec:FluoroAndBackground} details the Gibbs sampler for the intensity parameters, and Sections \ref{sec:Simulation} and \ref{sec:dnarulers} describe the simulation of traces and the collection of experimental data, respectively, for validation purposes. 

\section{Results}\label{Sec:Results}
The CRJMCMC algorithm is validated on both simulated and experimental datasets where 18,600 datasets were used to benchmark the CRJMCMC algorithm against state-of-the-art alternatives described in Bryan IV et al. (2022) \cite{bryan2022diffraction}, Garry et al. (2020) \cite{garry2020bayesian}, and Tsekouras et al. (2016) \cite{tsekouras2016novel}. The simulation study was designed to span a range of conditions by varying SNR, fluorophore number, single fluorophore photon count, and the frequencies and durations of blink and dark states. Default values included photon counts ranging from 500 to 2,000 per fluorophore, an SNR between 0.01 and 1, and fluorophore counts ranging from one to four. Emphasis was placed on lower fluorophore counts that are most commonly observed in FLImP experiments, with typical scenarios involving two to four fluorophores per complex, extended here to one to four, to include the study of monomer traces. This focus reflects the prevalence of small oligomeric states in proteins such as the EGFR, where dimers and tetramers dominate under physiological conditions \citep{needham2016egfr}. Each condition was then extended beyond typical experimental values to assess robustness under more extreme scenarios. The full simulation study is provided in the Supplementary Information (S11).

Experimental validation was performed using FLImP traces acquired from GATTAquant DNA origami rulers \citep{gattaquant2024gattastorm}, which consist of DNA rectangles containing up to four binding sites labelled with ATTO-647N fluorophores, and results were compared with those obtained from the existing analysis approach for FLImP data \citep{iyer2024drug} in the absence of ground truth.

\subsection{Comparison against other methods}
\begin{figure}[htbp]
    \centering
    \includegraphics[width=\linewidth]{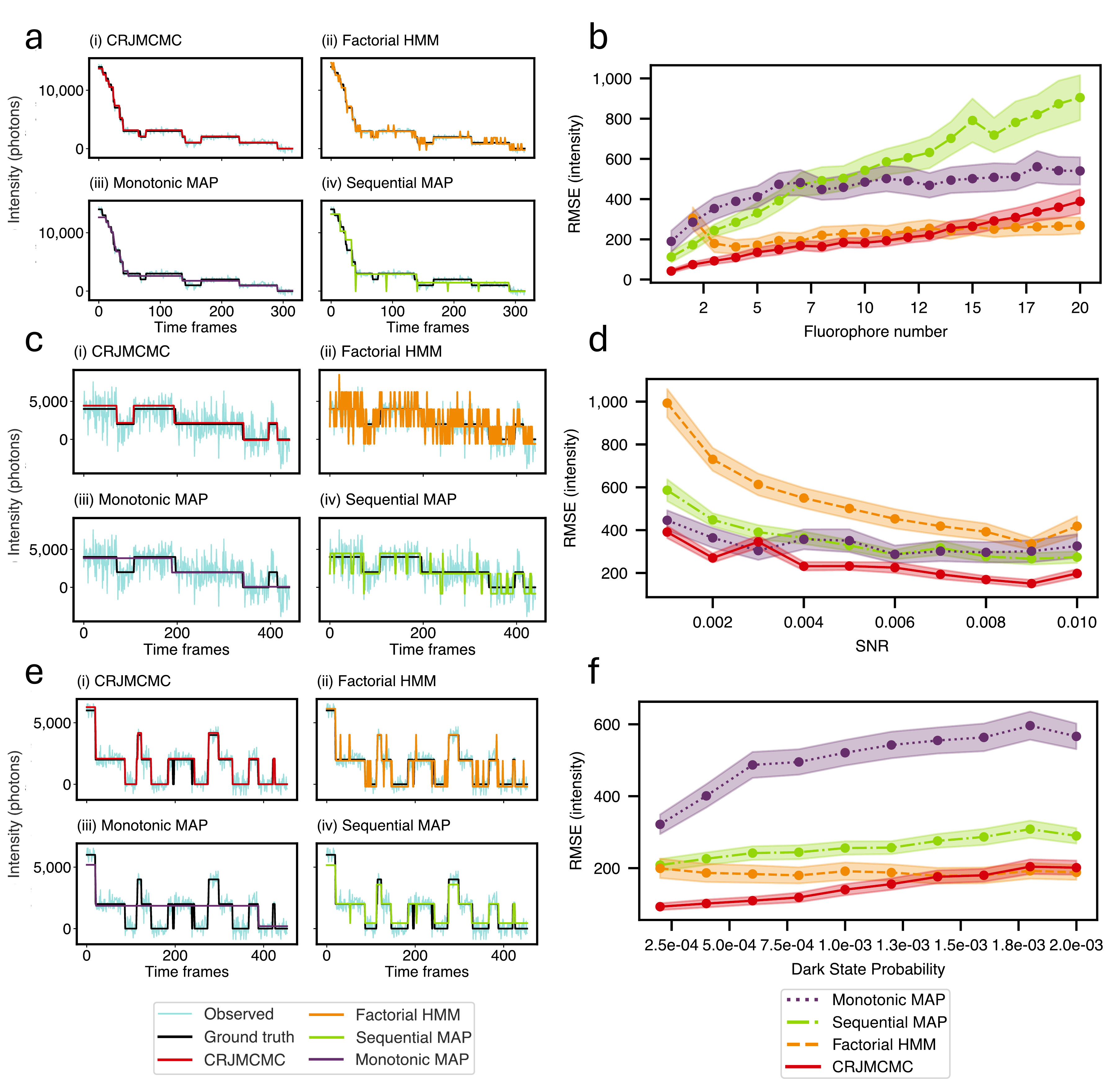}

\caption{\textbf{Performance of CRJMCMC across simulated integrated intensity traces.} CRJMCMC is compared with previously published methods by \cite{tsekouras2016novel} (green line), \cite{garry2020bayesian} (purple line), and \cite{bryan2022diffraction} (orange line). Panels \textbf{a}, \textbf{c}, and \textbf{e} show representative integrated intensity traces under increasing fluorophore number (\textbf{a}), decreasing SNR (\textbf{c}), and increasing frequency of short-lived states (\textbf{e}). Corresponding average root mean squared error (RMSE) values and 95\% confidence intervals are summarised in panels \textbf{b}, \textbf{d}, and \textbf{f} across each scenario.}
    \label{fig:SimulatedData}
\end{figure}

The performance of CRJMCMC is evaluated against three alternative approaches; the sequential MAP change point method from Tsekouras et al. (2016) \cite{tsekouras2016novel}; the monotonic decay MAP change point method from \cite{garry2020bayesian}, calibrated using labelled single-fluorophore traces; and the factorial hidden Markov model (HMM) implemented via Markov chain Monte Carlo (MCMC) as described in Bryan IV et al. (2022) \cite{bryan2022diffraction}, using the same initial estimates for the mean single-fluorophore intensity as calculated in CRJMCMC\footnote{The factorial HMM–MCMC method was applied to each dataset separately, rather than in parallel as multiple ROIs, as the traces are not constrained to be of equal length and therefore do not satisfy this requirement of \cite{bryan2022diffraction}.}\footnote{It is important to note that, at low fluorophore counts, the factorial HMM-MCMC approach frequently produced flat traces corresponding to zero fluorophores. These traces have been filtered from the analysis to avoid skewing the comparison, however this resulted in the exclusion of approximately 50\% of all simulated traces, including 98\% of one-fluorophore traces, 89\% of two-fluorophore traces, and 37\% of three-fluorophore traces. Given this high exclusion rate, the factorial HMM-MCMC method may be unsuitable for determining per-frame fluorophore counts in smaller oligomeric states, such as those encountered in the study of EGFR oligomerisation with imaging techniques such as FLImP.}.

CRJMCMC was implemented using three independent parallel chains for each dataset. Each chain was initialised with a unique seed and run for 20,000 iterations, with half of the iterations discarded as burn in. Convergence was assessed by considering pairs of chains and calculating the potential scale reduction factor for all parameters, using a threshold of $1.2$. If convergence was not achieved, a further 10,000 iterations were performed until two chains converged, up to a user defined maximum number of iterations. A detailed description of this convergence criteria and an analysis of convergence and other MCMC diagnostics can be found in the Supplementary Information (S8, S10). Estimation of intensity hyperparameters, such as single fluorophore and background intensity, was carried out by grouping all datasets with the same mean and SNR, in line with the assumption in photobleach step analysis that fluorophores within a single experiment possess similar photophysical properties\footnote{As a result of this, the calibration carried out on the monotonic MAP algorithm in Garry et al. (2020) \cite{garry2020bayesian} produces the same intensity parameters for all data in a pool.}. 

The performance metric, the root mean square error (RMSE) between the ground truth and estimated intensity traces was computed across all frames for each method, and the 95\% confidence intervals included for each. This metric was chosen to demonstrate performance over all parameters in each frame, rather than focusing solely on total fluorophore count. Additional performance metrics, such as per-frame active fluorophore accuracy and precision, are provided in the Supplementary Information (S11). Figure \ref{fig:SimulatedData} shows both the RMSE and representative example results from each method as fluorophore counts, SNR, and short-lived state frequency are varied. The graphs show intensity against time frames, alongside the ground truth and predicted intensity traces, based on the estimated mean single fluorophore intensity, $\mu_f$, the calculated number of active fluorophores in each frame, $n_i$, derived from estimated change point locations, and the estimated background, $\mu_b$.

As shown in Fig. \ref{fig:SimulatedData}(a) and \ref{fig:SimulatedData}(b), at higher fluorophore counts, the performance of both MAP-based methods deteriorates, whereas CRJMCMC perform comparably with the factorial HMM-MCMC approach, maintaining low RMSE up to nineteen fluorophores. The monotonic MAP and sequential MAP approaches tend to incorrectly estimate the mean fluorophore intensity, leading to overestimated counts and larger errors. Strict priors, controlled by the scaling factors on $\mu_f$ and $\mu_b$, ensure that CRJMCMC produces reliable estimates for these parameters, which guide convergence to the correct number of active fluorophores. At low fluorophore counts, the factorial HMM-MCMC sees an increase in RMSE and, for single-fluorophore traces, produces only flat traces corresponding to zero fluorophores in this region of the simulation study. Consequently, no results are obtained for such traces. This demonstrates its unsuitability for estimation in smaller structures. In contrast, the CRJMCMC approach maintains low RMSE under the same conditions.

Figures \ref{fig:SimulatedData}(c) and \ref{fig:SimulatedData}(d) show the results when SNR is varied, and it can be seen that CRJMCMC achieves the lowest RMSE, remaining robust down to an SNR of 0.001. The factorial HMM-MCMC showed substantial overfitting at low SNR, with excessive state transitions leading to poorer performance. In contrast, the CRJMCMC algorithm avoids overfitting by incorporating priors on the number of change points.

Figures \ref{fig:SimulatedData}(e) and \ref{fig:SimulatedData}(f) show the results of varying dark-state frequency, achieved by varying probability of transitioning from the active bright state into the dark state. Additional metrics for varying blink state frequency can be found in the Supplementary Information (S11). Across all tested probabilities, CRJMCMC achieved low RMSE, performing comparably with the factorial HMM-MCMC approach at higher frequencies. As expected, the monotonic decay MAP method performed poorly as frequency increased, due to its inability to detect reversible short-lived off states.

In terms of computational cost, Table \ref{tab:times} shows the average run times across all simulations for each of the considered methods, where the CRJMCMC algorithm takes on average 89.31s ($\pm$ 1.01) to complete a single trace, and the monotonic MAP and sequential MAP change point algorithms take on average 10.90s and 15.85s. However, the improvement in performance provided by the CRJMCMC method outweighs this additional computational cost. Furthermore, CRJMCMC remains over 3 times faster than the factorial HMM-MCMC method described in \citet{bryan2022diffraction}, which involves sampling across an expansive latent state space, and which has been previously observed to incur a high computational cost of up to 900s per run \cite{MATTAMIRA20253161}. All analyses were performed on Dell PowerEdge R6525 compute nodes, and the CRJMCMC method was performed using CPU-based parallelism, thus further optimisation through including GPU acceleration is expected to lead to considerable gains in speed and scalability.

\begin{table} 
\centering \caption{Average run times and 95\% confidence intervals (in seconds) across considered methods.}
\begin{tabular}{lc}
\toprule\label{tab:times}
Method & Time (s) \\
\midrule
CRJMCMC & 89.31 ($\pm$ 1.01) \\
Factorial HMM-MCMC & 324.81 ($\pm$ 7.50)\\
Monotonic MAP & 10.90 ($\pm$ 0.35) \\
Sequential MAP & 15.84 ($\pm$ 0.23)\\
\bottomrule
\end{tabular}
\end{table}

\subsection{Results on experimental data}
\begin{figure}[htbp]
    \centering
    \includegraphics[width=\linewidth]{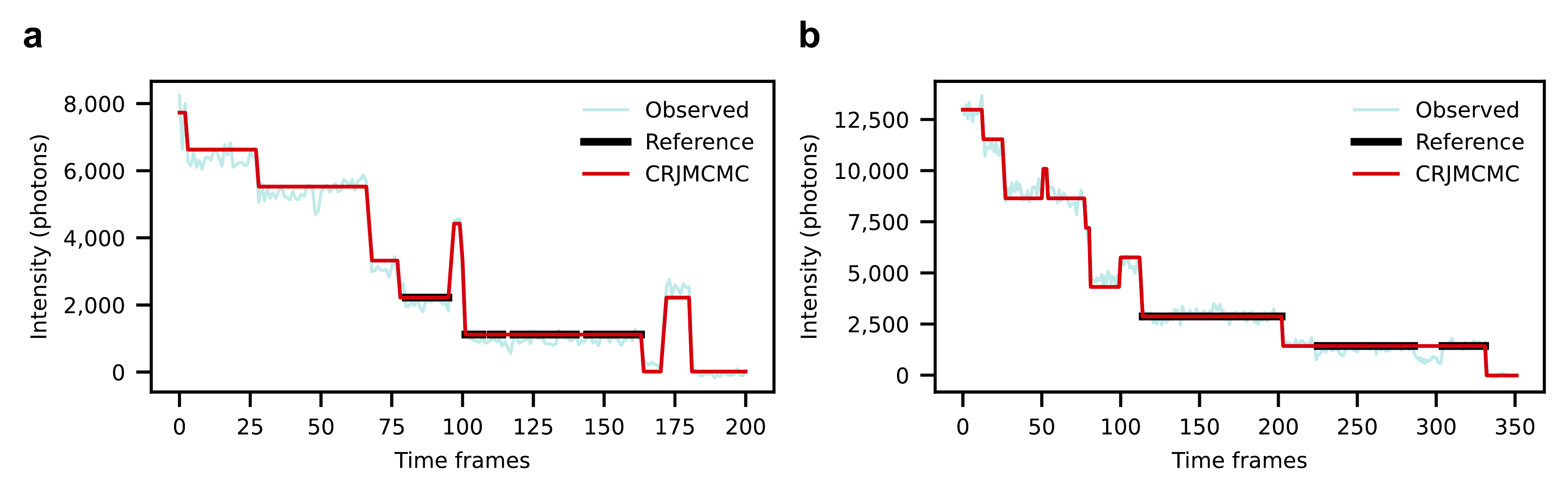}

    \caption{\textbf{Analysed intensity traces from GATTAquant DNA origami rulers}. (\textbf{a}) CRJMCMC (red) detects additional fluorophore levels exceeding those identified by \cite{iyer2024drug} (black). (\textbf{b}) CRJMCMC recovers state transitions excluded by the filtering in Iyer et al. (2024) \cite{iyer2024drug}.}
\label{fig:FLImPData}
\end{figure}

370 FLImP selected traces with their estimated fluorophore counts were used as ground truth for validation, where trace selection and fluorophore level identification were performed using the FLImP procedure described in Iyer et al. (2024) \cite{iyer2024drug}. Two examples from DNA origami ruler traces are shown in Fig. \ref{fig:FLImPData}, comparing results from the CRJMCMC algorithm to the labelled portions of the traces returned by the existing analysis for FLImP data described in Iyer et al. (2024) \cite{iyer2024drug}, used here as ground truth. To assess performance of these datasets, frame-wise accuracy, precision, and sensitivity were computed by comparing the estimated number of active fluorophores to the reference values provided by the FLImP track selection algorithm. True positives and true negatives represent frames in which the estimated and FLImP fluorophore counts match when they are greater than zero, or equal to zero, respectively, and false positives and false negatives correspond to over- or underestimation of fluorophore numbers, respectively\footnote{Specificity is omitted here as the FLImP algorithm currently does not store zero-level information and so true negatives are not able to be assessed.}.

As shown in Fig. \ref{fig:FLImPData} the CRJMCMC algorithm achieves high frame-wise accuracy relative to the method in Iyer et al. (2024) \cite{iyer2024drug}, even in complex scenarios such as staggered fluorophore state transitions or short-lived state transitions. In these cases, the method in Iyer et al. (2024) \cite{iyer2024drug} often discards substantial portions of the trace during filtering, whereas the CRJMCMC algorithm retains and models a larger fraction of the data, capturing approximately 30\% more frames. In addition, it can be seen that the compound reversible jump approach is able to identify additional levels, and thus, higher-order complexes than are currently captured by \cite{iyer2024drug}, suggesting the presence of additional spurious fluorophores. Table \ref{tab:FLImP_Metrics} shows accuracy, precision, and sensitivity of CRJMCMC estimates across FLImP-labelled portions of 370 integrated intensity traces with two, three, and four identified fluorophore levels. Owing to the absence of an independent ground truth, the method of \cite{iyer2024drug} was used as a reference for these metrics. As this reference is imperfect and is known to be less efficient and to exclude valid observations, the values in Table \ref{tab:FLImP_Metrics} underrepresent the full performance gains achieved by the CRJMCMC algorithm. Despite this, strong performance is observed across all tested fluorophore levels.

\begin{table}[]
    \centering

    \begin{tabular}{cccc}
\toprule
Fluorophore Level & 2 & 3 & 4 \\
\midrule
Accuracy 
& 0.901 ($\pm$ 0.036)
& 0.844 ($\pm$ 0.051) 
& 0.906 ($\pm$ 0.061) \\

Precision 
& 0.918 ($\pm$ 0.034) 
& 0.916 ($\pm$ 0.043)
& 0.982 ($\pm$ 0.020) \\

Sensitivity 
& 0.952 ($\pm$ 0.028) 
& 0.898 ($\pm$ 0.043)
& 0.923 ($\pm$ 0.059) \\

\bottomrule
\end{tabular}

    \caption{Summary of accuracy, precision, and sensitivity across traces with two, three, and four fluorophore levels identified, based on results from 370 FLImP-analysed tracks, with 95\% confidence intervals in brackets.}
    \label{tab:FLImP_Metrics}
\end{table}

\section{Discussion}\label{Sec:Discussion}
In this paper, an extended compound RJMCMC algorithm, termed CRJMCMC, was developed to enable robust estimation of change point locations in the presence of short-lived states. Central to this approach is the introduction of moves which propose or remove pairs of change points jointly to capture short-lived but meaningful changes. In the context of photobleach step analysis for the study of protein organisation at the nanoscale, using FLImP integrated intensity traces \citep{iyer2024drug}, this algorithm improves estimation of active fluorophore counts per-frame, reducing the need for prior calibration or heavy filtering.

The CRJMCMC method was benchmarked against several state-of-the-art alternatives in photobleach step analysis, including sequential MAP change point detection \citep{tsekouras2016novel}, MAP with a monotonic decay model \citep{garry2020bayesian}, and a factorial HMM-MCMC approach \citep{bryan2022diffraction}. Application to simulated integrated intensity traces showed improved performance across a range of fluorophore counts, SNR, and short-state transition frequencies. The CRJMCMC method remained robust up to nineteen fluorophores and retained high per-frame accuracy as SNR decreased as low as 0.001, where existing methods often overfit the number of fluorophore transitions. CRJMCMC notably outperformed methods at low fluorophore counts, demonstrating its suitability for the study of EGFR oligomerisation under physiological ligand concentrations, where smaller-order structures such as dimers and tetramers are most prevalent \cite{needham2016egfr}.

Further validation using DNA origami ruler data demonstrated that the CRJMCMC method recovered approximately 30\% more usable frames than the existing analysis approach for FLImP data \citep{iyer2024drug}, while maintaining performance under experimental variability. These gains were achieved without calibration traces, prior knowledge of fluorophore brightness, or extensive pre-filtering. More usable frames directly improves localisation precision by providing more data points from which spatial separations between fluorophores can be inferred.

In addition, the existing analysis approach for FLImP data supports only traces containing two, three, or four fluorophore levels \cite{iyer2024drug}. This CRJMCMC algorithm expands this range, enabling accurate estimation of higher-order fluorophore counts, and in single molecule localisation microscopy, each additional fluorophore doubles the amount of structural information recovered. This method thus improves data utilisation and increases the throughput of one-dimensional integrated intensity traces obtained from FLImP. The CRJMCMC method also enables analysis of single fluorophore traces, which are currently ignored in the existing analysis approach for FLImP data, as they do not provide fluorophore separation information \cite{iyer2024drug}. However, these traces still contain useful information about the contents of a sample and can be used to improve localisation precision in single molecule imaging over STORM, as shown through Resolution Enhancement by Sequential Imaging (RESI) described in \cite{reinhardt2023aangstrom}. A limitation of this approach arises, however, when multiple fluorophores are active in a single frame, a case where FLImP is particularly well-suited. This CRJMCMC algorithm thus presents an opportunity for seamless integration of FLImP and RESI, allowing increased precision over a wide range of fluorophore counts, with the inclusion of single fluorophore traces.

Automatic fluorophore count estimation from traces also enables more robust oligomer size determination across a population of traces from a single acquisition, a process that is typically performed manually. This capability is not currently provided by the existing analysis for FLImP data \cite{iyer2024drug}, yet is valuable in its own right, e.g. to inform diagnostic and therapeutic applications \cite{kulenkampff2021quantifying} or for understanding cell functionality \cite{danielli2020quantifying}, and as a preprocessing step to assess sample quality prior to downstream analysis, particularly when working with limited-access or expensive equipment.

The CRJMCMC algorithm provides a robust, automatic method for estimating change point locations and per-frame active fluorophore counts from complex integrated intensity traces. Key limitations of existing tools are addressed, including lack of consideration for short-lived dark states, reliance on calibration, extensive user input, and exclusion of informative data through heuristic filtering. CRJMCMC enables more complete and automated analysis of step-like time-series signals in photobleach step analysis, with RMSE over time-signal traces seeing on average a two-fold improvement compared to existing methods. More broadly, the introduction of compound change point moves provides an approach to better model short-lived states within RJMCMC to more precisely and more rapidly fit temporal data with heterogeneous change point distributions. With domain specific knowledge of processes underpinning the distribution of heterogeneous temporal dynamics, this method has broad applications including ECG cardiac arrhythmias \cite{hayter2021distinct} and ion channel recordings \cite{mortensen2007single} in biology, fault detection in engineering \cite{ethabet2025sensor}, traffic collision events \cite{voroneckaja2023automatic}, seismology \cite{trnkoczy2009understanding}, and speech signal segmentation \cite{protserov2022segmentation}.

\backmatter
\bmhead{Data and Code availability}
Code for trace simulation and implementation of CRJMCMC is available upon reasonable request, as are the experimental DNA origami datasets used in analysis.
\bmhead{Supplementary information}
Supplementary information is available, including tables detailing the full simulation study with additional metrics, MCMC diagnostics, sensitivity analyses, and further information to support implementation of CRJMCMC.

\bmhead{Acknowledgements}
This work was funded by a Northern Ireland Department for the Economy Collaborative Award in Science and Technology (DfE CAST) studentship, in collaboration with the Central Laser Facility OCTOPUS group at the Science and Technology Facilities Council (STFC). Special thanks are given to Dr Sarah Needham for the preparation of DNA origami samples for FLImP imaging. 

\section{Methods}\label{sec:Methods}
\subsection{Multiple change point model}\label{Sec:Multiplechange pointModel}
Suppose that there are $k$ change points at positions $s_1,\dots,s_k$, in time, where $s_0=0$ and $s_{k+1}=L$, the time of the final frame, so that $\mathbf{s}=(s_0,s_1,\dots,s_k,s_{k+1})$\footnote{Note that the time of the final frame, $L$, is distinct from the final frame, $N$, as it is possible change points occur at times in between frames.}. It is assumed that the number of change points, $k$, is Poisson distributed with mean $\lambda$, constrained to the range $1\leq k\leq k_{\max}$, where $k_{\max}$ is the maximum number of expected change points, chosen prior to analysis as a value large enough to allow full exploration of the change point parameter space.

Given $k$, the locations, $s_1,s_2\dots,s_k$ are distributed as the $k$ even-numbered order statistics from a sample of $2k+1$ points from the uniform distribution on $(s_0,s_{k+1})$ \cite{green1995reversible}. As described in \citep{green1995reversible} and \cite{benson2018adaptive}, this distribution minimises clustering of change points and penalises the appearance of intervals between change points where there are no data. The full derivation of the joint distribution of $s_1,s_2,\dots,s_k$ is described in the Supplementary Information (S1), and is given by
\begin{equation}
    f(s_1,s_2\dots,s_k) = ({2k+1})!\frac{1}{\left(s_{k+1}-s_0\right)^{2k+1}}\prod_{i=0}^{k}\left(s_{i+1}-s_{i}\right).
    \label{Eq:JointPDFEvenNumberedOrder}
\end{equation}

Considering now the photobleach step analysis problem, the observed intensity in each frame, denoted $y_i$ for $i \in \{1,\dots,N\}$, is modelled as a Gaussian random variable whose mean and variance depend on the number of active fluorophores, $n_i$ \citep{Lakowicz_2010, chen2014molecular, lee2017unraveling}. The Gaussian distribution is chosen as it provides a tractable approximation to the Poisson distribution of photon emission, particularly when dealing with high photon counts such as those encountered in photobleach step analysis, where the factorials involved in the Poisson distribution make calculations infeasible \citep{Lakowicz_2010}. The Gaussian distribution also allows for independent modelling of fluorophore intensity variance and background noise variance. Note that this approach could be extended to other noise models; however a Gaussian approximation to the noise is sufficient for the purposes of this analysis.

It is assumed that all fluorophores in a single trace contribute independently and identically; each emits the same average photon intensity, $\mu_f$, and variance, $\sigma^2_f$\footnote{Alexa Fluor 488 fluorophores used in FLImP display only a single bright state and so there is a single mean intensity $\mu_f$, rather than multiple bright states, as in Bryan IV et al. (2022) \cite{bryan2022diffraction}.}. Background noise is similarly assumed to be stationary, with constant mean, $\mu_b$, and variance, $\sigma^2_b$, throughout each trace. The intensity, $y_i$, is thus given by
\begin{equation*}
    y_i \sim N\big(\mu_i, \sigma^2_i\big), \quad \text{where} \quad \mu_i = \mu_f n_i + \mu_b \quad \text{and} \quad \sigma^2_i = \sigma^2_f n_i + \sigma^2_b,
\end{equation*}
where $n_i \in \mathbb{N}$ is the number of active fluorophores in frame $i$\footnote{Although fluorophores can transition over multiple frames, suggesting that fractional active counts may exist, here, the $n_i$ are fixed to non-negative integers, as interest lies in determining whether a fluorophore is detectably active or inactive. This binary classification is the more widely accepted definition of $n_i$, and it is the form of information that will prove useful in photobleach step analysis for both fluorophore counting and, in the case of FLImP, downstream localisation \citep{leake2006stoichiometry}.}. This intensity is piecewise constant in the dwelling, $j$, for $j = 1,\dots,k$, between two adjacent change points, $s_j$ and $s_{j+1}$, as the number of active fluorophores, $n_j$, is assumed constant in each dwelling, with change points representing the transition(s) between bright and dark states. The intensity level of each dwelling is therefore $\mu_f n_j + \mu_b$, with $n_j$ calculated to best fit the dwelling \citep{lee2017unraveling}\footnote{Note here that the number of active fluorophores is calculated rather than the height being randomly allocated using a weighted geometric mean, as is the case in Green et al. (1995) \cite{green1995reversible}, to fit within the constraint that $n_i$ are non-negative integers.}. In addition, it is not required that fluorophores all begin in an active state, but it is assumed that all fluorophores will have photobleached by the end of the trace, as this is part of the termination criteria in the FLImP image acquisition \cite{iyer2024drug}, and is widely accepted in the field of photobleach step analysis.

The mean single-fluorophore intensity, $\mu_f$,  and mean background intensity, $\mu_b$, are both assumed to be normally distributed with means $\eta_f$ and $\eta_b$ and variances $\nu_f$ and $\nu_b$, respectively, so that
\[ \mu_f \sim N (\eta_f,\nu_f) \quad \mu_b\sim N(\eta_b,\nu_b). \]

The variances of single-fluorophore and background intensity, $\sigma^2_f$ and $\sigma^2_b$, are both assumed to follow inverse-Gamma distributions, conjugate priors for the Gaussian distribution \cite{gelman2006prior}, with shape parameters $\alpha_f$ and $\alpha_b$, and scale parameters $\beta_f$ and $\beta_b$, respectively, so that
\begin{equation*}
    \sigma^2_f \sim \text{Inv-Gamma}(\alpha_f,\beta_f), \quad \sigma^2_b \sim \text{Inv-Gamma}(\alpha_b,\beta_b).
\end{equation*}

With all of the above parameters defined, the models are defined by the pair $(k,\boldsymbol{\theta}_k)$, where the vector of parameters to be estimated, $\boldsymbol{\theta}_k$, is given by \[\boldsymbol{\theta}_k=(\mathbf{s},\mu_f, \mu_b, \sigma^2_f, \sigma^2_b).\]

\subsection{Reversible jump for photobleach step analysis}\label{Sec:reversible jump MCMC}
Green (1995) \cite{green1995reversible} introduced RJMCMC analysis, presenting both the general framework and an application to one-dimensional multiple change point problems using the coal mining disasters dataset \cite{raftery1986bayesian}. RJMCMC extends the Metropolis–Hastings algorithm by enabling transitions between models whose parameter spaces are of different dimension while preserving detailed balance. Within a change point context, this corresponds to models with different numbers of change points.

Briefly, in Green (1995) \cite{green1995reversible}, at each iteration a move is randomly selected from four possible options: birth, death, shift, or a height change, with probabilities that depend on the current number of change points, $k$. Birth and death moves are dimension changing and respectively propose the addition or removal of a single change point, together with a transformation of the associated height parameters, such that detailed balance is preserved. Shift moves maintain the number of change points and are performed by selecting an existing change point at random and proposing a new location uniformly between the neighbouring change points. Height moves update the height of a randomly chosen segment between two change points, while keeping all locations fixed.

An adaptation of RJMCMC for photobleach step analysis is presented here, where the birth-death-shift structure developed by Green (1995) \cite{green1995reversible} is retained and the notation of Green (1995) \cite{green1995reversible} is followed throughout unless stated otherwise. As the number of active fluorophores, $n_i$, is calculated directly from the data, a move to modify the height of a randomly chosen change point is not required. Thus, this shift move probability, $\pi_k$, is simply defined as

\[\pi_k = 1- (b_k +d _k) \text{ for all } k=1,\dots,k_{\max},\]

where $b_k$ and $d_k$ are the probabilities of proposing birth and death moves respectively, both of which retain the definition in Green (1995) \cite{green1995reversible}, of

\begin{equation*}
    b_k = c\min\left(1,\frac{P(k+1)}{P(k)}\right)\quad \text{and} \quad d_{k+1} = c\min\left(1,\frac{P(k)}{P(k+1)}\right),
\end{equation*}
where $P(k)$ is the Poisson probability of there being $k$ change points with mean $\lambda$, and $c$ is the largest constant such that $b_k+d_k \leq 0.9 $ for all $ k=1,\dots,k_{\max}$. 

At each iteration, and for each possible change point configuration proposed, the heights between change points are determined by first calculating the number of active fluorophores in each dwelling, $n_j$, for $j=1,\dots,k$, and combining this with the current values for the mean fluorophore intensity and mean background intensity to get the height as $\mu_f n_j+\mu_b$. The number of fluorophores in each dwelling is calculated by determining the optimal $n_j$ so that $\mu_f n_j+\mu_b$ provides the best fit to the data in dwelling $j$, under the constraint that there is a genuine change at each $s_j$, i.e. $n_j \neq n_{j+1}$ for all $j=1,\dots,k$. A detailed description of the calculation of the number of active fluorophores in each dwelling is provided in the Supplementary Information (S3).

The shift move retains the structure described in Green (1995) \cite{green1995reversible}, but in this adaptation, the new change point location, $s_j^*$, is chosen randomly from a predefined custom proposal distribution, $q(s_j^*|s_j^*\in (s_{j-1},s_{j+1})$, rather than a uniform distribution. This proposal distribution is constructed in a preliminary analysis of the dataset by identifying regions where there is erratic behaviour in the intensity, and so is designed to improve the overall convergence of the sampler. A full description of the construction of this custom discrete distribution is provided in the Supplementary Information (S4)\footnote{A discretised distribution is chosen here for the change point locations to reflect the discrete time frames common in single molecule localisation microscopy, but note that the resolution can be refined to a user-required precision.}.

The adapted shift acceptance probability is then calculated as follows, with the new addition of the proposal ratio of location probabilities,
\begin{equation*}
    \alpha_{\text{shift}} = \min\left( 1,   \frac{(s_{j+1}-s_j^*)(s_{j}^*-s_{j-1})}{(s_{j+1}-s_j)(s_{j}-s_{j-1})} \times \text{likelihood ratio} \right).
\end{equation*}

 The likelihood ratio is determined using the Gaussian distribution over the data $\mathbf{y} = \{y_1,\dots,y_N\}$, and the number of active fluorophores for both $L(\mathbf{y}|k,\mathbb{\theta}_k^*)$ and $L(\mathbf{y}|k,\boldsymbol{\theta}_k)$ are calculated from $\mathbf{s}^*$ and $\mathbf{s}$, respectively.

The structure of the algorithm in Green (1995) \cite{green1995reversible} is retained for the birth move, with the following two main changes. 1) A new change point location, $s_j^*$, is proposed using the aforementioned custom distribution across all time frames. 2) Height is calculated by considering the best-fit number of active fluorophores, $n_j$, rather than randomly drawn. The prior ratio for the adapted birth move then becomes
\begin{equation*}
    \frac{P(k+1)}{P(k)} \times \frac{2(k+1)(2k+3)}{L^2} \times \frac{(s^* - s_j)(s_{j+1} - s^*)}{(s_{j+1} - s_j)},
\end{equation*}
with the new acceptance probability
\begin{equation*}
    \alpha_{\text{birth}}=\min\left(1,\,\,\text{prior ratio} \times \text{ likelihood ratio } \times \frac{d_k}{b_k q(s^*)(k+1)}\right),
\end{equation*}
where the uniform $1/L$ term in Green (1995) \cite{green1995reversible} is replaced with $q(s^*)$ from the custom proposal distribution. In addition, as no additional height parameter is introduced, no random variable, $u$ is required. The adapted reverse death move is simply performed by drawing one of the existing change points at random, and its acceptance probability is given by the reciprocal of the birth move with appropriate relabelling. Thus, detailed balance is therefore maintained solely through the proposed location, $s^*$, and the uniform selection of a change point to remove in the reverse death move. Consequently, the Jacobian reduces to one, simplifying the acceptance probabilities for both birth and death moves.

\subsection{Compound moves for short-lived states}\label{Sec:Compoundreversible jump MCMC}

Two compound change point moves are now introduced that simultaneously add or remove two change points, which together create a short-lived state, in a single iteration. A short-lived state is defined as a pair of change points, $s_{t_1}$ and $s_{t_2}$, which:
\begin{enumerate}
    \item Are on average a maximum duration, $\tau$, apart, such that $|s_{t_1}-s_{t_2}|\leq \tau$;
    
    \item Create a deviation from the current intensity level, rather than a monotonically increasing or decreasing pattern.
\end{enumerate}

For any two change points, the second requirement, that the pair creates a deviation, is simply determined by verifying that the number of active fluorophores before and after the change points are equal, but distinct from the number in between. However, the first requirement, that the two change points are sufficiently close together, is more difficult to verify without imposing a hard cut-off. To address this, a test analogous to the accept-reject step of a Metropolis-Hastings sampler is performed.

It is assumed that the duration between proposed change points is a random variable, denoted $D$, and follows an exponential distribution, $D \sim \exp(\lambda_D)$. The first requirement for a short-lived state is evaluated for an observed separation $d$ between change points under this model. To ensure that shorter durations, i.e. where $d < \tau$, are likely to be accepted, and longer durations are penalised while avoiding a hard threshold, the probability that $d$ is consistent with the assumed distribution of short-lived states is computed using the complement of the cumulative distribution function:
\[
\alpha_D = P(D > d) = \exp(-\lambda_D d),
\]
where the rate $\lambda_D$ is chosen so that $\alpha_D = p$ at a user-defined duration $\tau$, and for a user-defined probability $p$ which controls the rate of decay, where this $\tau$ is typically selected to represent the expected length of a short-lived state. This leads to:
\[
\lambda_D = -\frac{\log(p)}{\tau}, \quad \text{with} \quad p \in (0, 1).
\]

The probability $\alpha_D$ is then compared with a random draw $u \sim \text{Unif}(0, 1)$. If $u \leq \alpha_D$, the proposed pair of change points is accepted as satisfying the duration requirement of a short-lived state. Unlike hard thresholding, this approach introduces tunable parameters $p$ and $\tau$ to build-in prior knowledge while maintaining flexibility.

In order to add and remove short-lived states, a new parameter must be introduced which stores the number of change points associated with these states, denoted here as $k_t$. This parameter is a subset of $k$ so that $k_t\leq k$, and serves as an additional model indicator so that the models are now defined by $(k,k_t,\boldsymbol{\theta}_{k,k_t})$.

Assume, as before, that the number of change points, $k$, and the number of short-lived state change points, $k_t$, are both drawn from Poisson distributions with means $\lambda$ and $\lambda_t$, respectively, conditioned on  $k\leq k_{\max}$ and $k_t \leq k$.

Two new moves are defined which respectively add or remove short-lived states by adding or removing two change points simultaneously and thus have two new probabilities, $a_{k,k_t}$ and $r_{k,k_t}$, which represent the probability of proposing each of these moves. As proposed by \cite{ROTONDI2002633}, in order for detailed balance to hold in the case of short-lived state moves, the probabilities of proposing these moves must satisfy an analogous condition to that provided in Green (1995) \cite{green1995reversible}, 
\begin{equation}
    P(k)b_{k} = P(k+1)d_{k+1}.
    \label{Eq:GreenDetailedBalanced}
\end{equation}
For short-lived state moves this is given below as
\begin{equation}
    P(k)P_t(k_t)a_{k,k_t} = P(k+2)P_t(k_t+2)r_{k+2,k_t+2}
    \label{Eq:ShortStateDetailedBalance}
\end{equation}
It is assumed that $k$ and $k_t$ are independent, as $k_t$ can vary even when $k$ remains constant and vice versa; a change in one does not necessarily lead to a change in the other.
To ensure equation (\ref{Eq:ShortStateDetailedBalance}) holds, $a_{k,k_t}$ and $r_{k,k_t}$ are defined as
\begin{align*}
    a_{k,k_t} &= \gamma\min\left(1,\frac{P(k+2)P_t(k_t+2)}{P(k)P_t(k_t)}\right) \text{ and, }\\ r_{k+2,k_t+2} &= \gamma\min\left(1,\frac{P(k)P_t(k_t)}{P(k+2)P_t(k_t+2)}\right),
\end{align*}
where $\gamma$ is the largest constant such that $a_{k,k_t}+r_{k,k_t}\leq 0.1$ for all $k$ and all $k_t$. This definition ensures that equation (\ref{Eq:ShortStateDetailedBalance}) holds and so detailed balance is maintained. A full derivation of these definitions and this condition can be found in the Supplementary Information (S2).

To accommodate the two new proposal probabilities, the default maximum sum of $b_k + d_k$ is reduced from the value of $0.9$, defined in Green (1995) \cite{green1995reversible}, to $0.5$, and an additional constraint that $a_{k,k_t} + r_{k,k_t} \leq 0.1$ is introduced. Both of these thresholds are adjustable, depending on user requirements and the expected frequency of short-lived states, and in this case, were selected following sensitivity analysis detailed in the Supplementary Information (S9).

Finally, the probability of proposing a shift change point move, denoted $\pi_{k,k_t}$, now depends on both $k$ and $k_t$, and is defined so that
\begin{equation*}
    b_k+d_k+a_{k,k_t}+r_{k,k_t} + \pi_{k,k_t} = 1, \quad \text{for all } k \text{ and }
 k_t.
 \end{equation*}

 Adding two change points simultaneously is a two step process: drawing the location of the short-lived state, and drawing the distance between change points (i.e. the duration of the short-lived state). The locations of the change points are chosen by first drawing the centre of the short-lived state, denoted $\xi$, from the custom proposal distribution $q(0,L)$. The duration, $d$, between the two change points is then drawn from $\exp(\lambda_D)$. The locations of the new change points, denoted $s_{t_1}^*$ and $s_{t_2}^*$, are then defined as
\begin{equation}
    s_{t_1}^* = \xi - \frac{d}{2}, \quad \text{and} \quad s_{t_2}^* =\xi + \frac{d}{2}.
    \label{Eq:NewShortStatePos}
\end{equation}

Given the definition of a short-lived state, it is possible for multiple new short-lived states to emerge with non-zero probability. This outcome may occur when the newly introduced change points are located sufficiently close to existing change points and generate the appropriate pattern. Therefore, a check is performed at each stage to determine if any additional short-lived states have been created and determine the proposed value of $k_t^*$ accordingly.

To remove a short-lived state, two change points must be removed simultaneously, selected randomly to preserve dimension matching. The first change point, $s_{t_1}$, is drawn from the $k_t$ existing short-lived state change points. The number of valid second change points among the remaining $(k_t - 1)$; those that form a short-lived state with $s_{t_1}$, is denoted $\sum_\text{remove}$ and is computed using
\begin{equation}
    \sum_{\text{remove}}=\sum_{i=1}^{k_t-1}\mathbbm{1}\left(s_{t_1} \text{ and } s_{t_i} \text{ form a short-lived state}\right),
    \label{Eq:ProbChooseSecondTempDark}
\end{equation}
by checking all possibilities against the criteria for a short-lived state. The second change point is then chosen uniformly from these possibilities. In the add short-lived state move, the dimensionality increases from $k+6$ to $k+8$ with the addition of $s_{t_1}^*$ and $s_{t_2}^*$ and dimension matching is achieved by the inclusion of the two continuous random numbers, $\xi$ and $d$, and in the reverse direction, by drawing the locations of the two change points to be removed.

Suppose now without loss of generality, that $s_{t_1}$ and $s_{t_2}$ lie within the interval $(s_j,s_{j+1})$. The acceptance probability, $\alpha_{t}$, is therefore defined by
\begin{equation*}
    \alpha_{t} = \min\left( 1,  \text{prior ratio} \times \text{likelihood ratio} \times \text{proposal ratio} \right),
\end{equation*}
where the prior ratio is given by
\begin{equation*}
\text{prior ratio} = \frac{P(k+2)}{P(k)}\times \frac{P_t(k_t^*)}{P_t(k_t)} \times \frac{(2k+2)(2k+3)(2k+4)(2k+5)}{L^4} \times \frac{(s^*_{t_1}-s_{j})(s_{t_2}^*-s^*_{t_1})(s_{j}-s^*_{t_2})}{(s_{j+1}-s_{j})},
\end{equation*}
and the proposal ratio is defined by
\begin{equation*}
    \text{proposal ratio} = \frac{r_{k+2,k_t^*}}{a_{k,k_t}\left(\lambda_D e^{-\lambda_Dd}\right)q(\xi)k_t^*\sum_{\text{remove}}},
\end{equation*}
where $\left(\lambda_D e^{-\lambda_Dd}\right)$ is the probability of drawing the duration $d$, ${1}/{k_t^*}$ is the probability of choosing the first of the $k_t^*$ short-lived state change point to remove, and $1/\sum_{\text{remove}}$ is the probability of choosing the second to be removed. As in the single change point death move in Sec. \ref{Sec:reversible jump MCMC}, defining the remove short-lived state acceptance ratio is straightforward by simply taking the reciprocal of $\alpha_t$ with appropriate relabelling\footnote{The Jacobian in this case simplifies to one and therefore does not appear in the acceptance probability. To see this, consider only the parameters affected by the move and use the definitions of $s^*_{t_1}$ and $s^*_{t_2}$ from equation(\ref{Eq:NewShortStatePos}). The Jacobian is given by
\begin{equation*} 
    \frac{\partial(s^*_{t_1},s^*_{t_2})}{\partial(\xi,d)} =
    \left|\begin{array}{cc}
    \frac{\partial s^*_{t_1}}{\partial \xi} & \frac{\partial s^*_{t_1}}{\partial d} \\
    \frac{\partial s^*_{t_2}}{\partial \xi} & \frac{\partial s^*_{t_2}}{\partial d} 
    \end{array} \right| =
    \left|\begin{array}{cc}
    1 & -0.5\\
    1 & \,\,\,\,0.5
    \end{array} \right| = 1.
\end{equation*}
}.

Finally, short-lived state change points remain part of the full set of change points and can therefore be selected for removal in death moves, or relocation in shift moves, or created in a birth move. These moves may thus alter the number of short-lived states, and the acceptance probabilities must be adjusted accordingly. For example, a shift move may create or destroy a short-lived state depending on whether the new location of a change point satisfies the required proximity and pattern conditions. In such cases, the value of $k_t$ increases or decreases by one, without changing the overall dimensionality; only the labelling of change points as short-lived states is altered. The acceptance probabilities for birth, death, and shift moves are thus updated to include the ratio $P(k_t)/P(k_t^*)$ to account for these effects.

Hyperparameters $\lambda$, $\lambda_t$, and $\tau$ are dependent on the underlying system dynamics and were defined following sensitivity analysis within ranges reflective of the established behaviour from Alexa Fluor 488 fluorophores, with a maximum value of $k_{\max}=50$. These hyperparameters can all be adjusted accordingly to suit application, and the remaining tunable hyperparameters were selected based on the same structured sensitivity analysis, in which root mean square error on estimated intensity, accuracy, and precision were used as performance metrics. Full parameter ranges, acceptable subsets, and final chosen values are provided in the Supplementary Information (S9).

\subsection{Gibbs sampling for fluorophore and background intensity parameters}
\label{Sec:FluoroAndBackground}
At each iteration, following the chosen change point move, all intensity parameters, $\mu_f$, $\mu_b$, $\sigma_f^2$, and $\sigma_b^2$, are updated in a Gibbs sampling approach; updating the value of each parameter one by one while holding the others fixed, as is the case in \cite{bryan2022diffraction}.

The values of the hyperparameters $\eta_f, \nu_f^2, \eta_b, \nu_b^2, \alpha_f, \beta_f$ and $\alpha_b, \beta_b$ are linked to both the properties of the fluorophores and the experimental conditions of the imaging process. These can differ widely between experiments and are often difficult for non-technical users to determine or predict in order to set manually. It is therefore essential to produce reliable values for these intensity hyperparameters with minimal user input. This is addressed here by calculating these hyperparameters during a pre-processing step prior to analysis, detailed in the Supplementary Information (S5).

Intensity traces that exhibit high noise levels, frequent state transitions, or a large number of fluorophores are more likely to produce less reliable estimates for these hyperparameters. To address this issue, it is assumed, where appropriate, that experimental conditions remain constant either across an entire experimental setup or video, or within subsets of traces that share similar properties. Where sufficient data are available, the calculated values of the hyperparameters are pooled and a weighted average is then computed, where each trace is weighted in proportion with its noise level, so that traces with lower noise contribute more heavily to the estimate \citep{persson2013extracting}. Whether the noise be due to high background, high fluorophore count, or increased frequency of short-lived states, this pooling allows weaker or noisier intensity traces to be supported by information from more reliable traces.

\begin{figure}
    \centering
        \includegraphics[height =150mm]{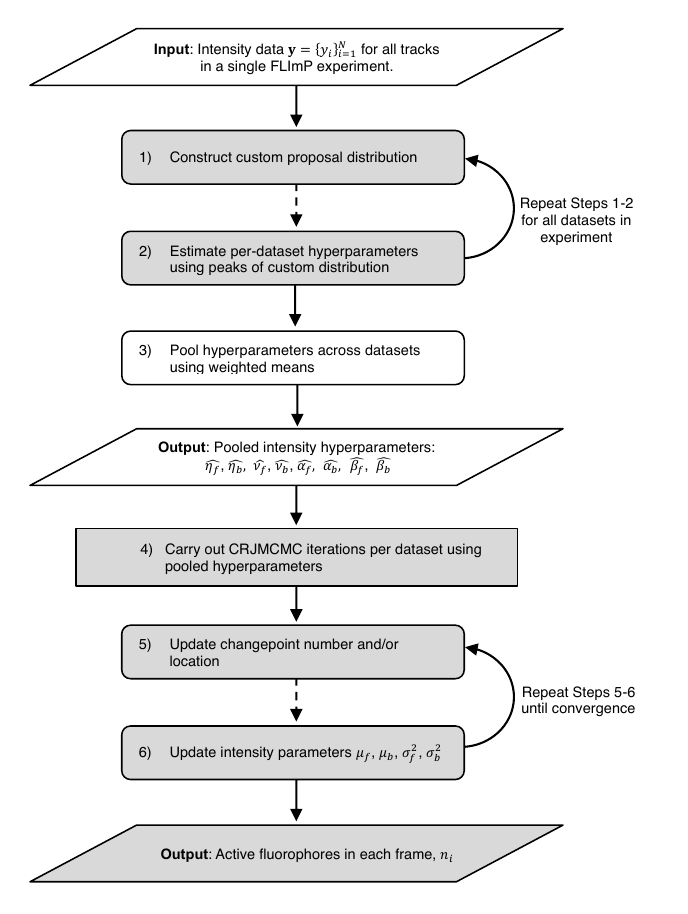}
    \caption{\textbf{Schematic of the pipeline used to estimate population-level hyperparameters and how it feeds into CRJMCMC iterations.} Hyperparameters governing the prior distributions are estimated during a pre-processing step and pooled across traces within each experiment to account for shared experimental conditions and mitigate the influence of noisy data. In CRJMCMC, at each iteration, a change point move is carried out, following by an update of the continuous parameters for fluorophore and background mean and variance ($\mu_f$, $\mu_b$, $\sigma_f^2$, $\sigma_b^2$) using Gibbs-sampling. \textbf{White}: Population-level, \textbf{Grey}: Dataset-level.}
    \label{fig:pre-process-pipeline}
\end{figure}

The entire sampling process, from estimating hyperparameters to carrying out the Gibbs sampling, is outlined in the diagram in Fig. \ref{fig:pre-process-pipeline}.

\subsection{Simulation}
Each fluorophore was modelled as a discrete time Markov process with fluorescent, dark, blink, and photobleached states. Default transition probabilities were defined using prior knowledge of the expected time spent in each state, and the full transition matrix for this is provided in the Supplementary Information (S7). Intensities were sampled from a Poisson distribution, and background noise was modelled as the sum of Poisson and Gaussian components with equal mean and variance. Following simulation, the mean background was subtracted to replicate the baseline correction applied during FLImP processing. Traces were truncated at a random time after all fluorophores photobleached to mirror the acquisition process used in FLImP experiments and frames were binned at a 20~$\mu$s resolution. A wide range of parameters were simulated to evaluate performance under different conditions. Within each parameter set, the following defaults were applied, reflecting typical ranges in FLImP: single fluorophore intensity levels of 500, 1,000, or 2,000 photons, fluorophore counts ranging from one to four, and background levels adjusted to achieve 0.01, 0.1, or 1.0 SNR. Ten replicates were generated for each parameter set, producing 18,600 traces in total. The full set of varied parameters, including blink and dark state properties can be found in the Supplementary Information (S7).
\label{sec:Simulation}

\subsection{Acquisition of DNA origami rulers}\label{sec:dnarulers}
Samples were imaged using an ONI Nanoimager S microscope equipped with a 1.49 NA oil immersion objective and operated using Nanoimager software using the previously described FLImP acquisition methodology \citep{iyer2024drug}. Briefly, the microscope was automatically focused and at each $(x,y)$ position in a 4mm square with 0.1mm step, a tile-scan a video acquisition was recorded with 20 ms exposure and Total Internal Reflection Fluorescence (TIRF) illumination. Additional details on sample labelling and tracking procedures are provided in Iyer et al. (2024) \cite{iyer2024drug}.

\bibliography{bibliography}

\begin{appendices}

\setcounter{section}{0}
\renewcommand{\thesection}{S}

\setcounter{subsection}{0}
\renewcommand{\thesubsection}{S\arabic{subsection}}

\section{Supplementary Information}\label{Sec:Appendix1}

\subsection{Even-numbered order statistics}\label{sec:even_order_stats}
It can be shown that, for a sample $ X_1, X_2, \ldots, X_n $ of independent and identically distributed random variables from a general continuous distribution, the joint probability density function of the order statistics is given by
\begin{equation}
    f_{X_{(1)}, \ldots, X_{(n)}}(x_1, \ldots, x_n) = n! \prod_{i=1}^{n} f_{X_i}(x_i),
    \label{eq:joint_pdf_orderstats}
\end{equation}
provided \( x_1 < x_2 < \cdots < x_n \), and that each \( x_i \) lies within the constraints of the original distribution from which the sample is drawn \citep{balakrishnan2014order}.

The change point locations, $s_1,\dots,s_k$, in both \cite{green1995reversible} and this compound change point adaptation, are chosen to be distributed as the even-numbered order statistics from a sample of $2k+1$ points, denoted $\{x_i\}_{i=1}^{2k+1}$, from the uniform distribution on $(s_0,s_{k+1})$, such that
\[s_1 = x_2,s_2 = x_4,\dots,s_k = x_{2k}.\]

To derive the joint probability distribution $f_{X_{(2)},\ldots,X_{(2k)}}(x_2,\ldots,x_{2k})$, denoted as $f(s_1,\dots,s_k)$, all the odd-numbered order statistics, $X_{(1)},X_{(3)},\ldots,X_{(2k+1)}$, must be integrated out to obtain
\[f(s_1,\dots,s_k) = \int_{-\infty}^{\infty}\dots\int_{-\infty}^{\infty}\int_{-\infty}^{\infty}f_{X_{(1)},X_{(3)}\ldots,X_{({2k+1})}}(x_1,\ldots,x_n)dx_1dx_3\dots dx_{2k+1}.\]
The original joint probability distribution function is constrained by $x_1<x_2<\cdots<x_{2k+1}$ and the $x_i$ are drawn from a uniform distribution on $(s_0,s_{k+1})$, meaning $s_0<x_1<x_2<\cdots<x_{2k+1}<s_{k+1}$. It follows that
\begin{gather*}
    s_0<x_1<x_2,\\
    x_2<x_3<x_4,\\
    \vdots\\
    x_{2k}<x_{2k+1}<s_{k+1}.
\end{gather*}
The limits of integration therefore become
\[f(s_1,\dots,s_k) = \int_{x_{2k}}^{s_{k+1}}\dots\int_{x_2}^{x_4}\int_{s_0}^{x_2}f_{X_{(1)},X_{(3)}\ldots,X_{({2k+1})}}(x_1,\ldots,x_{2k+1})dx_1dx_3\dots dx_{2k+1}.\]
Recall that $s_1 = x_2,s_2 = x_4,\dots,s_k = x_{2k}$, and so
\[f(s_1,\dots,s_k) = \int_{s_k}^{s_{k+1}}\dots\int_{s_1}^{s_2}\int_{s_0}^{s_1}f_{X_{(1)},X_{(3)}\ldots,X_{({2k+1})}}(x_1,\ldots,x_{2k+1})dx_1dx_3\dots dx_{2k+1}.\]
Substituting the expression for $f_{X_{(1)},\ldots,X_{({2k+1})}}(x_1,\ldots,x_{2k+1})$ from equation (\ref{eq:joint_pdf_orderstats}),
\[f(s_1,\dots,s_k) = \int_{s_k}^{s_{k+1}}\dots\int_{s_1}^{s_2}\int_{s_0}^{s_1}({2k+1})!f_{X_{(1)}},f_{X_{(3)}}\ldots,f_{X_{({2k+1})}}(x_{2k+1})dx_1dx_3\dots dx_{2k+1}.\]
As each $X_i$ is drawn from a uniform distribution on $(s_0,s_{k+1})$,
\[f_{X_i}(x_i)=\frac{1}{s_{k+1}-s_{0}} \quad \text{for all } i=1,\dots,(2k+1).\]
It follows that
\[f(s_1,\dots,s_k) = ({2k+1})!\frac{1}{\left(s_{k+1}-s_0\right)^{2k+1}}\int_{s_k}^{s_{k+1}}\dots\int_{s_1}^{s_2}\int_{s_0}^{s_1}dx_1dx_3\dots dx_{2k+1}.\]
Evaluating the integrals provides
\[f(s_1,\dots,s_k) = ({2k+1})!\frac{1}{\left(s_{k+1}-s_0\right)^{2k+1}}\left(s_{k+1}-s_{k}\right)\dots\left(s_{2}-s_{1}\right)\left(s_{1}-s_{0}\right),\]
Or more concisely,
\[f(s_1,\dots,s_k) = \frac{({2k+1})!}{\left(s_{k+1}-s_0\right)^{2k+1}}\prod_{i=0}^{k}\left(s_{i+1}-s_{i}\right),\]
as seen in Eq. (1) in Sec. 5.1.

In the case of a shift move in compound RJMCMC (CRJMCMC), where the location $s_j$ is proposed to shift to $s_j^*$, the prior ratio is constructed as
\[\frac{f(s_1,\dots s_{j-1},s^*_{j},s_{j+1},\dots,s_k)}{f(s_1,\dots s_{j-1},s_{j},s_{j+1},\dots,s_k)}.\]
Substituting in the expression for the joint distribution in each case gives, 
\[\frac{\displaystyle\frac{({2k+1})!}{\left(s_{k+1}-s_0\right)^{2k+1}}\left(s_{1}-s_{0}\right)\dots\left(s_{j}^*-s_{j-1}\right)\left(s_{j+1}-s_{j}^*\right)\dots\left(s_{k+1}-s_{k}\right)}  {\displaystyle\frac{({2k+1})!}{\left(s_{k+1}-s_0\right)^{2k+1}}\left(s_{1}-s_{0}\right)\dots\left(s_{j}-s_{j-1}\right)\left(s_{j+1}-s_{j}\right)\dots\left(s_{k+1}-s_{k}\right)}.\]
Therefore,
\[\frac{f(s_1,\dots s_{j-1},s^*_{j},s_{j+1},\dots,s_k)}{f(s_1,\dots s_{j-1},s_{j},s_{j+1},\dots,s_k)}=\frac{\left(s_{j}^*-s_{j-1}\right)\left(s_{j+1}-s_{j}^*\right)}{\left(s_{j}-s_{j-1}\right)\left(s_{j+1}-s_{j}\right)}.\]

Consider now the case of a birth move, where a new change point, $s^*$, has been added the number of change points has increased to $k+1$. Suppose, without loss of generality, that the new change point lies in the interval between $s_j$ and $s_{j+1}$. The joint distribution of the $k+1$ change points then becomes
\[f(s_1,\dots, s_j,s^*,s_{j+1},\dots,s_{k}) = \frac{(2k+3)!}{\left(s_{k+1}-s_0\right)^{2k+3}}\left(s_{1}-s_{0}\right)\dots\left(s^*-s_j\right)\left(s_{j+1}-s^*\right)\dots\left(s_{k+1}-s_{k}\right).\]
The prior ratio on change point locations for the birth move is therefore
\begin{multline*}
   \frac{f(s_1,\dots, s_j,s^*,s_{j+1},\dots,s_{k})}{f(s_1,\dots, s_j,s_{j+1},\dots,s_{k})} \\= \frac{\displaystyle\frac{(2k+3)!}{\left(s_{k+1}-s_0\right)^{2k+3}}\left(s_{1}-s_{0}\right)\dots\left(s^*-s_j\right)\left(s_{j+1}-s^*\right)\dots\left(s_{k+1}-s_{k}\right)}{\displaystyle\frac{(2k+1)!}{\left(s_{k+1}-s_0\right)^{2k+1}}\left(s_{1}-s_{0}\right)\dots\left(s_{j+1}-s_j\right)\dots\left(s_{k+1}-s_{k}\right)}. 
\end{multline*}
Simplifying this,
\[\frac{f(s_1,\dots, s_j,s^*,s_{j+1},\dots,s_{k})}{f(s_1,\dots, s_j,s_{j+1},\dots,s_{k})}
= \frac{(2k+3)(2k+2)\left(s^*-s_j\right)\left(s_{j+1}-s^*\right)}{\left(s_{k+1}-s_0\right)^{2}\left(s_{j+1}-s_j\right)}.\]
Recall that $s_{k+1}=L$ and $s_0=0$, and so
\[\frac{f(s_1,\dots, s_j,s^*,s_{j+1},\dots,s_{k})}{f(s_1,\dots, s_j,s_{j+1},\dots,s_{k})}
= \frac{(2k+3)(2k+2)}{L^2}\frac{\left(s^*-s_j\right)\left(s_{j+1}-s^*\right)}{\left(s_{j+1}-s_j\right)}.\]

Finally, consider the case of an add short-lived state move, where two consecutive change points, $s^*_1$ and $s^*_2$,  are added where $s^*_1<s^*_2$, and so the number of change points has increased from $k$ to $k+2$. Without loss of generality, suppose again these lie in the interval between $s_j$ and $s_{j+1}$. The joint distribution of the proposed $k+2$ change points then becomes

\begin{multline*}
f(s_1,\dots, s_j,s^*_{t_1},s^*_{t_2},s_{j+1},\dots,s_{k}) \\ = \frac{(2k+5)!}{\left(s_{k+1}-s_0\right)^{2k+5}} \left(s_{1}-s_{0}\right)\dots\left(s^*_{t_1}-s_j\right)\left(s^*_{t_2}-s^*_{t_1}\right)\left(s_{j+1}-s^*_{t_2}\right) \dots \left(s_{k+1}-s_{k}\right).
\end{multline*}

The prior ratio for the add short-lived state move is therefore
\begin{align*}
&\frac{f(s_1,\dots, s_j,s^*_{t_1},s^*_{t_2},s_{j+1},\dots,s_{k})}{f(s_1,\dots, s_j,s_{j+1},\dots,s_{k})} \\ &= \frac{\displaystyle\frac{(2k+5)!}{\left(s_{k+1}-s_0\right)^{2k+5}} \left(s_{1}-s_{0}\right)\dots\left(s^*_{t_1}-s_j\right)\left(s^*_{t_2}-s^*_{t_1}\right)\left(s_{j+1}-s^*_{t_2}\right) \dots \left(s_{k+1}-s_{k}\right)}{\displaystyle\frac{(2k+1)!}{\left(s_{k+1}-s_0\right)^{2k+1}} \left(s_{1}-s_{0}\right)\dots\left(s_{j+1}-s_j\right) \dots \left(s_{k+1}-s_{k}\right)}.
\end{align*}
Simplifying,
\begin{align*}
&\frac{f(s_1,\dots, s_j,s^*_{t_1},s^*_{t_2},s_{j+1},\dots,s_{k})}{f(s_1,\dots, s_j,s_{j+1},\dots,s_{k})} \\ &= \frac{(2k+5)(2k+4)(2k+3)(2k+2)}{\left(s_{k+1}-s_0\right)^{4}}\times\frac{\left(s^*_{t_1}-s_j\right)\left(s^*_{t_2}-s^*_{t_1}\right)\left(s_{j+1}-s^*_{t_2}\right)}{ \left(s_{j+1}-s_j\right)}.
\end{align*}
Using again the fact that $s_{k+1}=L$ and $s_0=0$,
\begin{align*}
&\frac{f(s_1,\dots, s_j,s_{t_1}^*,s^*_{t_2},s_{j+1},\dots,s_{k})}{f(s_1,\dots, s_j,s_{j+1},\dots,s_{k})} \\ &= \frac{(2k+5)(2k+4)(2k+3)(2k+2)}{L^{4}}\times\frac{\left(s^*_{t_1}-s_j\right)\left(s^*_{t_2}-s^*_{t_1}\right)\left(s_{j+1}-s^*_{t_2}\right)}{ \left(s_{j+1}-s_j\right)}.
\end{align*}

\subsection{Proof of detailed balance for short-lived states} \label{sec:detailed_balance}

In order for detailed balance to hold in the case of short-lived state moves, the probabilities of proposing the add and remove short-lived states moves when there are $k$ change points and $k_t$ short-lived state change points, denoted $a_{k,k_t}$ and $r_{k,k_t}$ respectively, must satisfy an analogous condition to \[P(k)b_{k} = P(k+1)d_{k+1}, \text{ where }
    b_k = c\min\left(1,\frac{P(k+1)}{P(k)}\right)\quad \text{and} \quad d_{k+1} = c\min\left(1,\frac{P(k)}{P(k+1)}\right),
    \] for two change point moves, given below as
\begin{equation}
    P(k)P_t(k_t)a_{k,k_t} = P(k+2)P_t(k_t+2)r_{k+2,k_t+2}
    \label{Eq:AppShortStateDetailedBalance}
\end{equation}
where $P_t(k_t)$ is the Poisson probability of $k_t$ short-lived state change points with mean $\lambda_t$. 
To ensure equation this holds, define the probabilities $a_{k,k_t}$ and $r_{k,k_t}$ as follows
\begin{equation*}
    a_{k,k_t} = \gamma\min\left(1,\frac{P(k+2)P_t(k_t+2)}{P(k)P_t(k_t)}\right)\quad \text{and} \quad r_{k+2,k_t+2} = \gamma\min\left(1,\frac{P(k)P_t(k_t)}{P(k+2)P_t(k_t+2)}\right),
\end{equation*}
where $\gamma$ is the largest constant such that $a_{k,k_t}+r_{k,k_t}\leq 0.1$ for all $k$ and all $k_t$. This definition ensures that equation (\ref{Eq:AppShortStateDetailedBalance}) holds and so detailed balance is maintained. To see this, consider, for example, first the case where $P(k+2)P_t(k_t+2)>P(k)P_t(k_t)$. The probabilities $a_{k,k_t}$ and $r_{k+2,k_t+2}$ then become
\begin{equation*}
    a_{k,k_t} = \gamma\times1,\quad \text{and} \quad r_{k+2,k_t+2} = \gamma\times\frac{P(k)P_t(k_t)}{P(k+2)P_t(k_t+2)}.
\end{equation*}
Similarly, in the case where $P(k+2)P_t(k_t+2)<P(k)P_t(k_t)$, it follows that 
\begin{equation*}
    a_{k,k_t} = \gamma\times\frac{P(k+2)P_t(k_t+2)}{P(k)P_t(k_t)},\quad \text{and} \quad r_{k+2,k_t+2} = \gamma\times1,
\end{equation*}
both of which satisfy equation (\ref{Eq:AppShortStateDetailedBalance}).

\subsection{Calculating the number of active fluorophores}
\label{sec:calc_fluorophores}
The number of active fluorophores in each frame is determined by calculating the mean intensity, $\bar{y_j}$, from the data within each section, $j$, between change points $s_j$ and $s_{j+1}$, for $j=0,\dots,k$, and finding the number of active fluorophores, $n_j$, which minimises the difference
\[\left|\bar{y}-\left(\mu_f n_j+\mu_b\right)\right|\]
such that $n_j \in \mathbb{N}$. It is assumed that the number of active fluorophores remains constant within each section, with $n_j$ representing all frames in section $j$. As $n_j$ represents a discrete count of fluorophores, it can only take non-negative integer values \citep{leake2006stoichiometry}. An issue that may arise is that the same `best fit' value of $n_j$ is calculated for consecutive sections, indicating no change in the underlying model has occurred, even when a change point is present. This issue can result in an overestimation of the number of change points, many of which would be redundant. To address this issue, a forced perturbation of $n_j$ is applied at each change point. In each section, $j$, the mean intensity is calculated, and the corresponding number of fluorophores, $n_j$, is determined, where it is noted that this process is carried out from the end of the trace to the beginning to prioritise sections of the data where there are less fluorophores, and so less variability, expected. If $n_j=n_{j-1}$, the number of active fluorophores in the current section, $n_j$, is either increased or decreased by one, depending on which option minimises $\left|\bar{y_j}-\left(\mu_f n_j+\mu_b\right)\right|$:
\[n_j = \begin{cases}
    n_j+1, &\text{if } \left|\bar{y_j}-\big(\mu_f (n_j+1)+\mu_b\big) \right| < \left|\bar{y_j}-\big(\mu_f (n_j-1)+\mu_b\big) \right|\\
    n_j-1, & \text{otherwise}
\end{cases}\]
\subsection{Custom proposal distribution}\label{sec:custom_proposal_dist}
A custom proposal distribution is designed to improve the overall convergence of change point moves in CRJMCMC and is constructed by performing a preliminary analysis of the data and identifying regions where there is erratic behaviour in the intensity. This process begins by estimating the SNR of the intensity profile using the mean intensity of the entire dataset, divided by its standard deviation. The dataset is then divided into windows, and the $z$-score of the difference in mean intensity between windows is determined, assuming a mean of zero and a variance given by the reciprocal of the signal-to-noise ratio. The use of data windows, rather than individual differences ensures that small, but significant, changes (e.g., photobleaching or dark states) are not smoothed out or merged with nearby minor fluctuations, as can be the case with edge-preserving smoothing and linear filters \citep{lee2017unraveling}. The proposal distribution is then formed by adding Gaussian probability distribution functions to a uniform distribution over all times, with Gaussian means centred at the midpoint of each window and variances proportional to the corresponding window $z$-scores. The distribution is then normalised. For computational efficiency, the proposal distribution is discretised at a user-defined time resolution, defining the possible change point locations. To improve the efficiency further, the probabilities for each of these possible values are pre-calculated for look-up purposes when calculating acceptance probabilities.

\subsection{Determining intensity mean and variance hyperparameters}
The hyperparameters $\eta_f,\nu_f^2, \eta_b,\nu_b^2,\alpha_f,\beta_f$ and $\alpha_b,\beta_b$, are calculated in a pre-processing step using the peaks of the change point location proposal distribution, $q(0,L)$, which identifies candidate locations for change points. A filter is applied at this stage to remove candidate change points where the difference in mean intensity between sections is insufficient. This filter is guided by a heuristic lower bound on fluorophore intensity, which may either be set by the user or estimated from properties of the data, for example using the mode intensity scaled by a multiplier taken here to be 0.9. In practice, any sensible method for identifying a lower bound would be suitable. With increased data, the influence of the specific choice becomes negligible.

Using these candidate locations the mean single fluorophore intensity, $\eta_f$, is calculated as the mean intensity difference between sections at these estimated change points. The mean background intensity hyperparameter, $\eta_b$, is calculated as the mean intensity of the final section, under the assumption that all fluorophores have photobleached by the end of the trace and only background noise remains \citep{chen2014molecular,rollins2015stochastic}. 

The shape and scale hyperparameters for single-fluorophore variance, $\alpha_f,$ and $\beta_f$, are calculated using the assumption that fluorophore intensity follows a Poisson distribution, and so variance should equal mean intensity. Therefore, $\alpha_f$ is set equal to $\eta_f$, and to ensure the mode of the distribution is at $\alpha_f$, $\beta_f = \eta_f(\alpha_f+1)$. For background variance hyperparameters, $\alpha_b$ and $\beta_b$, $\alpha_b$ is set as the final section variance, and $\beta_b$ is fixed so that this final section variance is the mode of this distribution.

As discussed in Sec. 5.4, the calculated hyperparameter values are pooled across all traces within a single video or experimental setup. A weighted average is then calculated based on the noise levels within these traces, so that less noisy traces are weighted more heavily than noisy traces \citep{persson2013extracting}. For relatively homogenous trace populations these weights can be set as the inverse of trace variance. For populations with a large amount of heterogeneity, the weight can be chosen to be the inverse of the average window variance, maximum photon intensity, and length of trace, to manage variability due to increased fluorophore number, increased dark states, and background noise.

The variance hyperparameters $\nu_f^2$ and $\nu_b^2$ and the proposal distribution standard deviations for $\mu_f$ and $\mu_b$ are calculated proportionally to the weighted averages of $\eta_f$ and $\eta_b$, respectively using a scaling factor to ensure the priors on these values are strong.

\subsection{Alternative Methods}\label{sec:alt_methods}
\subsubsection{Tsekouras \textit{et al.} (2016)}

\citet{tsekouras2016novel} models active fluorophore count by considering step counts, $K$, and event locations, $\mathbf{s}=\{s_1,\dots,s_K\}$, with Gaussian likelihood for intensity with mean and variance $\mu_i$ and $\sigma_i^2$ in each frame $i$ given by
\[
\mu_i=n_i\mu_f+\mu_b, \quad \sigma^2_i=n_i\sigma^2_f+\sigma^2_b,
\]
where $\mu_f$ and $\sigma^2_f$ represent mean and variance of single fluorophore intensity, and $\mu_b$ and $\sigma^2_b$ represent mean and variance of background intensity, respectively, and $n_i$ is the number of active fluorophores in frame $i$. This approach allows both mean and variance to scale with the number of active fluorophores, but all intensity parameters are fixed prior to analysis rather than updated dynamically.  

The parameter set is defined as
\[
\boldsymbol{\theta}=(\mathbf{s},K,m,arr,\mu_f,\mu_b,\sigma^2_f,\sigma^2_b,\gamma),
\]
where $m$ is the number of events and $arr$ is the arrangement count, which stores the possible ways that $m$ events can combine to produce $K$ observed steps. In this approach, an event corresponds to a single fluorophore transition from active to inactive or vice versa, while a step corresponds to any discrete intensity change. Several events may therefore occur simultaneously and yet appear as a single step. The inclusion of $arr$ ensures that overlapping or even opposing events are explicitly accounted for and prevents bias toward models with large numbers of simultaneous events which could otherwise cancel one another out. In theory, this event–step approach is capable of describing reversible dark state transitions as well as multiple photobleaching events, although the method is demonstrated on monotonic intensity traces \cite{tsekouras2016novel}.  

A hierarchical prior is applied to prevent overfitting, with Poisson and exponential distributions used for $K$ and $m$. The negative log posterior is calculated for all candidate models, with the minimum chosen as the optimal solution. The process is deterministic and requires brute force evaluation of multiple models, which limits computational scalability.  

To mitigate computational cost, a three stage approach is used: (i) estimation of $\mu_f$, $\mu_b$, $\sigma^2_f$, and $\sigma^2_b$ per trace using an adaptation of the change point method in \cite{kalafut2008objective}, (ii) sequential exclusion of implausible models by dividing traces into windows and estimating the most likely step number via Gaussian likelihood maximisation, and (iii) minimisation of the negative log posterior within each window to determine the optimal solution. The method was applied to synthetic and experimental datasets and was shown to recover fluorophore counts even at low signal-to-noise, although it tends to misidentify events near window boundaries.  

This approach is in principle capable of describing short-lived blinking and dark states, and overlapping events, but its reliance on fixed intensity parameters makes results dependent on the quality of the trace and the deterministic optimisation limits scalability across experimental conditions.

\subsubsection{Garry \textit{et al.} (2020)}

\citet{garry2020bayesian} introduces an approach where the variable to be estimated is the number of active fluorophores in each frame $n_i$, with a Gaussian likelihood given by
\[
\mu_i = v n_i + a, \quad \sigma^2_i = \sigma_1^2 n_i + \sigma_0^2,
\]
where $v$ is the single fluorophore mean intensity, $a$ is the background intensity, $\sigma_1^2$ and $\sigma_0^2$ are the corresponding variances. The prior distribution on the number of fluorophores takes a binomial form
\[
\mathbb{P}(\{n_i\})=\prod_{i=1}^{N-1}\frac{n_{i-1}!}{n_i!(n_{i-1}-n_i)!}\,q^{n_i}(1-q)^{n_{i-1}-n_i},
\]
with survival probability $q=\exp(-\Delta t/\tau)$, where $\tau$ is the single fluorophore lifetime. This prior enforces monotonic decay in the intensity model by considering only photobleaching events.  

Maximum a posteriori estimates are obtained via golden section search over candidate $n_0$ values. The parameters $v$, $a$, $q$, $\sigma_0^2$, and $\sigma_1^2$ are fixed for a dataset, requiring calibration either from extensive simulations or from experimental traces containing known single fluorophores. This calibration step is essential for analysis but may not be feasible in all experimental settings, where the required information is not available.

The method was tested on synthetic and experimental photobleaching data and showed high computational efficiency. However, it assumes all fluorophores are bright at the beginning of the trace, requires calibration data, and cannot account for short-lived blinking or reversible dark state transitions.

\subsubsection{Bryan \textit{et al.} (2022)}
\label{sec:bryan}
\citet{bryan2022diffraction} proposes a fully Bayesian approach formulated as a factorial hidden Markov model (FHMM). Each fluorophore is represented by a hidden Markov chain with states active ($\sigma_A$), dark ($\sigma_D$), or photobleached ($\sigma_B$). The state of fluorophore $k$ in region $r$ at time $n$ is denoted $s_n^{k,r}$, with transitions described by a transition matrix $\boldsymbol{\pi}$, and initial states drawn from $\boldsymbol{\pi}_0=(\pi_{0,A},\pi_{0,D})$. This FHMM structure enables simultaneous modelling of both the number of fluorophores present and their individual photophysical trajectories.  

The mean intensity at time $n$ in region $r$ is given by
\[
\mu_n^r=\mu_B^r+\sum_{k=1}^{K_r}\mu_{s_{n-1}^{k,r}},
\]
where $\mu_B^r$ is background and $\mu_{s_{n-1}^{k,r}}$ equals $\mu_A$ for the active state and zero otherwise. To address the unknown fluorophore count, a Bayesian nonparametric scheme is employed by introducing $K \gg K_r$ candidate fluorophores, each with load
\[
b^{k,r}\sim \mathrm{Bernoulli}\bigg(\frac{\gamma}{K+\gamma+1}\bigg),
\]
that indicates whether the fluorophore contributes to the observed intensity. Fluorophores with $b^{k,r}=0$ are known as virtual and do not contribute to the overall intensity, but they must still be tracked during inference. This treatment allows the true fluorophore number to be estimated but the increased state space of the FHMM, which is the main source of computational inefficiency.  

The observed brightness is modelled as
\[
w_n^r \mid s_n^{1:K,r},b^{1:K,r},\mu_A^r,\mu_B^r \sim \text{Gamma}\Bigg(\frac{1}{2}\Big(\mu_B^r+\sum_{k=1}^K b^{k,r}\mu_{s_n^{k,r}}\Big),2G\Bigg),
\]
where $G$ accounts for camera noise characteristics.  

Posterior inference is performed using Gibbs sampling. Each parameter is sampled conditionally on the others, enabling estimation of both the underlying fluorophore count and their state trajectories, as well as background, transition probabilities, and fluorophore brightness. This provides both point estimates and credible intervals, quantifying uncertainty in the number of fluorophores and their dynamics.  

This approach explicitly models dark state transitions and allows intensity parameters to be updated during analysis, in contrast to earlier fixed parameter methods. It does not require all fluorophores to start in the bright state and remains accurate up to a reported 100 fluorophores. However, the FHMM structure, combined with the large number of virtual fluorophores, creates a very high dimensional posterior, making inference substantially slower than deterministic approaches, even though many of the variables do not contribute directly to the solution.

\subsection{Simulating intensity profiles}
\label{sec:simulating}
The behaviour of fluorophores in FLImP can each be described by a Markov chain which transitions between the active fluorescent `on' state, $\sigma_A$, where they are emitting photons, the possible transient `off' blink and dark states, denoted $\sigma_B$ and $\sigma_D$ respectively, and eventually moves into the absorbing photobleached state, $\sigma_P$. A sequence of states is simulated for each fluorophore from this underlying discrete-time Markov chain with state space $S=\{\sigma_A,\sigma_B,\sigma_D,\sigma_P\}$ and a typical transition matrix, $\mathbf{P}$, given by
\[ \mathbf{P} =  \begin{bmatrix}
P_{AA} & P_{AB} & P_{AD} & P_{AP} \\
P_{BA} & P_{BB} & P_{BD} & P_{BP} \\
P_{DA} & P_{DB} & P_{DD} & P_{DP} \\
P_{PA} & P_{PB} & P_{PD} & P_{PP} 
\end{bmatrix}  =\begin{bmatrix}
0.9991 & 0.0002 & 0.0002 & 0.0005 \\
0.0500 & 0.9500 & 0.0000 & 0.0000 \\
0.0010 & 0.0000 & 0.9990 & 0.0000 \\
0.0000 & 0.0000 & 0.0000 & 1.0000 
\end{bmatrix}  \]
This transition matrix is altered to allow variation in the duration of blink and dark states, the frequency of blink and dark states, and the time to photobleach events. Duration of blink and dark states are controlled by varying the durations $\text{dur}_\text{blink}$ and $\text{dur}_\text{dark}$, multiplied by the time-binning resolution 20 $\mu$s, and calculating
\[P_{BB} = 1-\displaystyle\frac{1}{\text{dur}_\text{blink}} \quad \text{and} \quad 
P_{DD} = 1-\displaystyle\frac{1}{\text{dur}_\text{dark}},\]
and setting 
\[P_{AB} = \displaystyle\frac{1}{\text{dur}_\text{blink}} \quad \text{and} \quad 
P_{AD} = \displaystyle\frac{1}{\text{dur}_\text{dark}}.\]
The frequencies of blink and dark states are controlled by varying $P_{AB}$ and $P_{AD}$, and the time-to-photobleach is controlled by varying $P_{AP}$, so that
\[P_{AA} = 1- P_{AB}-P_{AD} - P_{AP}.\]
Once the states have been simulated, fluorophore intensity is independently drawn from a Poisson distribution for each active fluorophore in each time frame, and background noise compromised of Gaussian and Poisson components added for each time frame, measured in microseconds. Finally, 20 $\mu$s bins are used to mimic the time-resolution observed in FLImP. 

\subsection{Convergence testing}
To ensure convergence of chains for each dataset during CRJMCMC, the convergence of each parameter, or sets of parameters, is examined in blocks until all have reached an appropriate level of convergence. This is determined according to the potential scale reduction factor and the multivariate potential scale reduction factor, with a maximum value of 1.2 allowed in each case.

Parameters are sequentially tested, beginning with the number of change points, $k$, as this affects the length of the vector of change point locations, $s$, which complicates the sequential testing of convergence. Therefore, first the the potential scale reduction factor for parameter, $k$, is calculated, discarding half of the iterations as burn-in. If this parameter has reached a sufficient level of convergence, the mode number of change points, $k$, in each chain is then determined and the iterations which contain the corresponding number of change points retained\footnote{If the mode number of change points in each chain do not match, it is assumed that the chains have not appropriately converged, and iterations are continued.}. This allows comparison of traces for the first change point locations, second change point locations etc., so that they correspond across and within chains. As the number of iterations with the mode length of $\mathbf{s}$ may not be equal in each chain, the final minimum number of mode-length $\mathbf{s}$ from both chains is used. This ensures that the resulting set of traces have the same dimension, which allows calculation of the potential scale reduction factor for each of the change point locations in $\mathbf{s}$\footnote{By calculating convergence for both $k$ and $s$, there is no need to calculate convergence for $k_t$, as this will have converged if both $k$ and $s$ have converged.}. Finally the intensity parameters, $\mu_f$, $\mu_b$, $\sigma_f^2$ and $\sigma_2^b$, are tested, and provided all of the above meet the criteria, the chains are then terminated and the results saved, otherwise, the chains proceed from their last iteration and carry out an additional 10,000 iterations until convergence or user-defined maximum number of iterations.

\clearpage

\newpage

\subsection{Sensitivity analysis}\label{sec:sensitivity_anal}
A structured sensitivity analysis was performed to evaluate the robustness of the CRJMCMC algorithm to user-defined hyperparameters. Parameters were varied individually across broad ranges while holding all others fixed at pre-specified default values listed in Table \ref{tab:final_hyperparam}. Performance was measured using root mean square error (RMSE), accuracy, and precision, averaged over 100 simulated datasets spanning one to four fluorophores, a signal-to-noise ratio (SNR) of 0.1, and single fluorophore intensity of 1,000 photons.

The range of values evaluated and the average performance of each parameter can be found in Tables \ref{tab:rmse_custom_dist_base_var} to \ref{tab:rmse_short_state_prob_accept} with 95\% confidence intervals provided in brackets. For each parameter, a one way analysis of variance (ANOVA) was performed as a single test across all parameter values simultaneously, to determine whether variation in that hyperparameter produced any overall difference among group means in the averaged performance metrics (RMSE, accuracy, and precision). Statistical significance was assessed at the 95\% level ($\alpha = 0.05$). The null hypothesis in each case was of global equality among average RMSE for all parameter values, with an alternative that the average RMSEs are not all equal, e.g. allowing $\mu_i$ to be the average RMSE for $\lambda$ with value $i$, a test on the full set of parameter values has
\begin{align*}
    H_0&:\mu_1=\mu_{1.5}=\dots=\mu_{20}\\
    H_1&:\mu_i\text{'s not all equal.}
\end{align*}
Parameter sets showing a significant effect in RMSE were then examined by iteratively excluding the most poorly performing parameter values and repeating the same global ANOVA, until no further significant effect was detected. After removal of all parameter values associated with a statistically significant degradation in performance, the remaining subsets were retained as acceptable ranges. These represent configurations for which no statistically significant difference in group means was observed at the 95\% level in RMSE. The $F$-statistic and corresponding $p$-values produced for the full and reduced parameter sets are included in Tables \ref{tab:rmse_custom_dist_base_var} to \ref{tab:rmse_short_state_prob_accept}. Table \ref{tab:final_hyperparam} presents the final reduced ranges for all parameters, and the specific values used in analysis for this paper\footnote{The scaling factors for $\nu_f$ and $\nu_b$ were allowed to decrease within the allowed ranges to $\nu_f=0.0025$ and $\nu_b=0.005$ when analysing DNA origami to enforce stricter prior distributions on the intensity parameters.}\footnote{$\lambda$ and $\lambda_t$ were allowed to increase within the allowed ranges to $\lambda=10$ and $\lambda_t=5$ when varying frequency of blink and dark states to encourage more flexible proposals of short-lived states.}. 

\begin{table}
\centering
\caption{Final non-significant parameter ranges retained after sensitivity analysis.}
\label{tab:final_hyperparam}
\begin{tabular}{lcc}
\toprule
\textbf{Parameter} & \textbf{Possible Range} &  \textbf{Chosen Value} \\
\midrule
Custom Distribution Base Variance & 250 - 60,000  & \textbf{10,000}\\

Custom Distribution Window Size & 10 - 50 & \textbf{10} \\

Scaling Factor for $\nu_f$ & 0.001 - 0.100 & \textbf{0.005} \\

Scaling Factor for $\nu_b$ & 0.001 - 1.0 & \textbf{1.0} \\

Birth-Death Max. Probability ($c$) & 0.01 - 0.90 & \textbf{0.5} \\

Short-Lived State Max. Probability ($\gamma$) & 0.01 - 
0.50 & \textbf{0.1} \\

Change Point Number Poisson Mean ($\lambda$) & 1 - 20 & \textbf{2.5}\\

Short-Lived State Poisson Mean ($\lambda_t$)& 0.0005 - 10 & \textbf{0.001} \\

Short-Lived State Max. Duration ($\tau$)& 1 - 15 & \textbf{10} \\

Short-Lived Accept Probability ($p$) & 0.1 - 0.9 & \textbf{0.5} \\

\bottomrule
\end{tabular}
\end{table}
\begin{table}

\caption{\textbf{Custom Proposal Distribution Base Variance} Sensitivity analysis results reporting RMSE, precision, and accuracy across a range of values for the base variance of the custom proposal distribution. ANOVA over RMSE for the full parameter set produced $F(12, 1287)= 5.64$, $p < 1.59\text{e-}9$; for the reduced parameter set in Table \ref{tab:final_hyperparam}, $F(9, 990) = 1.06$, $p = 0.39$.}

\label{tab:rmse_custom_dist_base_var}
\begin{tabular}{cccc}
\toprule
Value & RMSE (Intensity) & Accuracy & Precision \\
\midrule
10.0 & 87.620 $(\pm10.96)$ & 0.989 $(\pm2.24\text{e-}3)$ & 0.978 $(\pm1.95\text{e-}2)$ \\
50.0 & 78.049 $(\pm10.19)$ & 0.991 $(\pm1.84\text{e-}3)$ & 0.981 $(\pm1.95\text{e-}2)$ \\
100.0 & 72.433 $(\pm9.69)$ & 0.992 $(\pm1.55\text{e-}3)$ & 0.982 $(\pm1.95\text{e-}2)$ \\
250.0 & 63.935 $(\pm9.28)$ & 0.994 $(\pm1.44\text{e-}3)$ & 0.983 $(\pm1.95\text{e-}2)$ \\
500.0 & 58.569 $(\pm9.37)$ & 0.994 $(\pm1.46\text{e-}3)$ & 0.985 $(\pm1.96\text{e-}2)$ \\
1000.0 & 50.937 $(\pm8.76)$ & 0.995 $(\pm1.19\text{e-}3)$ & 0.986 $(\pm1.96\text{e-}2)$ \\
2000.0 & 51.241 $(\pm9.27)$ & 0.995 $(\pm1.45\text{e-}3)$ & 0.985 $(\pm1.96\text{e-}2)$ \\
3000.0 & 52.026 $(\pm8.74)$ & 0.995 $(\pm1.19\text{e-}3)$ & 0.985 $(\pm1.96\text{e-}2)$ \\
5000.0 & 53.568 $(\pm9.27)$ & 0.995 $(\pm1.43\text{e-}3)$ & 0.985 $(\pm1.96\text{e-}2)$ \\
10000.0 & 56.674 $(\pm8.27)$ & 0.995 $(\pm1.30\text{e-}3)$ & 0.985 $(\pm1.96\text{e-}2)$ \\
20000.0 & 59.488 $(\pm8.10)$ & 0.995 $(\pm1.49\text{e-}3)$ & 0.985 $(\pm1.96\text{e-}2)$ \\
40000.0 & 59.415 $(\pm8.82)$ & 0.995 $(\pm1.81\text{e-}3)$ & 0.984 $(\pm1.96\text{e-}2)$ \\
60000.0 & 62.055 $(\pm8.29)$ & 0.995 $(\pm1.49\text{e-}3)$ & 0.984 $(\pm1.96\text{e-}2)$ \\
\bottomrule
\end{tabular}

\end{table}

\begin{table}
\begin{tabular}{cccc}
\toprule
Value & RMSE (Intensity) & Accuracy & Precision \\
\midrule
5.0 & 86.989 $(\pm7.92)$ & 0.994 $(\pm1.93\text{e-}3)$ & 0.985 $(\pm1.96\text{e-}2)$ \\
10.0 & 55.318 $(\pm8.16)$ & 0.996 $(\pm1.27\text{e-}3)$ & 0.985 $(\pm1.96\text{e-}2)$ \\
15.0 & 54.849 $(\pm9.54)$ & 0.995 $(\pm1.67\text{e-}3)$ & 0.985 $(\pm1.96\text{e-}2)$ \\
20.0 & 52.733 $(\pm8.96)$ & 0.995 $(\pm1.28\text{e-}3)$ & 0.985 $(\pm1.96\text{e-}2)$ \\
30.0 & 50.747 $(\pm9.25)$ & 0.995 $(\pm1.27\text{e-}3)$ & 0.985 $(\pm1.96\text{e-}2)$ \\
40.0 & 56.506 $(\pm11.27)$ & 0.993 $(\pm2.67\text{e-}3)$ & 0.984 $(\pm1.96\text{e-}2)$ \\
50.0 & 56.627 $(\pm11.38)$ & 0.993 $(\pm2.67\text{e-}3)$ & 0.984 $(\pm1.96\text{e-}2)$ \\
\bottomrule
\caption{\textbf{Custom Distribution Window Size} Sensitivity analysis results reporting RMSE, precision, and accuracy across a range of values for the window size of the custom proposal distribution. ANOVA over RMSE for the full parameter set produced $F(6, 693) = 6.51$, $p = 1.07\text{e-}6$; for the reduced parameter set in Table \ref{tab:final_hyperparam}, $F(5, 594) = 0.21$, $p = 0.96$.}
\end{tabular}

\end{table}

\begin{table}
\begin{tabular}{cccc}
\toprule
Value & RMSE (Intensity) & Accuracy & Precision \\
\midrule
0.001 & 70.471 $(\pm7.93)$ & 0.995 $(\pm1.29\text{e-}3)$ & 0.985 $(\pm1.96\text{e-}2)$ \\
0.002 & 65.527 $(\pm7.72)$ & 0.995 $(\pm1.19\text{e-}3)$ & 0.985 $(\pm1.96\text{e-}2)$ \\
0.005 & 55.664 $(\pm8.21)$ & 0.995 $(\pm1.29\text{e-}3)$ & 0.985 $(\pm1.96\text{e-}2)$ \\
0.01 & 52.674 $(\pm8.90)$ & 0.995 $(\pm1.27\text{e-}3)$ & 0.985 $(\pm1.96\text{e-}2)$ \\
0.02 & 51.658 $(\pm9.21)$ & 0.983 $(\pm1.44\text{e-}2)$ & 0.965 $(\pm3.00\text{e-}2)$ \\
0.05 & 55.736 $(\pm12.63)$ & 0.898 $(\pm3.88\text{e-}2)$ & 0.788 $(\pm7.49\text{e-}2)$ \\
0.1 & 68.499 $(\pm21.19)$ & 0.682 $(\pm5.54\text{e-}2)$ & 0.402 $(\pm9.26\text{e-}2)$ \\
0.2 & 81.154 $(\pm23.94)$ & 0.533 $(\pm5.21\text{e-}2)$ & 0.191 $(\pm7.56\text{e-}2)$ \\
0.5 & 121.796 $(\pm31.31)$ & 0.476 $(\pm4.34\text{e-}2)$ & 0.097 $(\pm5.68\text{e-}2)$ \\
1.0 & 113.911 $(\pm29.16)$ & 0.479 $(\pm4.34\text{e-}2)$ & 0.093 $(\pm5.63\text{e-}2)$ \\
\bottomrule
\caption{\textbf{Scaling Factor for $\boldsymbol{\nu}_\mathbf{f}$} Sensitivity analysis results reporting RMSE, precision, and accuracy across a range of values for the $\nu_f$ scaling factor. ANOVA over RMSE for the full parameter set produced $F(9, 990) = 7.19$, $p = 3.75\text{e-}10$; for the reduced parameter set in Table \ref{tab:final_hyperparam}, $F(6, 693) = 1.73$, $p = 0.11$.}
\end{tabular}

\end{table}

\begin{table}
\begin{tabular}{cccc}
\toprule
Value & RMSE (Intensity) & Accuracy & Precision \\
\midrule
0.001 & 56.378 $(\pm7.84)$ & 0.995 $(\pm1.17\text{e-}3)$ & 0.985 $(\pm1.96\text{e-}2)$ \\
0.002 & 56.865 $(\pm8.18)$ & 0.995 $(\pm1.23\text{e-}3)$ & 0.985 $(\pm1.96\text{e-}2)$ \\
0.005 & 56.324 $(\pm8.10)$ & 0.995 $(\pm1.21\text{e-}3)$ & 0.985 $(\pm1.96\text{e-}2)$ \\
0.01 & 56.338 $(\pm8.23)$ & 0.995 $(\pm1.28\text{e-}3)$ & 0.985 $(\pm1.96\text{e-}2)$ \\
0.02 & 56.033 $(\pm8.09)$ & 0.995 $(\pm1.27\text{e-}3)$ & 0.985 $(\pm1.96\text{e-}2)$ \\
0.05 & 56.666 $(\pm8.37)$ & 0.995 $(\pm1.31\text{e-}3)$ & 0.985 $(\pm1.96\text{e-}2)$ \\
0.1 & 57.151 $(\pm8.40)$ & 0.995 $(\pm1.37\text{e-}3)$ & 0.985 $(\pm1.96\text{e-}2)$ \\
0.2 & 54.793 $(\pm8.21)$ & 0.995 $(\pm1.47\text{e-}3)$ & 0.985 $(\pm1.96\text{e-}2)$ \\
0.5 & 55.117 $(\pm8.10)$ & 0.996 $(\pm1.27\text{e-}3)$ & 0.985 $(\pm1.96\text{e-}2)$ \\
1.0 & 56.016 $(\pm8.29)$ & 0.995 $(\pm1.30\text{e-}3)$ & 0.985 $(\pm1.96\text{e-}2)$ \\
\bottomrule
\caption{\textbf{Scaling Factor for $\boldsymbol{\nu}_\mathbf{b}$} Sensitivity analysis results reporting RMSE, precision, and accuracy across a range of values for the $\nu_b$ scaling factor. ANOVA over RMSE for the full parameter set produced $F(9, 990) =  0.03$, $p = 1.00$.}
\end{tabular}

\end{table}

\begin{table}
\begin{tabular}{cccc}
\toprule
Value & RMSE (Intensity) & Accuracy & Precision \\
\midrule
0.01 & 58.088 $(\pm9.08)$ & 0.995 $(\pm1.62\text{e-}3)$ & 0.989 $(\pm1.31\text{e-}2)$ \\
0.05 & 56.577 $(\pm8.33)$ & 0.995 $(\pm1.39\text{e-}3)$ & 0.988 $(\pm1.31\text{e-}2)$ \\
0.1 & 55.972 $(\pm8.58)$ & 0.995 $(\pm1.54\text{e-}3)$ & 0.989 $(\pm1.31\text{e-}2)$ \\
0.25 & 62.205 $(\pm9.98)$ & 0.994 $(\pm2.08\text{e-}3)$ & 0.983 $(\pm1.96\text{e-}2)$ \\
0.5 & 55.309 $(\pm8.14)$ & 0.996 $(\pm1.27\text{e-}3)$ & 0.985 $(\pm1.96\text{e-}2)$ \\
0.75 & 55.271 $(\pm8.14)$ & 0.996 $(\pm1.29\text{e-}3)$ & 0.985 $(\pm1.96\text{e-}2)$ \\
0.9 & 56.372 $(\pm8.01)$ & 0.995 $(\pm1.20\text{e-}3)$ & 0.985 $(\pm1.96\text{e-}2)$ \\
\bottomrule
\caption{\textbf{Birth-Death Max. Probability ($\mathbf{c}$)} Sensitivity analysis results reporting RMSE, precision, and accuracy across a range of values for the maximum allowed move probability for birth and death moves, $c$. ANOVA over RMSE for the full parameter set produced $F(6, 693) = 0.31$, $p = 0.93$.}
\end{tabular}

\end{table}

\begin{table}
\begin{tabular}{cccc}
\toprule
Value & RMSE (Intensity) & Accuracy & Precision \\
\midrule
0.001 & 101.597 $(\pm16.86)$ & 0.982 $(\pm4.96\text{e-}3)$ & 0.973 $(\pm2.02\text{e-}2)$ \\
0.005 & 73.829 $(\pm10.77)$ & 0.992 $(\pm2.12\text{e-}3)$ & 0.981 $(\pm1.98\text{e-}2)$ \\
0.01 & 67.788 $(\pm9.88)$ & 0.993 $(\pm1.72\text{e-}3)$ & 0.982 $(\pm1.98\text{e-}2)$ \\
0.025 & 61.007 $(\pm8.90)$ & 0.995 $(\pm1.40\text{e-}3)$ & 0.984 $(\pm1.96\text{e-}2)$ \\
0.05 & 60.196 $(\pm8.83)$ & 0.995 $(\pm1.40\text{e-}3)$ & 0.985 $(\pm1.96\text{e-}2)$ \\
0.1 & 55.829 $(\pm8.24)$ & 0.995 $(\pm1.29\text{e-}3)$ & 0.985 $(\pm1.96\text{e-}2)$ \\
0.2 & 55.362 $(\pm8.94)$ & 0.995 $(\pm1.82\text{e-}3)$ & 0.985 $(\pm1.96\text{e-}2)$ \\
0.3 & 57.913 $(\pm9.09)$ & 0.995 $(\pm1.70\text{e-}3)$ & 0.985 $(\pm1.96\text{e-}2)$ \\
0.4 & 53.869 $(\pm8.10)$ & 0.996 $(\pm1.23\text{e-}3)$ & 0.986 $(\pm1.96\text{e-}2)$ \\
0.5 & 55.216 $(\pm7.94)$ & 0.996 $(\pm1.18\text{e-}3)$ & 0.986 $(\pm1.96\text{e-}2)$ \\
\bottomrule
\caption{\textbf{Short-Lived State Max. Probability ($\boldsymbol{\gamma}$)} Sensitivity analysis results reporting RMSE, precision, and accuracy across a range of values for the maximum allowed move probability for add and remove short-lived state moves, $\gamma$. ANOVA over RMSE for the full parameter set produced $F(9, 990) = 8.02$, $p = 1.58\text{e-}11$; for the reduced parameter set in Table \ref{tab:final_hyperparam}, $F(7, 792) = 1.04$, $p = 0.40$.}
\end{tabular}

\end{table}

\begin{table}
\captionof{table}{\textbf{Change Point Number Poisson Mean ($\lambda$)} Sensitivity analysis results reporting RMSE, precision, and accuracy across a range of values for the mean number of change points, $\lambda$. ANOVA over RMSE for the full parameter set produced $F(12, 1287) = 0.15$, $p = 0.31$.}
\begin{tabular}{cccc}
\toprule
Value & RMSE (Intensity) & Accuracy & Precision \\
\midrule
1.0 & 55.269 $(\pm7.92)$ & 0.996 $(\pm1.18\text{e-}3)$ & 0.986 $(\pm1.96\text{e-}2)$ \\
1.5 & 57.079 $(\pm8.22)$ & 0.995 $(\pm1.30\text{e-}3)$ & 0.985 $(\pm1.96\text{e-}2)$ \\
2.0 & 55.382 $(\pm8.56)$ & 0.995 $(\pm1.49\text{e-}3)$ & 0.985 $(\pm1.96\text{e-}2)$ \\
2.5 & 55.477 $(\pm8.23)$ & 0.995 $(\pm1.29\text{e-}3)$ & 0.985 $(\pm1.96\text{e-}2)$ \\
3.0 & 57.727 $(\pm9.20)$ & 0.995 $(\pm1.91\text{e-}3)$ & 0.985 $(\pm1.96\text{e-}2)$ \\
4.0 & 54.455 $(\pm8.33)$ & 0.996 $(\pm1.32\text{e-}3)$ & 0.985 $(\pm1.96\text{e-}2)$ \\
5.0 & 55.164 $(\pm8.08)$ & 0.995 $(\pm1.38\text{e-}3)$ & 0.986 $(\pm1.96\text{e-}2)$ \\
6.0 & 57.027 $(\pm8.03)$ & 0.995 $(\pm1.20\text{e-}3)$ & 0.985 $(\pm1.96\text{e-}2)$ \\
8.0 & 57.060 $(\pm8.89)$ & 0.995 $(\pm1.58\text{e-}3)$ & 0.985 $(\pm1.96\text{e-}2)$ \\
10.0 & 57.688 $(\pm8.81)$ & 0.995 $(\pm1.48\text{e-}3)$ & 0.985 $(\pm1.96\text{e-}2)$ \\
12.0 & 56.920 $(\pm8.49)$ & 0.995 $(\pm1.39\text{e-}3)$ & 0.985 $(\pm1.96\text{e-}2)$ \\
15.0 & 52.580 $(\pm7.52)$ & 0.996 $(\pm1.05\text{e-}3)$ & 0.986 $(\pm1.96\text{e-}2)$ \\
20.0 & 53.444 $(\pm7.59)$ & 0.996 $(\pm1.05\text{e-}3)$ & 0.985 $(\pm1.96\text{e-}2)$ \\
\bottomrule
\end{tabular}

\end{table}

\begin{table}
\captionof{table}{\textbf{Short-Lived State Poisson Mean ($\lambda_t$)} Sensitivity analysis results reporting RMSE, precision, and accuracy across a range of values for the mean number of short-lived state change points, $\lambda_t$. ANOVA over RMSE for the full parameter set produced $F(18, 1881) = 1.14$, $p = 0.31$.}
\begin{tabular}{cccc}
\toprule
Value & RMSE (Intensity) & Accuracy & Precision \\
\midrule
0.0005 & 55.830 $(\pm8.26)$ & 0.995 $(\pm1.29\text{e-}3)$ & 0.985 $(\pm1.96\text{e-}2)$ \\
0.001 & 55.600 $(\pm8.21)$ & 0.995 $(\pm1.29\text{e-}3)$ & 0.985 $(\pm1.96\text{e-}2)$ \\
0.005 & 57.385 $(\pm8.49)$ & 0.995 $(\pm1.37\text{e-}3)$ & 0.985 $(\pm1.96\text{e-}2)$ \\
0.01 & 58.411 $(\pm8.58)$ & 0.995 $(\pm1.39\text{e-}3)$ & 0.985 $(\pm1.96\text{e-}2)$ \\
0.025 & 55.625 $(\pm8.35)$ & 0.995 $(\pm1.35\text{e-}3)$ & 0.985 $(\pm1.96\text{e-}2)$ \\
0.05 & 55.204 $(\pm8.28)$ & 0.995 $(\pm1.34\text{e-}3)$ & 0.986 $(\pm1.96\text{e-}2)$ \\
0.1 & 57.310 $(\pm8.98)$ & 0.995 $(\pm1.62\text{e-}3)$ & 0.984 $(\pm1.97\text{e-}2)$ \\
0.25 & 57.759 $(\pm9.04)$ & 0.995 $(\pm1.68\text{e-}3)$ & 0.984 $(\pm1.97\text{e-}2)$ \\
0.5 & 59.792 $(\pm9.50)$ & 0.994 $(\pm1.93\text{e-}3)$ & 0.985 $(\pm1.96\text{e-}2)$ \\
1.0 & 57.265 $(\pm9.18)$ & 0.995 $(\pm1.75\text{e-}3)$ & 0.984 $(\pm1.96\text{e-}2)$ \\
2.0 & 55.776 $(\pm8.80)$ & 0.995 $(\pm1.60\text{e-}3)$ & 0.985 $(\pm1.96\text{e-}2)$ \\
3.0 & 52.181 $(\pm7.46)$ & 0.996 $(\pm1.06\text{e-}3)$ & 0.986 $(\pm1.96\text{e-}2)$ \\
4.0 & 49.547 $(\pm6.74)$ & 0.997 $(\pm8.27\text{e-}4)$ & 0.987 $(\pm1.96\text{e-}2)$ \\
5.0 & 51.328 $(\pm7.40)$ & 0.996 $(\pm9.81\text{e-}4)$ & 0.986 $(\pm1.96\text{e-}2)$ \\
6.0 & 48.227 $(\pm6.80)$ & 0.997 $(\pm8.17\text{e-}4)$ & 0.987 $(\pm1.96\text{e-}2)$ \\
7.0 & 49.705 $(\pm7.02)$ & 0.997 $(\pm8.95\text{e-}4)$ & 0.986 $(\pm1.96\text{e-}2)$ \\
8.0 & 47.797 $(\pm6.76)$ & 0.997 $(\pm8.25\text{e-}4)$ & 0.987 $(\pm1.96\text{e-}2)$ \\
9.0 & 46.727 $(\pm6.23)$ & 0.997 $(\pm7.19\text{e-}4)$ & 0.987 $(\pm1.96\text{e-}2)$ \\
10.0 & 46.432 $(\pm6.43)$ & 0.997 $(\pm7.79\text{e-}4)$ & 0.987 $(\pm1.96\text{e-}2)$ \\
\bottomrule
\end{tabular}

\end{table}

\begin{table}
\captionof{table}{\textbf{Short-Lived State Max. Duration ($\tau$)} Sensitivity analysis results reporting RMSE, precision, and accuracy across a range of values for the mean number of short-lived state change points, $\tau$. ANOVA over RMSE for the full parameter set produced $F(19, 1980) = 10.94$, $p = 6.70\text{e-}32$; for the reduced parameter set in Table \ref{tab:final_hyperparam}, $F(8, 891) = 1.89$, $p = 0.06$}
\begin{tabular}{cccc}
\toprule
Value & RMSE (Intensity) & Accuracy & Precision \\
\midrule
1.0 & 43.652 $(\pm6.56)$ & 0.997 $(\pm8.03\text{e-}4)$ & 0.986 $(\pm1.96\text{e-}2)$ \\
2.0 & 43.280 $(\pm6.38)$ & 0.997 $(\pm7.56\text{e-}4)$ & 0.987 $(\pm1.96\text{e-}2)$ \\
3.0 & 46.468 $(\pm6.43)$ & 0.997 $(\pm7.67\text{e-}4)$ & 0.987 $(\pm1.96\text{e-}2)$ \\
4.0 & 48.856 $(\pm7.61)$ & 0.996 $(\pm1.31\text{e-}3)$ & 0.986 $(\pm1.96\text{e-}2)$ \\
5.0 & 49.168 $(\pm6.97)$ & 0.997 $(\pm8.87\text{e-}4)$ & 0.986 $(\pm1.96\text{e-}2)$ \\
6.0 & 52.318 $(\pm7.41)$ & 0.996 $(\pm1.03\text{e-}3)$ & 0.986 $(\pm1.96\text{e-}2)$ \\
8.0 & 50.952 $(\pm7.13)$ & 0.996 $(\pm9.37\text{e-}4)$ & 0.986 $(\pm1.96\text{e-}2)$ \\
10.0 & 55.722 $(\pm8.21)$ & 0.995 $(\pm1.28\text{e-}3)$ & 0.985 $(\pm1.96\text{e-}2)$ \\
12.0 & 58.604 $(\pm8.33)$ & 0.995 $(\pm1.27\text{e-}3)$ & 0.985 $(\pm1.96\text{e-}2)$ \\
15.0 & 58.369 $(\pm8.59)$ & 0.995 $(\pm1.46\text{e-}3)$ & 0.985 $(\pm1.96\text{e-}2)$ \\
20.0 & 67.077 $(\pm10.14)$ & 0.993 $(\pm1.88\text{e-}3)$ & 0.981 $(\pm1.98\text{e-}2)$ \\
25.0 & 66.878 $(\pm10.01)$ & 0.993 $(\pm1.77\text{e-}3)$ & 0.981 $(\pm1.98\text{e-}2)$ \\
30.0 & 81.707 $(\pm14.29)$ & 0.988 $(\pm4.03\text{e-}3)$ & 0.976 $(\pm2.02\text{e-}2)$ \\
40.0 & 86.274 $(\pm15.25)$ & 0.987 $(\pm4.46\text{e-}3)$ & 0.975 $(\pm2.03\text{e-}2)$ \\
50.0 & 85.242 $(\pm15.13)$ & 0.987 $(\pm4.38\text{e-}3)$ & 0.975 $(\pm2.03\text{e-}2)$ \\
60.0 & 90.231 $(\pm16.32)$ & 0.985 $(\pm4.97\text{e-}3)$ & 0.974 $(\pm2.04\text{e-}2)$ \\
70.0 & 93.415 $(\pm16.51)$ & 0.984 $(\pm4.96\text{e-}3)$ & 0.973 $(\pm2.04\text{e-}2)$ \\
80.0 & 93.873 $(\pm16.57)$ & 0.984 $(\pm5.00\text{e-}3)$ & 0.973 $(\pm2.03\text{e-}2)$ \\
90.0 & 96.957 $(\pm17.43)$ & 0.983 $(\pm5.44\text{e-}3)$ & 0.973 $(\pm2.04\text{e-}2)$ \\
100.0 & 99.073 $(\pm17.55)$ & 0.982 $(\pm5.40\text{e-}3)$ & 0.972 $(\pm2.04\text{e-}2)$ \\
\bottomrule
\end{tabular}

\end{table}

\begin{table}
\caption{\textbf{Short-Lived State Accept Probability ($\mathbf{p}$)} Sensitivity analysis results reporting RMSE, precision, and accuracy across a range of values for the target acceptance probability for short-lived state duration, $p$. ANOVA over RMSE for the full parameter set produced $F(8, 891) = 1.91$, $p = 0.05$.}
\label{tab:rmse_short_state_prob_accept}
\begin{tabular}{cccc}
\toprule
Value & RMSE (Intensity) & Accuracy & Precision \\
\midrule
0.1 & 52.538 $(\pm7.75)$ & 0.996 $(\pm1.21\text{e-}3)$ & 0.986 $(\pm1.96\text{e-}2)$ \\
0.2 & 54.714 $(\pm8.47)$ & 0.995 $(\pm1.50\text{e-}3)$ & 0.986 $(\pm1.96\text{e-}2)$ \\
0.3 & 53.408 $(\pm7.72)$ & 0.996 $(\pm1.17\text{e-}3)$ & 0.986 $(\pm1.96\text{e-}2)$ \\
0.4 & 56.151 $(\pm8.42)$ & 0.995 $(\pm1.52\text{e-}3)$ & 0.985 $(\pm1.96\text{e-}2)$ \\
0.5 & 55.766 $(\pm8.25)$ & 0.995 $(\pm1.29\text{e-}3)$ & 0.985 $(\pm1.96\text{e-}2)$ \\
0.6 & 61.439 $(\pm10.14)$ & 0.994 $(\pm2.62\text{e-}3)$ & 0.985 $(\pm1.96\text{e-}2)$ \\
0.7 & 61.788 $(\pm9.13)$ & 0.994 $(\pm1.53\text{e-}3)$ & 0.984 $(\pm1.96\text{e-}2)$ \\
0.8 & 64.856 $(\pm10.09)$ & 0.993 $(\pm1.91\text{e-}3)$ & 0.982 $(\pm1.97\text{e-}2)$ \\
0.9 & 72.496 $(\pm12.19)$ & 0.991 $(\pm2.96\text{e-}3)$ & 0.982 $(\pm1.96\text{e-}2)$ \\
\bottomrule
\end{tabular}

\end{table}

\clearpage
\subsection{MCMC diagnostics}

\begin{center}
\begin{table}[htbp]
\centering
\begin{tabular}{cccc}
\toprule
Parameter & MCSE & ESS & PSRF \\
\midrule

$k$ & 0.023 ($\pm$2.22e-03) & 54.524 ($\pm$1.16e+01) & 1.017 ($\pm$4.22e-03) \\

$\mathbf{s}$ & 0.042 ($\pm$3.33e-03) & 22.380 ($\pm$2.25e+00) & 1.004 ($\pm$3.17e-04) \\

$\mu_f$ & 0.160 ($\pm$9.96e-03) & 1522.296 ($\pm$1.65e+01) & 1.097 ($\pm$2.81e-02) \\
$\mu_b$ & 0.074 ($\pm$3.45e-03) & 1531.740 ($\pm$7.56e+00) & 1.013 ($\pm$6.49e-03) \\

$\sigma_f^2$ & 0.020 ($\pm$1.48e-03) & 1194.836 ($\pm$5.91e+00) & 1.079 ($\pm$4.00e-02) \\
$\sigma_b^2$ & 0.037 ($\pm$2.82e-03) & 1229.286 ($\pm$6.62e+00) & 1.277 ($\pm$8.91e-02) \\
\bottomrule
\end{tabular}
\caption{MCMC diagnostics summary per parameter. Values show the mean and 95\% confidence interval across runs.}
\label{tab:diagnostics-summary}
\end{table}

\end{center}

Table \ref{tab:diagnostics-summary} shows diagnostic results for the estimated parameters from 3,600 simulated intensity traces spanning one to four fluorophores, mean intensities of 500, 1,000, and 2,000 photons, and SNR of 0.01, 0.1, and 1, all of which are typical for data derived from FLImP. Chains were terminated at 20,000 iterations to enable consistent evaluation and comparison of convergence behaviour across parameters.

The intensity parameters $\mu_f$, $\mu_b$, $\sigma_f^2$, and $\sigma_b^2$, denoting fluorophore and background intensity mean and variance, exhibit higher effective sample sizes (ESS $> 1,000$) in comparison to the change point parameters. The number of change points, $k$, and their locations, $s$ (averaged across all locations), showed substantially lower ESS (54.5 and 22.4, respectively) suggesting posterior stickiness. However, this behaviour is consistent with the expected structure of change point models in MCMC, where the likelihood tends to dominate the posterior, and proposed changes to $k$ or $\mathbf{s}$ often lead to large shifts in likelihood; once the correct change point configuration is identified, the sampler rarely deviates, resulting in high posterior stability but limited exploration in later iterations. As such, the decreased ESS for change point parameters follows the expected pattern for multiple change point models. In light of this posterior stickiness, a 5\% trimmed mean and 95\% confidence intervals based on 1,000 bootstrap resamples were used for the average PSRF for $k$ and $\mathbf{s}$. All parameters estimated have PSRF values close to 1.0, indicating efficient mixing and stable posterior estimates. Finally, Monte Carlo standard errors (MCSEs) remained low for all parameters, further supporting the numerical stability of the posterior estimates, even where sampling efficiency was reduced for change point-related variables.

\begin{figure}[h!]
\centering
        \includegraphics[width = 0.9
    \linewidth]{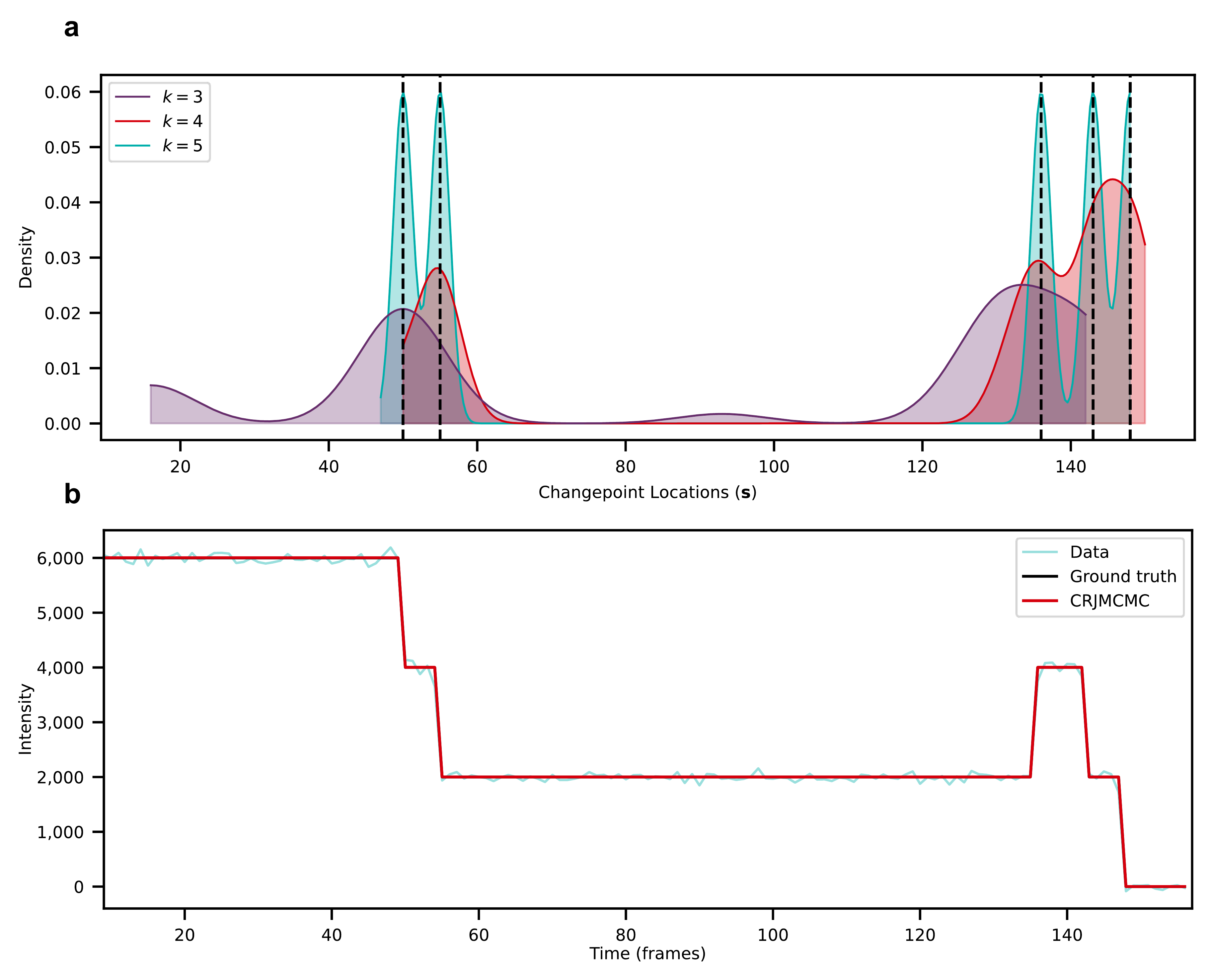}

    \caption{\textbf{Posterior estimates of change point locations and corresponding simulated intensity trace.} (\textbf{a}) Posterior kernel density estimates (standard deviation of 2) for change point locations, conditioned on the number of change points, from a single Markov chain with 20,000 iterations following a burn-in of 10{,}000 iterations. (\textbf{b}) Corresponding simulated intensity trace (turquoise), CRJMCMC model estimation (red line), with three fluorophores and an SNR of 1.}

    \label{fig:kde_cp}
\end{figure}

\begin{figure}[h!]
    \centering
\includegraphics[width=0.9\linewidth]{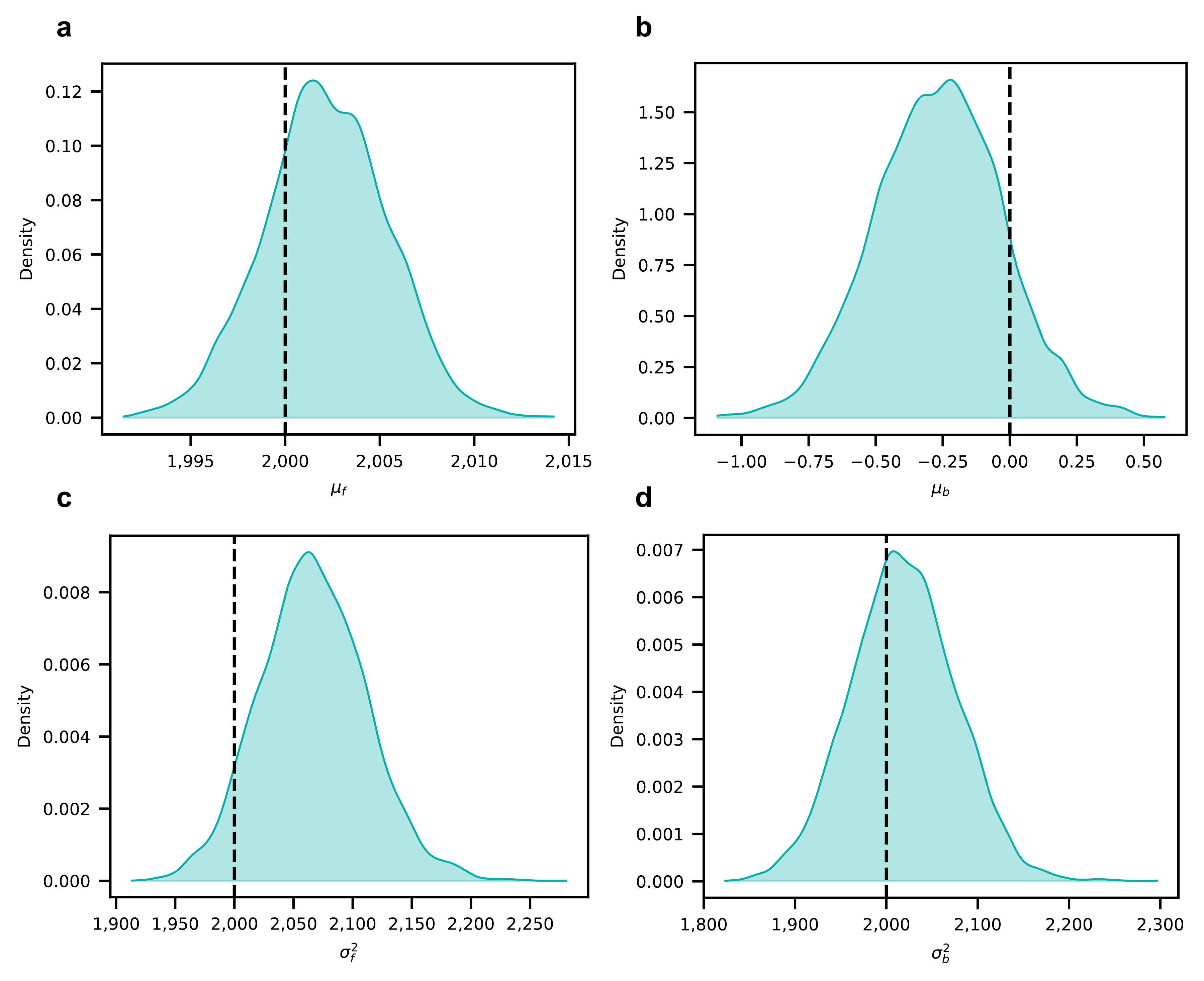}

    \caption{ \textbf{Posterior kernel density estimates (standard deviation of 2)  for the intensity parameters (a) $\mu_f$, (b) $\mu_b$, (c) $\sigma^2_f$, and (d) $\sigma^2_b$} from a single Markov chain with 20{,}000 iterations following a burn-in of 10{,}000 iterations. Parameters were estimated from a simulated integrated intensity trace with three fluorophores, mean single fluorophore intensity of 2,000 photons, and an SNR of 1.}

    \label{fig:kde_traces_combined}
\end{figure}

\begin{figure}[h!]
    \centering
    \includegraphics[width=0.9\linewidth]{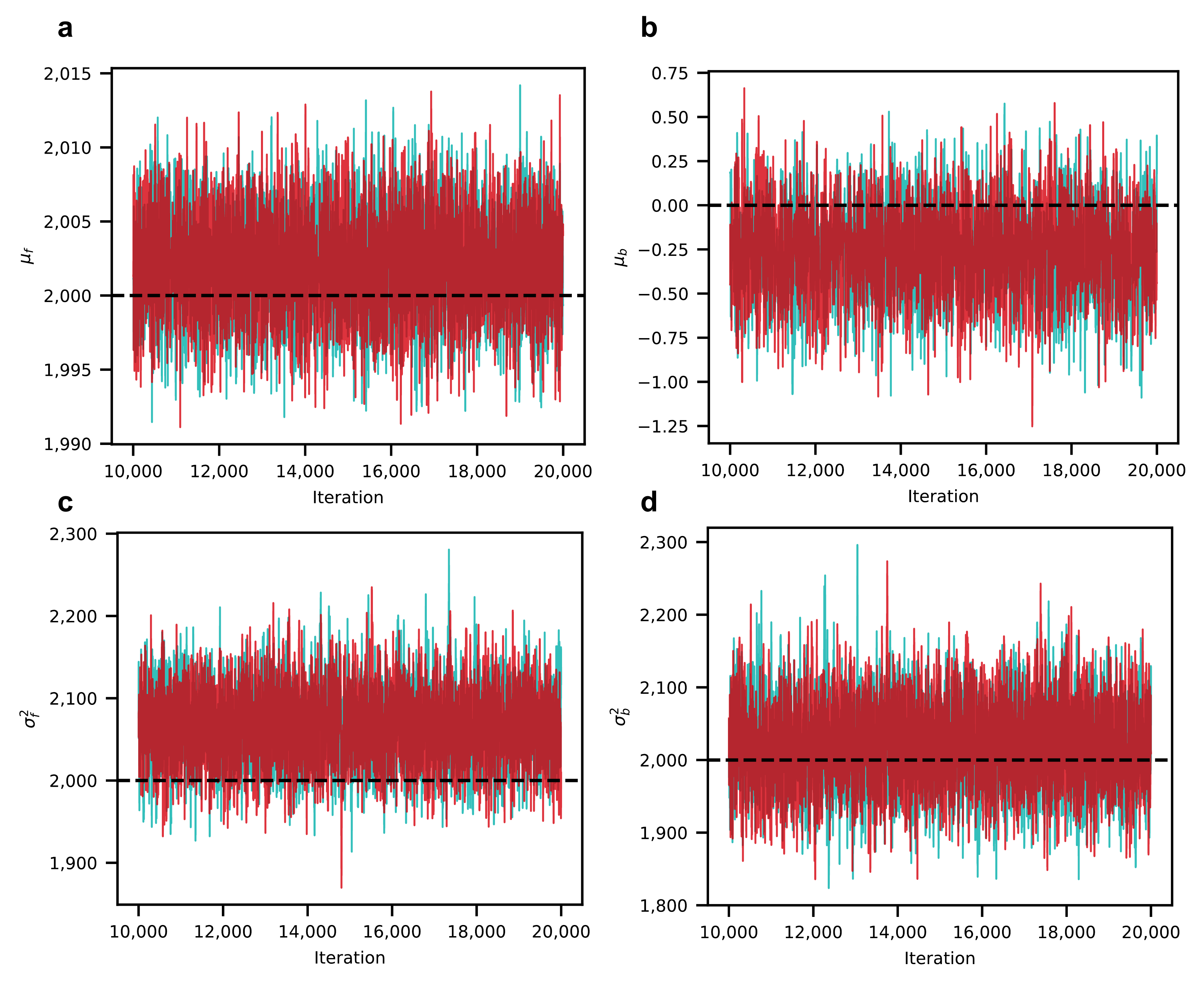}
    
\caption{\textbf{Trace plots for the intensity parameters (a) $\mu_f$, (b) $\mu_b$, (c) $\sigma^2_f$, and (d) $\sigma^2_b$} obtained from two independent, parallel Markov chains (red and turquoise), each run for 20{,}000 iterations following a burn-in of 10{,}000 iterations. Parameters were estimated from a simulated integrated intensity trace with three fluorophores, mean single fluorophore intensity of 2,000 photons, and an SNR of 1.}
    \label{fig:trace_plots}
\end{figure}

To demonstrate the performance of CRJMCMC in terms of convergence metrics, Fig. \ref{fig:kde_cp}, Fig. \ref{fig:kde_traces_combined}, and Fig. \ref{fig:trace_plots} respectively present an example of the posterior kernel density estimates of change point locations with the corresponding simulated integrated intensity trace provided for comparison, and the posterior kernel density estimates and trace plots for the intensity parameters. The trace plots exhibit good mixing and the posterior distributions show clear peaks, reflective of the convergence diagnostics in Table \ref{tab:diagnostics-summary}.

\clearpage
\subsection{Additional simulation results}

To evaluate performance of the CRJMCMC algorithm across a wide range of conditions, an extensive simulation study of approximately 18,600 independent simulated intensity traces was performed by varying fluorophore number, SNR, single-fluorophore intensity ($\mu_f$), dark-state frequency and duration, blink-state frequency and duration.

Performance was assessed using multiple metrics: average accuracy, precision, sensitivity, and specificity, and Cohen's kappa of the estimated active fluorophore count in each frame; RMSE of total estimated intensity; and absolute error in intensity parameters $\mu_f$, $\mu_b$, $\sigma_f^2$, and $\sigma_b^2$.

The methods compared include the CRJMCMC, monotonic decay MAP \citep{garry2020bayesian}\footnote{As a result of pooling intensity parameters with the same mean and SNR, the calibration carried out on the monotonic MAP algorithm in Garry et al. (2020) \cite{garry2020bayesian} produces the same intensity parameters for all data in a pool.}, sequential MAP estimation \citep{tsekouras2016novel}, and factorial HMM-MCMC \citep{bryan2022diffraction}\footnote{Due to the distribution used to model intensity in Bryan IV et al. (2022) \cite{bryan2022diffraction}, this method does not produce directly comparable estimates of $\sigma_f^2$ and $\sigma_b^2$; variance-based performance metrics are therefore reported only for the remaining Gaussian-based methods.}. For each considered parameter value, the best-performing method is indicated in bold and 95\% confidence intervals are provided in brackets.

\subsubsection{Varying SNR}

\begin{table}[htbp]
\captionof{table}{Accuracy}
\begin{tabular}{ccccc}
\toprule
SNR & CRJMCMC & Factorial HMM & Monotonic MAP & Sequential MAP \\
\midrule
0.001 & \textbf{0.834 (±2.54\text{e-}2)} & 0.423 (±2.74\text{e-}2) & 0.765 (±3.64\text{e-}2) & 0.640 (±4.29\text{e-}2) \\
0.002 & \textbf{0.911 (±1.76\text{e-}2)} & 0.503 (±3.29\text{e-}2) & 0.856 (±2.57\text{e-}2) & 0.719 (±4.45\text{e-}2) \\
0.003 & \textbf{0.871 (±2.36\text{e-}2)} & 0.551 (±3.85\text{e-}2) & \textbf{0.903 (±1.79\text{e-}2)} & 0.775 (±4.24\text{e-}2) \\
0.004 & \textbf{0.957 (±9.44\text{e-}3)} & 0.582 (±4.04\text{e-}2) & 0.874 (±2.06\text{e-}2) & 0.787 (±4.45\text{e-}2) \\
0.005 & \textbf{0.930 (±1.79\text{e-}2)} & 0.635 (±4.04\text{e-}2) & 0.877 (±2.14\text{e-}2) & 0.810 (±4.18\text{e-}2) \\
0.006 & \textbf{0.938 (±1.65\text{e-}2)} & 0.629 (±4.66\text{e-}2) & 0.884 (±2.17\text{e-}2) & 0.840 (±3.84\text{e-}2) \\
0.007 & \textbf{0.976 (±5.01\text{e-}3)} & 0.681 (±4.76\text{e-}2) & 0.911 (±1.77\text{e-}2) & 0.834 (±4.33\text{e-}2) \\
0.008 & \textbf{0.964 (±1.23\text{e-}2)} & 0.648 (±5.06\text{e-}2) & 0.904 (±1.87\text{e-}2) & 0.854 (±4.21\text{e-}2) \\
0.009 & \textbf{0.981 (±3.95\text{e-}3)} & 0.732 (±5.14\text{e-}2) & 0.891 (±1.98\text{e-}2) & 0.891 (±3.23\text{e-}2) \\
0.01 & \textbf{0.972 (±1.09\text{e-}2)} & 0.741 (±5.22\text{e-}2) & 0.892 (±2.08\text{e-}2) & 0.896 (±3.32\text{e-}2) \\
\bottomrule
\end{tabular}
\end{table}

\begin{table}[htbp]
\captionof{table}{Precision}
\begin{tabular}{ccccc}
\toprule
SNR & CRJMCMC & Factorial HMM & Monotonic MAP & Sequential MAP \\
\midrule
0.001 & \textbf{0.963 (±1.29\text{e-}2)} & 0.540 (±5.08\text{e-}2) & 0.910 (±2.79\text{e-}2) & 0.857 (±3.55\text{e-}2) \\
0.002 & \textbf{0.979 (±5.23\text{e-}3)} & 0.663 (±4.38\text{e-}2) & 0.916 (±2.11\text{e-}2) & 0.886 (±2.95\text{e-}2) \\
0.003 & \textbf{0.990 (±2.64\text{e-}3)} & 0.795 (±4.87\text{e-}2) & 0.933 (±1.93\text{e-}2) & 0.890 (±3.44\text{e-}2) \\
0.004 & \textbf{0.985 (±3.68\text{e-}3)} & 0.773 (±5.34\text{e-}2) & 0.910 (±1.99\text{e-}2) & 0.901 (±3.13\text{e-}2) \\
0.005 & \textbf{0.989 (±3.30\text{e-}3)} & 0.806 (±5.29\text{e-}2) & 0.904 (±2.25\text{e-}2) & 0.886 (±3.98\text{e-}2) \\
0.006 & \textbf{0.991 (±2.68\text{e-}3)} & 0.853 (±4.12\text{e-}2) & 0.898 (±2.48\text{e-}2) & 0.923 (±3.17\text{e-}2) \\
0.007 & \textbf{0.990 (±2.72\text{e-}3)} & 0.849 (±5.01\text{e-}2) & 0.926 (±2.03\text{e-}2) & 0.901 (±3.83\text{e-}2) \\
0.008 & \textbf{0.993 (±1.71\text{e-}3)} & 0.884 (±3.39\text{e-}2) & 0.919 (±2.02\text{e-}2) & 0.928 (±2.93\text{e-}2) \\
0.009 & \textbf{0.992 (±1.99\text{e-}3)} & 0.856 (±5.06\text{e-}2) & 0.897 (±2.26\text{e-}2) & 0.944 (±2.62\text{e-}2) \\
0.01 & \textbf{0.993 (±2.03\text{e-}3)} & 0.930 (±2.97\text{e-}2) & 0.910 (±2.18\text{e-}2) & 0.919 (±3.24\text{e-}2) \\
\bottomrule
\end{tabular}
\end{table}

\begin{table}[htbp]
\captionof{table}{Sensitivity}
\begin{tabular}{ccccc}
\toprule
SNR & CRJMCMC & Factorial HMM & Monotonic MAP & Sequential MAP \\
\midrule
0.001 & \textbf{0.790 (±3.51\text{e-}2)} & 0.424 (±4.81\text{e-}2) & \textbf{0.756 (±4.68\text{e-}2)} & 0.594 (±5.69\text{e-}2) \\
0.002 & \textbf{0.898 (±2.23\text{e-}2)} & 0.527 (±5.52\text{e-}2) & \textbf{0.885 (±3.15\text{e-}2)} & 0.707 (±5.62\text{e-}2) \\
0.003 & 0.830 (±3.17\text{e-}2) & 0.484 (±5.61\text{e-}2) & \textbf{0.930 (±2.12\text{e-}2)} & 0.765 (±5.48\text{e-}2) \\
0.004 & \textbf{0.952 (±1.30\text{e-}2)} & 0.566 (±6.50\text{e-}2) & 0.898 (±3.21\text{e-}2) & 0.788 (±5.64\text{e-}2) \\
0.005 & \textbf{0.917 (±2.25\text{e-}2)} & 0.626 (±6.41\text{e-}2) & \textbf{0.921 (±2.79\text{e-}2)} & 0.836 (±4.89\text{e-}2) \\
0.006 & \textbf{0.922 (±2.17\text{e-}2)} & 0.591 (±7.05\text{e-}2) & \textbf{0.948 (±2.10\text{e-}2)} & 0.823 (±5.47\text{e-}2) \\
0.007 & \textbf{0.976 (±6.68\text{e-}3)} & 0.674 (±6.67\text{e-}2) & 0.946 (±2.06\text{e-}2) & 0.860 (±4.81\text{e-}2) \\
0.008 & \textbf{0.958 (±1.62\text{e-}2)} & 0.625 (±7.17\text{e-}2) & \textbf{0.945 (±2.29\text{e-}2)} & 0.839 (±5.79\text{e-}2) \\
0.009 & \textbf{0.982 (±4.58\text{e-}3)} & 0.728 (±7.10\text{e-}2) & \textbf{0.960 (±1.79\text{e-}2)} & 0.880 (±4.87\text{e-}2) \\
0.01 & \textbf{0.965 (±1.47\text{e-}2)} & 0.692 (±6.68\text{e-}2) & \textbf{0.930 (±2.59\text{e-}2)} & 0.908 (±4.13\text{e-}2) \\
\bottomrule
\end{tabular}
\end{table}

\begin{table}[htbp]
\captionof{table}{Specificity}
\begin{tabular}{ccccc}
\toprule
SNR & CRJMCMC & Factorial HMM & Monotonic MAP & Sequential MAP \\
\midrule
0.001 & \textbf{0.922 (±1.73\text{e-}2)} & 0.509 (±5.19\text{e-}2) & 0.794 (±5.00\text{e-}2) & 0.810 (±3.70\text{e-}2) \\
0.002 & \textbf{0.929 (±1.83\text{e-}2)} & 0.568 (±5.47\text{e-}2) & 0.760 (±4.70\text{e-}2) & 0.821 (±3.40\text{e-}2) \\
0.003 & \textbf{0.966 (±1.01\text{e-}2)} & 0.737 (±5.00\text{e-}2) & 0.796 (±4.81\text{e-}2) & 0.819 (±3.73\text{e-}2) \\
0.004 & \textbf{0.945 (±1.56\text{e-}2)} & 0.700 (±5.97\text{e-}2) & 0.744 (±5.09\text{e-}2) & 0.849 (±3.14\text{e-}2) \\
0.005 & \textbf{0.962 (±1.00\text{e-}2)} & 0.711 (±5.70\text{e-}2) & 0.745 (±4.90\text{e-}2) & 0.834 (±3.50\text{e-}2) \\
0.006 & \textbf{0.969 (±9.89\text{e-}3)} & 0.760 (±6.10\text{e-}2) & 0.730 (±5.60\text{e-}2) & 0.878 (±2.78\text{e-}2) \\
0.007 & \textbf{0.953 (±1.44\text{e-}2)} & 0.726 (±5.63\text{e-}2) & 0.780 (±4.70\text{e-}2) & 0.839 (±4.00\text{e-}2) \\
0.008 & \textbf{0.975 (±6.74\text{e-}3)} & 0.792 (±5.42\text{e-}2) & 0.759 (±5.16\text{e-}2) & 0.885 (±2.78\text{e-}2) \\
0.009 & \textbf{0.972 (±7.13\text{e-}3)} & 0.752 (±6.04\text{e-}2) & 0.713 (±5.35\text{e-}2) & 0.898 (±2.17\text{e-}2) \\
0.01 & \textbf{0.979 (±5.97\text{e-}3)} & 0.878 (±4.39\text{e-}2) & 0.767 (±5.20\text{e-}2) & 0.885 (±2.65\text{e-}2) \\
\bottomrule
\end{tabular}
\end{table}

\begin{table}[htbp]
\captionof{table}{Cohen's Kappa}
\begin{tabular}{ccccc}
\toprule
SNR & CRJMCMC & Factorial HMM & Monotonic MAP & Sequential MAP \\
\midrule
0.001 & \textbf{0.724 (±3.57\text{e-}2)} & 0.176 (±2.47\text{e-}2) & 0.597 (±5.48\text{e-}2) & 0.436 (±5.47\text{e-}2) \\
0.002 & \textbf{0.841 (±2.67\text{e-}2)} & 0.267 (±3.15\text{e-}2) & 0.733 (±4.11\text{e-}2) & 0.552 (±5.82\text{e-}2) \\
0.003 & \textbf{0.783 (±3.34\text{e-}2)} & 0.326 (±4.32\text{e-}2) & \textbf{0.812 (±3.36\text{e-}2)} & 0.628 (±5.45\text{e-}2) \\
0.004 & \textbf{0.926 (±1.36\text{e-}2)} & 0.376 (±5.01\text{e-}2) & 0.767 (±3.73\text{e-}2) & 0.666 (±5.95\text{e-}2) \\
0.005 & \textbf{0.883 (±2.55\text{e-}2)} & 0.448 (±5.34\text{e-}2) & 0.763 (±4.10\text{e-}2) & 0.702 (±5.44\text{e-}2) \\
0.006 & \textbf{0.896 (±2.46\text{e-}2)} & 0.463 (±5.67\text{e-}2) & 0.790 (±3.90\text{e-}2) & 0.741 (±5.22\text{e-}2) \\
0.007 & \textbf{0.956 (±8.79\text{e-}3)} & 0.517 (±6.16\text{e-}2) & 0.833 (±3.27\text{e-}2) & 0.734 (±6.01\text{e-}2) \\
0.008 & \textbf{0.941 (±1.65\text{e-}2)} & 0.493 (±5.90\text{e-}2) & 0.806 (±3.92\text{e-}2) & 0.767 (±5.64\text{e-}2) \\
0.009 & \textbf{0.965 (±6.47\text{e-}3)} & 0.591 (±6.58\text{e-}2) & 0.790 (±3.80\text{e-}2) & 0.804 (±4.76\text{e-}2) \\
0.01 & \textbf{0.956 (±1.39\text{e-}2)} & 0.632 (±6.34\text{e-}2) & 0.809 (±3.70\text{e-}2) & 0.831 (±4.33\text{e-}2) \\
\bottomrule
\end{tabular}
\end{table}

\begin{table}[htbp]
\captionof{table}{RMSE (Intensity)}
\begin{tabular}{ccccc}
\toprule
SNR & CRJMCMC & Factorial HMM & Monotonic MAP & Sequential MAP \\
\midrule
0.001 & \textbf{391.029 (±27.95)} & 993.066 (±65.28) & \textbf{445.655 (±46.08)} & 586.331 (±50.33) \\
0.002 & \textbf{270.335 (±21.23)} & 730.395 (±52.94) & 363.919 (±45.71) & 447.926 (±30.42) \\
0.003 & \textbf{345.410 (±28.45)} & 613.066 (±50.06) & \textbf{304.318 (±42.30)} & 390.594 (±32.57) \\
0.004 & \textbf{231.382 (±19.87)} & 549.730 (±46.14) & 357.094 (±45.97) & 364.192 (±28.71) \\
0.005 & \textbf{232.233 (±18.57)} & 500.846 (±45.73) & 350.732 (±52.15) & 328.572 (±24.69) \\
0.006 & \textbf{225.045 (±24.82)} & 452.682 (±44.44) & \textbf{286.954 (±38.66)} & 285.851 (±22.00) \\
0.007 & \textbf{193.572 (±22.05)} & 418.621 (±39.78) & 301.904 (±44.89) & 317.468 (±32.74) \\
0.008 & \textbf{168.974 (±16.16)} & 392.309 (±38.03) & 296.145 (±44.70) & 277.189 (±24.38) \\
0.009 & \textbf{150.439 (±17.03)} & 335.897 (±26.94) & 301.649 (±48.37) & 266.356 (±26.99) \\
0.01 & \textbf{197.536 (±20.15)} & 418.327 (±45.33) & 325.663 (±51.87) & 273.902 (±26.63) \\
\bottomrule
\end{tabular}
\end{table}

\begin{table}[htbp]
\captionof{table}{Absolute Error ($\mu_f$)}
\begin{tabular}{ccccc}
\toprule
SNR & CRJMCMC & Factorial HMM & Monotonic MAP & Sequential MAP \\
\midrule
0.001 & 198.313 (±14.53) & 198.000 (±14.99) & \textbf{122.405 (±3.59)} & 355.402 (±70.19) \\
0.002 & \textbf{130.545 (±8.29)} & \textbf{132.549 (±8.81)} & \textbf{145.176 (±18.35)} & 259.664 (±46.21) \\
0.003 & 252.762 (±11.68) & 244.347 (±14.71) & \textbf{128.588 (±15.14)} & 210.426 (±41.38) \\
0.004 & \textbf{135.493 (±7.53)} & \textbf{134.661 (±10.13)} & 167.031 (±19.89) & 214.199 (±41.62) \\
0.005 & \textbf{138.743 (±11.80)} & \textbf{136.743 (±14.21)} & \textbf{127.419 (±14.58)} & \textbf{161.402 (±29.53)} \\
0.006 & 134.897 (±17.77) & \textbf{121.066 (±21.41)} & \textbf{99.658 (±14.98)} & \textbf{155.509 (±45.84)} \\
0.007 & \textbf{124.511 (±10.46)} & \textbf{116.954 (±16.40)} & \textbf{141.237 (±16.25)} & \textbf{150.308 (±29.66)} \\
0.008 & \textbf{83.718 (±13.85)} & \textbf{76.478 (±17.60)} & 130.927 (±14.64) & 157.364 (±34.85) \\
0.009 & \textbf{76.962 (±5.27)} & \textbf{77.303 (±8.68)} & 133.531 (±8.76) & 134.735 (±36.14) \\
0.01 & \textbf{132.805 (±10.99)} & \textbf{150.469 (±20.56)} & \textbf{141.545 (±15.50)} & \textbf{113.062 (±21.74)} \\
\bottomrule
\end{tabular}
\end{table}

\begin{table}[htbp]
\captionof{table}{Absolute Error ($\mu_b$)}
\begin{tabular}{ccccc}
\toprule
SNR & CRJMCMC & Factorial HMM & Monotonic MAP & Sequential MAP \\
\midrule
0.001 & \textbf{42.903 (±5.91)} & 578.705 (±95.64) & 112.093 (±5.75) & 284.914 (±54.89) \\
0.002 & \textbf{33.665 (±4.23)} & 402.923 (±69.73) & 94.819 (±14.19) & 195.645 (±29.90) \\
0.003 & \textbf{33.752 (±6.64)} & 246.414 (±48.56) & 95.116 (±9.74) & 149.195 (±28.44) \\
0.004 & \textbf{29.107 (±4.97)} & 265.140 (±46.76) & 133.497 (±17.85) & 156.892 (±26.32) \\
0.005 & \textbf{16.749 (±2.12)} & 259.837 (±48.19) & 120.460 (±12.23) & 107.387 (±17.37) \\
0.006 & \textbf{10.011 (±2.46)} & 213.746 (±40.08) & 48.175 (±6.20) & 91.021 (±17.97) \\
0.007 & \textbf{43.516 (±5.42)} & 160.372 (±31.70) & 100.532 (±11.28) & 104.415 (±19.87) \\
0.008 & \textbf{11.807 (±2.33)} & 177.442 (±42.37) & 86.568 (±9.17) & 89.376 (±17.27) \\
0.009 & \textbf{9.206 (±3.03)} & 132.045 (±34.32) & 94.023 (±6.23) & 79.379 (±17.85) \\
0.01 & \textbf{32.360 (±5.63)} & 132.318 (±26.75) & 78.550 (±5.72) & 81.518 (±17.52) \\
\bottomrule
\end{tabular}
\end{table}

\begin{table}[htbp]
\captionof{table}{Absolute Error ($\sigma_f^2$)}
\begin{tabular}{cccc}
\toprule
SNR & CRJMCMC & Monotonic MAP & Sequential MAP \\
\midrule
0.001 & 515.803 (±40.69) & 1.20\text{e+}6 (±1.14\text{e+}5) & \textbf{320.083 (±52.21)} \\
0.002 & \textbf{351.360 (±13.18)} & 6.40\text{e+}5 (±6.44\text{e+}4) & 460.007 (±73.23) \\
0.003 & \textbf{494.416 (±33.26)} & 4.27\text{e+}5 (±4.42\text{e+}4) & \textbf{543.884 (±81.58)} \\
0.004 & \textbf{328.296 (±9.99)} & 3.39\text{e+}5 (±3.62\text{e+}4) & 645.404 (±91.01) \\
0.005 & \textbf{377.706 (±16.03)} & 2.65\text{e+}5 (±2.80\text{e+}4) & 668.577 (±87.90) \\
0.006 & \textbf{309.035 (±39.17)} & 2.42\text{e+}5 (±2.68\text{e+}4) & 662.368 (±92.25) \\
0.007 & \textbf{344.830 (±32.24)} & 1.95\text{e+}5 (±2.09\text{e+}4) & 743.042 (±94.08) \\
0.008 & \textbf{285.537 (±21.30)} & 1.89\text{e+}5 (±2.17\text{e+}4) & 760.997 (±99.90) \\
0.009 & \textbf{241.601 (±23.19)} & 1.67\text{e+}5 (±1.82\text{e+}4) & 725.815 (±94.78) \\
0.01 & \textbf{296.616 (±29.01)} & 1.78\text{e+}5 (±2.25\text{e+}4) & 804.878 (±98.32) \\
\bottomrule
\end{tabular}
\end{table}

\begin{table}[htbp]
\captionof{table}{Absolute Error ($\sigma_b^2$)}
\begin{tabular}{cccc}
\toprule
SNR & CRJMCMC & Monotonic MAP & Sequential MAP \\
\midrule
0.001 & \textbf{4.92\text{e+}4 (±4.75\text{e+}3)} & 8.31\text{e+}4 (±2.43\text{e+}3) & 1.17\text{e+}6 (±1.13\text{e+}5) \\
0.002 & \textbf{3.35\text{e+}4 (±2.37\text{e+}3)} & 6.34\text{e+}4 (±5.82\text{e+}3) & 5.83\text{e+}5 (±5.60\text{e+}4) \\
0.003 & \textbf{5.04\text{e+}4 (±9.17\text{e+}3)} & 8.55\text{e+}4 (±1.71\text{e+}4) & 3.79\text{e+}5 (±3.77\text{e+}4) \\
0.004 & \textbf{2.82\text{e+}4 (±4.61\text{e+}3)} & 9.25\text{e+}4 (±1.89\text{e+}4) & 2.90\text{e+}5 (±2.82\text{e+}4) \\
0.005 & \textbf{3.85\text{e+}4 (±3.33\text{e+}3)} & 9.07\text{e+}4 (±1.60\text{e+}4) & 2.29\text{e+}5 (±2.26\text{e+}4) \\
0.006 & \textbf{1.56\text{e+}4 (±1.14\text{e+}3)} & 4.06\text{e+}4 (±9.83\text{e+}3) & 1.79\text{e+}5 (±1.86\text{e+}4) \\
0.007 & \textbf{4.43\text{e+}4 (±1.50\text{e+}3)} & 7.80\text{e+}4 (±1.60\text{e+}4) & 1.62\text{e+}5 (±1.64\text{e+}4) \\
0.008 & \textbf{9931.415 (±1094.86)} & 6.22\text{e+}4 (±8.49\text{e+}3) & 1.42\text{e+}5 (±1.49\text{e+}4) \\
0.009 & \textbf{2.59\text{e+}4 (±6.13\text{e+}3)} & 8.13\text{e+}4 (±1.45\text{e+}4) & 1.19\text{e+}5 (±1.24\text{e+}4) \\
0.01 & \textbf{1.69\text{e+}4 (±1.38\text{e+}3)} & 6.45\text{e+}4 (±9.81\text{e+}3) & 1.15\text{e+}5 (±1.17\text{e+}4) \\
\bottomrule
\end{tabular}
\end{table}

\clearpage
\subsubsection{Varying Fluorophore Number}

\begin{table}[htbp]
\captionof{table}{Accuracy}
\begin{tabular}{ccccc}
\toprule
Fluorophore number & CRJMCMC & Factorial HMM & Monotonic MAP & Sequential MAP \\
\midrule
1 & \textbf{0.998 (±9.59\text{e-}4)} &  & 0.951 (±1.85\text{e-}2) & 0.964 (±2.09\text{e-}2) \\
2 & \textbf{0.992 (±2.92\text{e-}3)} & 0.718 (±7.07\text{e-}2) & 0.910 (±2.19\text{e-}2) & 0.917 (±4.04\text{e-}2) \\
3 & \textbf{0.985 (±6.25\text{e-}3)} & 0.782 (±6.63\text{e-}2) & 0.892 (±2.08\text{e-}2) & 0.879 (±5.26\text{e-}2) \\
4 & \textbf{0.983 (±5.60\text{e-}3)} & 0.834 (±4.38\text{e-}2) & 0.844 (±2.19\text{e-}2) & 0.870 (±5.08\text{e-}2) \\
5 & \textbf{0.974 (±1.02\text{e-}2)} & 0.793 (±5.11\text{e-}2) & 0.798 (±2.43\text{e-}2) & 0.808 (±6.17\text{e-}2) \\
6 & \textbf{0.981 (±6.49\text{e-}3)} & 0.850 (±4.12\text{e-}2) & 0.736 (±2.19\text{e-}2) & 0.769 (±6.05\text{e-}2) \\
7 & \textbf{0.974 (±6.08\text{e-}3)} & 0.866 (±3.76\text{e-}2) & 0.704 (±2.25\text{e-}2) & 0.598 (±7.79\text{e-}2) \\
\bottomrule
\end{tabular}
\end{table}

\begin{table}[htbp]
\captionof{table}{Precision}
\begin{tabular}{ccccc}
\toprule
Fluorophore number & CRJMCMC & Factorial HMM & Monotonic MAP & Sequential MAP \\
\midrule
1 & \textbf{0.998 (±1.10\text{e-}3)} &  & 0.968 (±1.60\text{e-}2) & 0.963 (±2.62\text{e-}2) \\
2 & \textbf{0.995 (±1.62\text{e-}3)} & 0.914 (±5.02\text{e-}2) & 0.934 (±2.17\text{e-}2) & 0.967 (±2.21\text{e-}2) \\
3 & \textbf{0.994 (±1.77\text{e-}3)} & 0.933 (±4.17\text{e-}2) & 0.912 (±1.94\text{e-}2) & 0.937 (±4.72\text{e-}2) \\
4 & \textbf{0.993 (±1.85\text{e-}3)} & 0.971 (±1.92\text{e-}2) & 0.850 (±2.54\text{e-}2) & 0.933 (±4.33\text{e-}2) \\
5 & \textbf{0.990 (±2.99\text{e-}3)} & \textbf{0.975 (±1.35\text{e-}2)} & 0.790 (±2.73\text{e-}2) & 0.907 (±5.37\text{e-}2) \\
6 & \textbf{0.991 (±2.03\text{e-}3)} & \textbf{0.985 (±7.92\text{e-}3)} & 0.715 (±2.69\text{e-}2) & 0.828 (±6.59\text{e-}2) \\
7 & \textbf{0.985 (±3.23\text{e-}3)} & \textbf{0.979 (±1.00\text{e-}2)} & 0.683 (±2.61\text{e-}2) & 0.767 (±8.72\text{e-}2) \\
\bottomrule
\end{tabular}
\end{table}

\begin{table}[htbp]
\captionof{table}{Sensitivity}
\begin{tabular}{ccccc}
\toprule
Fluorophore number & CRJMCMC & Factorial HMM & Monotonic MAP & Sequential MAP \\
\midrule
1 & \textbf{0.999 (±7.44\text{e-}4)} &  & 0.946 (±3.35\text{e-}2) & 0.986 (±7.47\text{e-}3) \\
2 & \textbf{0.993 (±4.45\text{e-}3)} & 0.652 (±0.14) & 0.920 (±3.54\text{e-}2) & 0.911 (±5.36\text{e-}2) \\
3 & \textbf{0.986 (±7.94\text{e-}3)} & 0.752 (±8.55\text{e-}2) & 0.936 (±2.84\text{e-}2) & 0.851 (±6.98\text{e-}2) \\
4 & \textbf{0.986 (±6.68\text{e-}3)} & 0.814 (±5.51\text{e-}2) & 0.937 (±2.52\text{e-}2) & 0.853 (±6.92\text{e-}2) \\
5 & \textbf{0.978 (±1.18\text{e-}2)} & 0.765 (±6.23\text{e-}2) & \textbf{0.959 (±1.88\text{e-}2)} & 0.783 (±8.08\text{e-}2) \\
6 & \textbf{0.986 (±7.13\text{e-}3)} & 0.830 (±4.89\text{e-}2) & 0.936 (±2.14\text{e-}2) & 0.773 (±8.10\text{e-}2) \\
7 & \textbf{0.983 (±6.95\text{e-}3)} & 0.862 (±4.29\text{e-}2) & 0.954 (±1.64\text{e-}2) & 0.515 (±0.10) \\
\bottomrule
\end{tabular}
\end{table}

\begin{table}[htbp]
\captionof{table}{Specificity}
\begin{tabular}{ccccc}
\toprule
Fluorophore number & CRJMCMC & Factorial HMM & Monotonic MAP & Sequential MAP \\
\midrule
1 & \textbf{0.993 (±3.66\text{e-}3)} &  & 0.905 (±4.38\text{e-}2) & 0.936 (±2.79\text{e-}2) \\
2 & \textbf{0.981 (±6.21\text{e-}3)} & 0.821 (±0.15) & 0.823 (±5.10\text{e-}2) & 0.936 (±2.77\text{e-}2) \\
3 & \textbf{0.974 (±7.74\text{e-}3)} & 0.906 (±5.32\text{e-}2) & 0.710 (±5.68\text{e-}2) & \textbf{0.962 (±9.92\text{e-}3)} \\
4 & \textbf{0.964 (±1.17\text{e-}2)} & \textbf{0.929 (±3.54\text{e-}2)} & 0.601 (±5.36\text{e-}2) & 0.916 (±2.90\text{e-}2) \\
5 & \textbf{0.949 (±1.43\text{e-}2)} & \textbf{0.931 (±3.17\text{e-}2)} & 0.440 (±4.79\text{e-}2) & 0.905 (±2.74\text{e-}2) \\
6 & \textbf{0.949 (±1.23\text{e-}2)} & \textbf{0.947 (±2.16\text{e-}2)} & 0.386 (±3.64\text{e-}2) & 0.834 (±4.26\text{e-}2) \\
7 & \textbf{0.918 (±1.72\text{e-}2)} & \textbf{0.908 (±3.73\text{e-}2)} & 0.295 (±3.50\text{e-}2) & \textbf{0.898 (±3.31\text{e-}2)} \\
\bottomrule
\end{tabular}
\end{table}

\begin{table}[htbp]
\captionof{table}{Cohen's Kappa}
\begin{tabular}{ccccc}
\toprule
Fluorophore number & CRJMCMC & Factorial HMM & Monotonic MAP & Sequential MAP \\
\midrule
1 & \textbf{0.994 (±3.17\text{e-}3)} &  & 0.868 (±4.70\text{e-}2) & 0.926 (±3.06\text{e-}2) \\
2 & \textbf{0.985 (±5.75\text{e-}3)} & 0.523 (±8.33\text{e-}2) & 0.823 (±4.32\text{e-}2) & 0.872 (±5.52\text{e-}2) \\
3 & \textbf{0.975 (±9.70\text{e-}3)} & 0.694 (±8.50\text{e-}2) & 0.804 (±3.82\text{e-}2) & 0.824 (±7.20\text{e-}2) \\
4 & \textbf{0.973 (±8.65\text{e-}3)} & 0.763 (±5.79\text{e-}2) & 0.761 (±3.25\text{e-}2) & 0.818 (±6.72\text{e-}2) \\
5 & \textbf{0.963 (±1.35\text{e-}2)} & 0.733 (±6.27\text{e-}2) & 0.700 (±3.39\text{e-}2) & 0.742 (±7.95\text{e-}2) \\
6 & \textbf{0.973 (±9.34\text{e-}3)} & 0.808 (±5.03\text{e-}2) & 0.639 (±2.85\text{e-}2) & 0.691 (±7.78\text{e-}2) \\
7 & \textbf{0.964 (±8.59\text{e-}3)} & 0.829 (±4.47\text{e-}2) & 0.594 (±2.83\text{e-}2) & 0.482 (±9.62\text{e-}2) \\
\bottomrule
\end{tabular}
\end{table}
\begin{table}[htbp]
\captionof{table}{RMSE (Intensity)}
\begin{tabular}{ccccc}
\toprule
Fluorophore number & CRJMCMC & Factorial HMM & Monotonic MAP & Sequential MAP \\
\midrule
1 & \textbf{42.167 (±8.61)} &  & 189.850 (±49.43) & 112.787 (±27.15) \\
2 & \textbf{74.371 (±12.78)} & 305.033 (±47.16) & 285.115 (±47.12) & 172.395 (±28.05) \\
3 & \textbf{92.875 (±15.92)} & 179.156 (±42.61) & 353.147 (±52.12) & 243.873 (±29.68) \\
4 & \textbf{109.624 (±19.56)} & \textbf{162.411 (±34.63)} & 388.849 (±48.86) & 284.536 (±38.81) \\
5 & \textbf{133.467 (±22.21)} & \textbf{174.303 (±33.63)} & 408.991 (±48.70) & 333.959 (±50.07) \\
6 & \textbf{149.540 (±25.69)} & \textbf{193.627 (±40.90)} & 473.683 (±52.20) & 391.680 (±48.37) \\
7 & \textbf{168.251 (±23.19)} & \textbf{192.829 (±36.95)} & 482.028 (±59.55) & 472.735 (±57.42) \\
\bottomrule
\end{tabular}
\end{table}

\begin{table}[htbp]
\captionof{table}{Absolute Error ($\mu_f$)}
\begin{tabular}{ccccc}
\toprule
Fluorophore number & CRJMCMC & Factorial HMM & Monotonic MAP & Sequential MAP \\
\midrule
1 & \textbf{37.993 (±7.55)} &  & 197.612 (±22.96) & \textbf{51.798 (±22.15)} \\
2 & \textbf{28.422 (±5.88)} & 73.995 (±2.92) & 197.612 (±22.96) & 83.872 (±26.58) \\
3 & \textbf{26.118 (±5.61)} & \textbf{38.489 (±9.54)} & 197.612 (±22.96) & 107.006 (±47.56) \\
4 & \textbf{25.066 (±5.66)} & \textbf{36.439 (±7.71)} & 197.612 (±22.96) & 127.883 (±40.66) \\
5 & \textbf{24.594 (±5.62)} & 38.997 (±7.87) & 197.612 (±22.96) & 188.834 (±76.14) \\
6 & \textbf{21.432 (±5.07)} & 42.530 (±8.52) & 197.612 (±22.96) & 175.517 (±46.39) \\
7 & \textbf{20.237 (±4.56)} & 41.637 (±8.33) & 197.612 (±22.96) & 388.738 (±117.41) \\
8 & \textbf{20.553 (±4.90)} & 43.648 (±8.58) & 197.612 (±22.96) & 357.131 (±95.50) \\
9 & \textbf{22.141 (±4.81)} & 43.996 (±8.43) & 197.612 (±22.96) & 501.248 (±134.63) \\
10 & \textbf{19.378 (±4.73)} & 42.542 (±8.30) & 197.612 (±22.96) & 543.341 (±140.97) \\
11 & \textbf{23.409 (±5.49)} & 45.273 (±8.50) & 197.612 (±22.96) & 585.182 (±109.60) \\
12 & \textbf{19.795 (±5.07)} & 43.458 (±8.38) & 197.612 (±22.96) & 654.598 (±148.55) \\
13 & \textbf{25.283 (±6.45)} & 46.801 (±8.44) & 197.612 (±22.96) & 652.375 (±134.85) \\
14 & \textbf{27.402 (±6.37)} & 46.423 (±8.36) & 197.612 (±22.96) & 938.353 (±163.95) \\
15 & \textbf{38.943 (±11.55)} & \textbf{46.993 (±8.52)} & 197.612 (±22.96) & 1160.265 (±208.73) \\
16 & \textbf{32.930 (±8.61)} & \textbf{45.124 (±8.57)} & 197.612 (±22.96) & 1068.107 (±192.89) \\
17 & \textbf{31.944 (±7.14)} & \textbf{44.609 (±8.55)} & 197.612 (±22.96) & 1260.526 (±249.64) \\
18 & \textbf{48.292 (±14.76)} & \textbf{46.352 (±8.53)} & 197.612 (±22.96) & 1220.279 (±189.72) \\
19 & \textbf{44.371 (±11.08)} & \textbf{45.431 (±8.60)} & 197.612 (±22.96) & 1429.268 (±234.50) \\
20 & \textbf{44.744 (±10.63)} & \textbf{45.635 (±8.58)} & 197.612 (±22.96) & 1594.474 (±227.83) \\
\bottomrule
\end{tabular}
\end{table}

\begin{table}[htbp]
\captionof{table}{Absolute Error ($\mu_b$)}
\begin{tabular}{ccccc}
\toprule
Fluorophore number & CRJMCMC & Factorial HMM & Monotonic MAP & Sequential MAP \\
\midrule
1 & \textbf{1.728 (±0.28)} &  & 125.419 (±15.70) & 27.313 (±9.29) \\
2 & \textbf{1.655 (±0.26)} & 129.775 (±75.96) & 125.419 (±15.70) & 60.052 (±23.10) \\
3 & \textbf{1.784 (±0.30)} & 99.955 (±36.06) & 125.419 (±15.70) & 74.743 (±29.34) \\
4 & \textbf{1.793 (±0.30)} & 83.259 (±25.44) & 125.419 (±15.70) & 81.648 (±29.87) \\
5 & \textbf{1.963 (±0.34)} & 71.480 (±21.51) & 125.419 (±15.70) & 87.576 (±32.62) \\
6 & \textbf{2.313 (±0.41)} & 59.497 (±17.29) & 125.419 (±15.70) & 115.490 (±37.16) \\
7 & \textbf{2.282 (±0.40)} & 70.537 (±22.62) & 125.419 (±15.70) & 210.809 (±52.62) \\
8 & \textbf{2.167 (±0.39)} & 76.375 (±23.43) & 125.419 (±15.70) & 208.710 (±48.71) \\
9 & \textbf{2.151 (±0.39)} & 62.361 (±19.49) & 125.419 (±15.70) & 263.138 (±69.43) \\
10 & \textbf{2.043 (±0.37)} & 60.638 (±18.58) & 125.419 (±15.70) & 309.680 (±74.12) \\
11 & \textbf{2.079 (±0.37)} & 62.435 (±15.79) & 125.419 (±15.70) & 360.141 (±88.21) \\
12 & \textbf{1.998 (±0.38)} & 66.292 (±21.25) & 125.419 (±15.70) & 370.654 (±79.37) \\
13 & \textbf{1.948 (±0.35)} & 81.209 (±27.41) & 125.419 (±15.70) & 433.025 (±79.51) \\
14 & \textbf{2.065 (±0.39)} & 59.903 (±19.59) & 125.419 (±15.70) & 637.122 (±97.40) \\
15 & \textbf{2.251 (±0.42)} & 73.307 (±17.71) & 125.419 (±15.70) & 622.329 (±99.38) \\
16 & \textbf{2.115 (±0.45)} & 67.744 (±20.31) & 125.419 (±15.70) & 599.205 (±109.82) \\
17 & \textbf{2.003 (±0.36)} & 67.271 (±20.72) & 125.419 (±15.70) & 698.196 (±109.94) \\
18 & \textbf{2.272 (±0.45)} & 56.432 (±13.21) & 125.419 (±15.70) & 681.930 (±103.76) \\
19 & \textbf{2.210 (±0.41)} & 72.711 (±19.74) & 125.419 (±15.70) & 721.007 (±114.81) \\
20 & \textbf{2.018 (±0.37)} & 58.016 (±16.74) & 125.419 (±15.70) & 925.731 (±139.66) \\
\bottomrule
\end{tabular}
\end{table}

\begin{table}[htbp]
\captionof{table}{Absolute Error ($\sigma_f^2$)}
\begin{tabular}{cccc}
\toprule
Fluorophore number & CRJMCMC & Monotonic MAP & Sequential MAP \\
\midrule
1 & \textbf{129.260 (±23.12)} & 1.00\text{e+}5 (±2.10\text{e+}4) & 942.511 (±123.80) \\
2 & \textbf{124.314 (±15.84)} & 1.00\text{e+}5 (±2.10\text{e+}4) & 889.681 (±126.86) \\
3 & \textbf{133.590 (±17.32)} & 1.00\text{e+}5 (±2.10\text{e+}4) & 874.594 (±124.39) \\
4 & \textbf{152.431 (±19.70)} & 1.00\text{e+}5 (±2.10\text{e+}4) & 872.731 (±120.81) \\
5 & \textbf{195.479 (±35.18)} & 1.00\text{e+}5 (±2.10\text{e+}4) & 874.343 (±124.81) \\
6 & \textbf{192.596 (±26.34)} & 1.00\text{e+}5 (±2.10\text{e+}4) & 941.371 (±118.61) \\
7 & \textbf{242.428 (±38.18)} & 1.00\text{e+}5 (±2.10\text{e+}4) & 829.175 (±115.16) \\
8 & \textbf{235.710 (±40.05)} & 1.00\text{e+}5 (±2.10\text{e+}4) & 867.139 (±127.17) \\
9 & \textbf{277.034 (±49.78)} & 1.00\text{e+}5 (±2.10\text{e+}4) & 804.660 (±134.95) \\
10 & \textbf{273.883 (±47.14)} & 1.00\text{e+}5 (±2.10\text{e+}4) & 786.539 (±115.52) \\
11 & \textbf{311.767 (±59.75)} & 1.00\text{e+}5 (±2.10\text{e+}4) & 675.926 (±119.43) \\
12 & \textbf{370.805 (±70.30)} & 1.00\text{e+}5 (±2.10\text{e+}4) & 693.976 (±119.79) \\
13 & \textbf{326.473 (±56.06)} & 1.00\text{e+}5 (±2.10\text{e+}4) & 698.252 (±118.38) \\
14 & \textbf{483.489 (±90.86)} & 1.00\text{e+}5 (±2.10\text{e+}4) & \textbf{603.859 (±115.88)} \\
15 & \textbf{481.501 (±90.49)} & 1.00\text{e+}5 (±2.10\text{e+}4) & \textbf{566.225 (±99.00)} \\
16 & \textbf{563.907 (±107.99)} & 1.00\text{e+}5 (±2.10\text{e+}4) & \textbf{507.284 (±91.94)} \\
17 & \textbf{700.466 (±174.85)} & 1.00\text{e+}5 (±2.10\text{e+}4) & \textbf{510.094 (±97.03)} \\
18 & 1070.659 (±332.93) & 1.00\text{e+}5 (±2.10\text{e+}4) & \textbf{511.750 (±92.24)} \\
19 & 1079.482 (±302.20) & 1.00\text{e+}5 (±2.10\text{e+}4) & \textbf{499.474 (±96.94)} \\
20 & 1568.056 (±509.13) & 1.00\text{e+}5 (±2.10\text{e+}4) & \textbf{449.235 (±89.97)} \\
\bottomrule
\end{tabular}
\end{table}

\begin{table}[htbp]
\captionof{table}{Absolute Error ($\sigma_b^2$)}
\begin{tabular}{cccc}
\toprule
Fluorophore number & CRJMCMC & Monotonic MAP & Sequential MAP \\
\midrule
1 & \textbf{4495.961 (±1261.12)} & 1.05\text{e+}5 (±2.11\text{e+}4) & 3.82\text{e+}4 (±1.30\text{e+}4) \\
2 & \textbf{3276.893 (±701.23)} & 1.05\text{e+}5 (±2.11\text{e+}4) & 2.56\text{e+}4 (±8.30\text{e+}3) \\
3 & \textbf{3408.678 (±824.48)} & 1.05\text{e+}5 (±2.11\text{e+}4) & 2.74\text{e+}4 (±9.29\text{e+}3) \\
4 & \textbf{3863.234 (±1024.26)} & 1.05\text{e+}5 (±2.11\text{e+}4) & 3.17\text{e+}4 (±1.09\text{e+}4) \\
5 & \textbf{4125.150 (±1184.17)} & 1.05\text{e+}5 (±2.11\text{e+}4) & 3.60\text{e+}4 (±1.23\text{e+}4) \\
6 & \textbf{4886.944 (±1443.76)} & 1.05\text{e+}5 (±2.11\text{e+}4) & 4.28\text{e+}4 (±1.46\text{e+}4) \\
7 & \textbf{4576.617 (±1385.49)} & 1.05\text{e+}5 (±2.11\text{e+}4) & 4.09\text{e+}4 (±1.41\text{e+}4) \\
\bottomrule
\end{tabular}
\end{table}

\clearpage

\subsubsection{Varying Single Fluorophore Intensity}

\begin{table}[htbp]
\captionof{table}{Accuracy}
\begin{tabular}{ccccc}
\toprule
Mean Fluorophore Intensity ($\mu_f$) & CRJMCMC & Factorial HMM & Monotonic MAP & Sequential MAP \\
\midrule
2500 & \textbf{0.991 (±4.29\text{e-}3)} & \textbf{0.950 (±4.94\text{e-}2)} & 0.912 (±1.91\text{e-}2) & 0.964 (±1.91\text{e-}2) \\
3000 & \textbf{0.992 (±2.67\text{e-}3)} & 0.927 (±5.78\text{e-}2) & 0.903 (±2.09\text{e-}2) & 0.965 (±2.02\text{e-}2) \\
3500 & \textbf{0.994 (±2.19\text{e-}3)} & 0.914 (±5.85\text{e-}2) & 0.905 (±2.09\text{e-}2) & 0.975 (±1.23\text{e-}2) \\
4000 & \textbf{0.995 (±1.72\text{e-}3)} & 0.834 (±8.25\text{e-}2) & 0.927 (±1.77\text{e-}2) & 0.969 (±1.91\text{e-}2) \\
4500 & \textbf{0.995 (±1.56\text{e-}3)} & \textbf{0.982 (±2.06\text{e-}2)} & 0.892 (±2.30\text{e-}2) & \textbf{0.977 (±1.65\text{e-}2)} \\
5000 & \textbf{0.992 (±4.08\text{e-}3)} & \textbf{0.969 (±5.40\text{e-}2)} & 0.891 (±1.86\text{e-}2) & 0.976 (±1.13\text{e-}2) \\
6000 & \textbf{0.993 (±3.42\text{e-}3)} & \textbf{0.963 (±4.47\text{e-}2)} & 0.892 (±2.25\text{e-}2) & 0.949 (±2.80\text{e-}2) \\
7000 & \textbf{0.994 (±3.26\text{e-}3)} & 0.908 (±7.11\text{e-}2) & 0.909 (±1.90\text{e-}2) & 0.960 (±2.57\text{e-}2) \\
8000 & \textbf{0.997 (±1.30\text{e-}3)} & 0.826 (±7.30\text{e-}2) & 0.915 (±1.80\text{e-}2) & 0.960 (±2.38\text{e-}2) \\
9000 & \textbf{0.991 (±1.22\text{e-}2)} & \textbf{0.976 (±4.29\text{e-}2)} & 0.901 (±1.89\text{e-}2) & \textbf{0.966 (±2.14\text{e-}2)} \\
10000 & \textbf{0.993 (±3.68\text{e-}3)} & \textbf{0.985 (±2.03\text{e-}2)} & 0.905 (±2.10\text{e-}2) & 0.966 (±1.73\text{e-}2) \\
\bottomrule
\end{tabular}
\end{table}

\begin{table}[htbp]
\captionof{table}{Precision}
\begin{tabular}{ccccc}
\toprule
Mean Fluorophore Intensity ($\mu_f$) & CRJMCMC & Factorial HMM & Monotonic MAP & Sequential MAP \\
\midrule
2500 & \textbf{0.991 (±4.85\text{e-}3)} & \textbf{0.995 (±5.03\text{e-}3)} & 0.920 (±2.01\text{e-}2) & \textbf{0.969 (±2.56\text{e-}2)} \\
3000 & \textbf{0.993 (±2.34\text{e-}3)} & \textbf{0.990 (±7.20\text{e-}3)} & 0.912 (±2.19\text{e-}2) & \textbf{0.968 (±2.52\text{e-}2)} \\
3500 & \textbf{0.995 (±2.48\text{e-}3)} & \textbf{0.993 (±2.78\text{e-}3)} & 0.915 (±2.14\text{e-}2) & \textbf{0.974 (±1.93\text{e-}2)} \\
4000 & \textbf{0.995 (±2.40\text{e-}3)} & \textbf{0.995 (±3.45\text{e-}3)} & 0.939 (±1.66\text{e-}2) & \textbf{0.978 (±1.71\text{e-}2)} \\
4500 & 0.994 (±2.44\text{e-}3) & \textbf{0.998 (±1.11\text{e-}3)} & 0.888 (±2.57\text{e-}2) & \textbf{0.980 (±1.84\text{e-}2)} \\
5000 & \textbf{0.992 (±4.69\text{e-}3)} & \textbf{0.997 (±1.94\text{e-}3)} & 0.922 (±1.72\text{e-}2) & 0.985 (±7.85\text{e-}3) \\
6000 & \textbf{0.993 (±4.68\text{e-}3)} & \textbf{0.996 (±3.57\text{e-}3)} & 0.920 (±2.10\text{e-}2) & \textbf{0.962 (±3.12\text{e-}2)} \\
7000 & \textbf{0.994 (±3.79\text{e-}3)} & \textbf{0.997 (±1.51\text{e-}3)} & 0.926 (±1.91\text{e-}2) & \textbf{0.983 (±1.37\text{e-}2)} \\
8000 & \textbf{0.997 (±2.17\text{e-}3)} & \textbf{0.999 (±1.08\text{e-}3)} & 0.936 (±1.73\text{e-}2) & 0.968 (±2.67\text{e-}2) \\
9000 & \textbf{0.988 (±1.81\text{e-}2)} & \textbf{0.998 (±1.09\text{e-}3)} & 0.922 (±1.86\text{e-}2) & \textbf{0.975 (±2.42\text{e-}2)} \\
10000 & \textbf{0.995 (±2.66\text{e-}3)} & \textbf{0.998 (±1.42\text{e-}3)} & 0.916 (±2.28\text{e-}2) & 0.964 (±2.75\text{e-}2) \\
\bottomrule
\end{tabular}
\end{table}

\begin{table}[htbp]
\captionof{table}{Sensitivity}
\begin{tabular}{ccccc}
\toprule
Mean Fluorophore Intensity ($\mu_f$) & CRJMCMC & Factorial HMM & Monotonic MAP & Sequential MAP \\
\midrule
2500 & \textbf{0.996 (±2.10\text{e-}3)} & \textbf{0.936 (±5.92\text{e-}2)} & 0.949 (±2.25\text{e-}2) & 0.957 (±3.17\text{e-}2) \\
3000 & \textbf{0.996 (±2.49\text{e-}3)} & 0.911 (±7.80\text{e-}2) & 0.960 (±1.90\text{e-}2) & \textbf{0.976 (±1.93\text{e-}2)} \\
3500 & \textbf{0.997 (±2.09\text{e-}3)} & 0.887 (±7.98\text{e-}2) & 0.950 (±2.46\text{e-}2) & \textbf{0.980 (±1.81\text{e-}2)} \\
4000 & \textbf{0.997 (±1.15\text{e-}3)} & 0.809 (±9.55\text{e-}2) & 0.948 (±2.50\text{e-}2) & \textbf{0.980 (±1.76\text{e-}2)} \\
4500 & \textbf{0.999 (±5.91\text{e-}4)} & \textbf{0.981 (±2.45\text{e-}2)} & 0.972 (±1.36\text{e-}2) & 0.983 (±1.51\text{e-}2) \\
5000 & \textbf{0.998 (±7.98\text{e-}4)} & \textbf{0.965 (±6.61\text{e-}2)} & 0.915 (±2.96\text{e-}2) & 0.978 (±1.85\text{e-}2) \\
6000 & \textbf{0.997 (±1.64\text{e-}3)} & \textbf{0.958 (±5.71\text{e-}2)} & 0.917 (±2.94\text{e-}2) & 0.939 (±3.87\text{e-}2) \\
7000 & \textbf{0.999 (±7.13\text{e-}4)} & 0.888 (±8.83\text{e-}2) & 0.952 (±2.21\text{e-}2) & 0.966 (±2.70\text{e-}2) \\
8000 & \textbf{0.999 (±6.04\text{e-}4)} & 0.752 (±0.11) & 0.939 (±2.39\text{e-}2) & 0.954 (±3.46\text{e-}2) \\
9000 & \textbf{0.999 (±4.52\text{e-}4)} & \textbf{0.972 (±5.34\text{e-}2)} & 0.943 (±2.03\text{e-}2) & 0.970 (±2.27\text{e-}2) \\
10000 & \textbf{0.995 (±3.90\text{e-}3)} & \textbf{0.982 (±2.84\text{e-}2)} & 0.949 (±1.94\text{e-}2) & \textbf{0.964 (±2.80\text{e-}2)} \\
\bottomrule
\end{tabular}
\end{table}

\begin{table}
\captionof{table}{Specificity}
\begin{tabular}{ccccc}
\toprule
Mean Fluorophore Intensity ($\mu_f$) & CRJMCMC & Factorial HMM & Monotonic MAP & Sequential MAP \\
\midrule
2500 & \textbf{0.971 (±1.16\text{e-}2)} & \textbf{0.981 (±9.12\text{e-}3)} & 0.782 (±4.70\text{e-}2) & \textbf{0.967 (±9.31\text{e-}3)} \\
3000 & \textbf{0.972 (±1.36\text{e-}2)} & \textbf{0.964 (±1.99\text{e-}2)} & 0.749 (±5.45\text{e-}2) & \textbf{0.941 (±2.28\text{e-}2)} \\
3500 & \textbf{0.976 (±9.43\text{e-}3)} & \textbf{0.968 (±1.78\text{e-}2)} & 0.768 (±4.91\text{e-}2) & \textbf{0.949 (±2.01\text{e-}2)} \\
4000 & \textbf{0.981 (±7.78\text{e-}3)} & \textbf{0.983 (±1.06\text{e-}2)} & 0.824 (±4.32\text{e-}2) & \textbf{0.953 (±2.14\text{e-}2)} \\
4500 & 0.978 (±6.73\text{e-}3) & \textbf{0.991 (±5.43\text{e-}3)} & 0.713 (±5.48\text{e-}2) & 0.956 (±1.91\text{e-}2) \\
5000 & \textbf{0.968 (±1.45\text{e-}2)} & \textbf{0.986 (±8.33\text{e-}3)} & 0.763 (±4.84\text{e-}2) & 0.955 (±1.81\text{e-}2) \\
6000 & \textbf{0.980 (±7.86\text{e-}3)} & \textbf{0.989 (±6.63\text{e-}3)} & 0.794 (±4.77\text{e-}2) & 0.953 (±2.26\text{e-}2) \\
7000 & \textbf{0.976 (±9.63\text{e-}3)} & \textbf{0.987 (±7.52\text{e-}3)} & 0.772 (±5.30\text{e-}2) & 0.961 (±1.80\text{e-}2) \\
8000 & 0.988 (±4.45\text{e-}3) & \textbf{0.996 (±3.54\text{e-}3)} & 0.803 (±4.52\text{e-}2) & 0.964 (±1.66\text{e-}2) \\
9000 & \textbf{0.982 (±1.33\text{e-}2)} & \textbf{0.989 (±6.16\text{e-}3)} & 0.760 (±5.12\text{e-}2) & \textbf{0.965 (±1.84\text{e-}2)} \\
10000 & \textbf{0.981 (±1.28\text{e-}2)} & \textbf{0.991 (±5.35\text{e-}3)} & 0.796 (±4.94\text{e-}2) & 0.966 (±1.38\text{e-}2) \\
\bottomrule
\end{tabular}
\end{table}

\begin{table}[htbp]
\captionof{table}{Cohen's Kappa}
\begin{tabular}{ccccc}
\toprule
Mean Fluorophore Intensity ($\mu_f$) & CRJMCMC & Factorial HMM & Monotonic MAP & Sequential MAP \\
\midrule
2500 & \textbf{0.983 (±6.20\text{e-}3)} & \textbf{0.930 (±5.83\text{e-}2)} & 0.829 (±3.71\text{e-}2) & 0.941 (±2.63\text{e-}2) \\
3000 & \textbf{0.985 (±4.64\text{e-}3)} & 0.900 (±7.64\text{e-}2) & 0.810 (±4.09\text{e-}2) & 0.946 (±2.48\text{e-}2) \\
3500 & \textbf{0.989 (±3.66\text{e-}3)} & 0.881 (±7.66\text{e-}2) & 0.827 (±3.69\text{e-}2) & 0.957 (±1.56\text{e-}2) \\
4000 & \textbf{0.991 (±2.93\text{e-}3)} & 0.781 (±0.10) & 0.861 (±3.43\text{e-}2) & 0.952 (±2.36\text{e-}2) \\
4500 & \textbf{0.989 (±3.44\text{e-}3)} & \textbf{0.974 (±3.04\text{e-}2)} & 0.811 (±3.87\text{e-}2) & 0.964 (±1.73\text{e-}2) \\
5000 & \textbf{0.986 (±6.01\text{e-}3)} & \textbf{0.961 (±6.51\text{e-}2)} & 0.793 (±3.59\text{e-}2) & 0.957 (±1.67\text{e-}2) \\
6000 & \textbf{0.988 (±4.68\text{e-}3)} & \textbf{0.950 (±5.65\text{e-}2)} & 0.804 (±4.01\text{e-}2) & 0.926 (±3.58\text{e-}2) \\
7000 & \textbf{0.990 (±4.63\text{e-}3)} & 0.876 (±9.00\text{e-}2) & 0.818 (±4.05\text{e-}2) & 0.943 (±3.10\text{e-}2) \\
8000 & \textbf{0.994 (±2.25\text{e-}3)} & 0.757 (±9.82\text{e-}2) & 0.834 (±3.60\text{e-}2) & 0.942 (±2.93\text{e-}2) \\
9000 & \textbf{0.988 (±1.49\text{e-}2)} & \textbf{0.970 (±5.22\text{e-}2)} & 0.807 (±3.83\text{e-}2) & \textbf{0.949 (±2.77\text{e-}2)} \\
10000 & \textbf{0.987 (±6.29\text{e-}3)} & \textbf{0.980 (±2.64\text{e-}2)} & 0.836 (±3.49\text{e-}2) & 0.949 (±2.28\text{e-}2) \\
\bottomrule
\end{tabular}
\end{table}
\begin{table}
\captionof{table}{RMSE (Intensity)}
\begin{tabular}{ccccc}
\toprule
Mean Fluorophore Intensity ($\mu_f$) & CRJMCMC & Factorial HMM & Monotonic MAP & Sequential MAP \\
\midrule
2500 & \textbf{192.671 (±27.85)} & 362.015 (±103.43) & 630.363 (±71.36) & 391.719 (±52.84) \\
3000 & \textbf{198.495 (±35.41)} & 337.941 (±65.30) & 829.617 (±94.75) & 470.297 (±49.80) \\
3500 & \textbf{184.374 (±36.36)} & 312.230 (±56.81) & 911.034 (±105.76) & 503.828 (±60.61) \\
4000 & \textbf{214.380 (±38.59)} & 437.893 (±81.83) & 914.579 (±111.94) & 578.494 (±72.50) \\
4500 & \textbf{234.492 (±42.56)} & \textbf{251.823 (±44.86)} & 1173.936 (±135.15) & 609.569 (±76.91) \\
5000 & \textbf{269.949 (±52.95)} & \textbf{359.809 (±49.39)} & 1418.806 (±151.33) & 645.141 (±72.68) \\
6000 & \textbf{328.789 (±52.95)} & \textbf{456.954 (±75.31)} & 1643.841 (±186.25) & 959.759 (±134.62) \\
7000 & \textbf{317.656 (±64.92)} & \textbf{374.293 (±76.08)} & 1743.279 (±218.63) & 968.737 (±156.01) \\
8000 & \textbf{272.036 (±57.29)} & \textbf{424.197 (±131.44)} & 1912.649 (±236.62) & 1099.648 (±154.47) \\
9000 & \textbf{312.429 (±63.93)} & 628.352 (±109.69) & 2387.365 (±266.55) & 1153.934 (±195.98) \\
10000 & \textbf{599.823 (±119.03)} & \textbf{913.568 (±228.73)} & 2462.077 (±293.87) & 1410.130 (±209.28) \\
\bottomrule
\end{tabular}
\end{table}

\begin{table}
\captionof{table}{Absolute Error ($\mu_f$)}
\begin{tabular}{ccccc}
\toprule
Mean Fluorophore Intensity ($\mu_f$) & CRJMCMC & Factorial HMM & Monotonic MAP & Sequential MAP \\
\midrule
2500 & \textbf{79.100 (±15.09)} & 146.574 (±43.15) & 307.156 (±11.67) & \textbf{173.454 (±83.97)} \\
3000 & \textbf{34.157 (±6.91)} & \textbf{78.026 (±42.51)} & 337.979 (±17.37) & 153.203 (±58.72) \\
3500 & \textbf{30.226 (±5.48)} & \textbf{43.733 (±15.20)} & 464.887 (±12.57) & 134.555 (±53.82) \\
4000 & \textbf{66.391 (±10.60)} & 106.996 (±28.97) & 453.002 (±12.39) & 202.420 (±87.84) \\
4500 & \textbf{38.290 (±5.55)} & 63.451 (±10.61) & 683.396 (±39.74) & 142.536 (±54.65) \\
5000 & \textbf{72.863 (±12.39)} & 132.563 (±18.39) & 530.292 (±24.68) & 223.139 (±94.73) \\
6000 & \textbf{78.464 (±14.35)} & 173.596 (±33.29) & 730.679 (±12.64) & 451.369 (±221.98) \\
7000 & \textbf{52.996 (±9.93)} & \textbf{89.986 (±27.50)} & 694.530 (±8.09) & 355.012 (±205.96) \\
8000 & \textbf{39.525 (±10.92)} & 71.400 (±17.91) & 772.964 (±21.51) & 480.202 (±256.34) \\
9000 & \textbf{103.276 (±79.29)} & \textbf{139.821 (±53.23)} & 1230.972 (±43.25) & 420.727 (±232.29) \\
10000 & \textbf{156.759 (±27.56)} & 372.921 (±78.51) & 1168.568 (±85.44) & 552.130 (±303.58) \\
\bottomrule
\end{tabular}
\end{table}

\begin{table}
\captionof{table}{Absolute Error ($\mu_b$)}
\begin{tabular}{ccccc}
\toprule
Mean Fluorophore Intensity ($\mu_f$) & CRJMCMC & Factorial HMM & Monotonic MAP & Sequential MAP \\
\midrule
2500 & \textbf{10.528 (±1.70)} & 122.982 (±44.43) & 224.912 (±11.11) & 52.973 (±14.18) \\
3000 & \textbf{9.701 (±1.23)} & 113.878 (±38.21) & 272.931 (±14.64) & 66.360 (±18.03) \\
3500 & \textbf{14.072 (±2.41)} & 131.282 (±44.75) & 387.326 (±10.30) & 66.719 (±25.03) \\
4000 & \textbf{22.313 (±4.26)} & 235.143 (±66.44) & 355.590 (±17.73) & 78.767 (±18.17) \\
4500 & \textbf{9.438 (±2.06)} & 59.830 (±18.58) & 518.807 (±30.25) & 73.769 (±18.09) \\
5000 & \textbf{22.939 (±3.63)} & 80.213 (±22.14) & 453.217 (±24.65) & 90.451 (±34.05) \\
6000 & \textbf{5.117 (±0.83)} & 105.215 (±37.78) & 627.836 (±8.89) & 113.146 (±34.27) \\
7000 & \textbf{32.928 (±5.66)} & 148.148 (±64.15) & 615.900 (±20.53) & 84.441 (±20.39) \\
8000 & \textbf{15.763 (±2.72)} & 301.868 (±117.54) & 648.068 (±21.30) & 117.994 (±34.28) \\
9000 & \textbf{10.864 (±1.91)} & 260.592 (±121.31) & 1055.900 (±44.34) & 97.625 (±19.12) \\
10000 & \textbf{9.632 (±2.42)} & 212.941 (±81.91) & 1012.948 (±86.00) & 122.072 (±27.77) \\
\bottomrule
\end{tabular}
\end{table}

\begin{table}
\captionof{table}{Absolute Error ($\sigma_f^2$)}
\begin{tabular}{cccc}
\toprule
Mean Fluorophore Intensity ($\mu_f$) & CRJMCMC & Monotonic MAP & Sequential MAP \\
\midrule
2500 & \textbf{277.450 (±60.76)} & 2.21\text{e+}5 (±3.39\text{e+}4) & 2184.125 (±41.02) \\
3000 & \textbf{468.956 (±57.12)} & 2.55\text{e+}5 (±3.66\text{e+}4) & 2576.461 (±56.81) \\
3500 & \textbf{223.983 (±26.87)} & 3.05\text{e+}5 (±3.23\text{e+}4) & 3091.339 (±53.21) \\
4000 & \textbf{341.744 (±35.84)} & 4.17\text{e+}5 (±6.50\text{e+}4) & 3528.171 (±61.40) \\
4500 & \textbf{184.510 (±22.56)} & 6.39\text{e+}5 (±4.12\text{e+}4) & 4064.180 (±65.13) \\
5000 & \textbf{266.776 (±38.58)} & 4.62\text{e+}5 (±3.59\text{e+}4) & 4488.932 (±73.23) \\
6000 & \textbf{245.586 (±33.77)} & 5.77\text{e+}5 (±5.03\text{e+}4) & 5379.078 (±86.23) \\
7000 & \textbf{342.481 (±46.18)} & 5.39\text{e+}5 (±8.43\text{e+}4) & 6378.420 (±86.85) \\
8000 & \textbf{536.228 (±39.16)} & 9.25\text{e+}5 (±1.07\text{e+}5) & 7258.317 (±111.38) \\
9000 & \textbf{350.976 (±10.74)} & 1.54\text{e+}6 (±1.33\text{e+}5) & 8194.348 (±117.37) \\
10000 & \textbf{500.303 (±140.91)} & 1.48\text{e+}6 (±2.10\text{e+}5) & 9139.014 (±130.86) \\
\bottomrule
\end{tabular}
\end{table}

\begin{table}
    
\captionof{table}{Absolute Error ($\sigma_b^2$)}
\begin{tabular}{cccc}
\toprule
Mean Fluorophore Intensity ($\mu_f$) & CRJMCMC & Monotonic MAP & Sequential MAP \\
\midrule
2500 & \textbf{4.47\text{e+}4 (±1.11\text{e+}4)} & 3.04\text{e+}5 (±6.08\text{e+}3) & 7.68\text{e+}4 (±1.99\text{e+}4) \\
3000 & \textbf{1.94\text{e+}4 (±1.56\text{e+}3)} & 4.18\text{e+}5 (±1.61\text{e+}4) & 1.17\text{e+}5 (±2.60\text{e+}4) \\
3500 & \textbf{1.68\text{e+}4 (±2.36\text{e+}3)} & 6.37\text{e+}5 (±1.92\text{e+}4) & 1.22\text{e+}5 (±2.92\text{e+}4) \\
4000 & \textbf{2.25\text{e+}4 (±3.28\text{e+}3)} & 7.58\text{e+}5 (±1.49\text{e+}4) & 1.60\text{e+}5 (±3.42\text{e+}4) \\
4500 & \textbf{3.50\text{e+}4 (±6.79\text{e+}3)} & 1.12\text{e+}6 (±4.45\text{e+}4) & 1.38\text{e+}5 (±3.58\text{e+}4) \\
5000 & \textbf{6.60\text{e+}4 (±1.33\text{e+}4)} & 1.10\text{e+}6 (±5.26\text{e+}4) & 1.62\text{e+}5 (±4.06\text{e+}4) \\
6000 & \textbf{2.49\text{e+}4 (±2.81\text{e+}3)} & 1.91\text{e+}6 (±3.55\text{e+}4) & 1.95\text{e+}5 (±4.88\text{e+}4) \\
7000 & \textbf{1.47\text{e+}5 (±2.60\text{e+}4)} & 2.20\text{e+}6 (±5.10\text{e+}4) & 2.39\text{e+}5 (±5.81\text{e+}4) \\
8000 & \textbf{1.53\text{e+}5 (±1.98\text{e+}4)} & 2.76\text{e+}6 (±9.80\text{e+}4) & 2.95\text{e+}5 (±6.77\text{e+}4) \\
9000 & \textbf{2.10\text{e+}5 (±3.09\text{e+}4)} & 4.82\text{e+}6 (±1.20\text{e+}5) & 3.32\text{e+}5 (±7.62\text{e+}4) \\
10000 & \textbf{1.16\text{e+}5 (±7.47\text{e+}3)} & 4.75\text{e+}6 (±2.67\text{e+}5) & 2.99\text{e+}5 (±7.88\text{e+}4) \\
\bottomrule
\end{tabular}
\end{table}

\clearpage

\subsubsection{Varying Dark State Transition Probability}

\begin{table}[htbp]
\captionof{table}{Accuracy}
\begin{tabular}{ccccc}
\toprule
Dark State Probability & CRJMCMC & Factorial HMM & Monotonic MAP & Sequential MAP \\
\midrule
0.0002 & \textbf{0.983 (±4.26\text{e-}3)} & 0.813 (±3.57\text{e-}2) & 0.895 (±1.17\text{e-}2) & 0.920 (±1.75\text{e-}2) \\
0.0004 & \textbf{0.982 (±4.15\text{e-}3)} & 0.852 (±2.90\text{e-}2) & 0.842 (±1.45\text{e-}2) & 0.883 (±2.13\text{e-}2) \\
0.0006 & \textbf{0.981 (±3.88\text{e-}3)} & 0.850 (±2.84\text{e-}2) & 0.785 (±1.42\text{e-}2) & 0.880 (±2.09\text{e-}2) \\
0.0008 & \textbf{0.977 (±4.53\text{e-}3)} & 0.854 (±2.96\text{e-}2) & 0.772 (±1.41\text{e-}2) & 0.877 (±2.03\text{e-}2) \\
0.001 & \textbf{0.971 (±5.07\text{e-}3)} & 0.875 (±2.19\text{e-}2) & 0.748 (±1.45\text{e-}2) & 0.862 (±2.06\text{e-}2) \\
0.0012 & \textbf{0.964 (±6.03\text{e-}3)} & 0.870 (±2.55\text{e-}2) & 0.731 (±1.49\text{e-}2) & 0.871 (±1.95\text{e-}2) \\
0.0014 & \textbf{0.955 (±6.69\text{e-}3)} & 0.854 (±2.71\text{e-}2) & 0.726 (±1.38\text{e-}2) & 0.867 (±1.88\text{e-}2) \\
0.0016 & \textbf{0.956 (±5.86\text{e-}3)} & 0.850 (±2.92\text{e-}2) & 0.726 (±1.45\text{e-}2) & 0.871 (±1.81\text{e-}2) \\
0.0018 & \textbf{0.947 (±6.74\text{e-}3)} & 0.884 (±2.18\text{e-}2) & 0.702 (±1.34\text{e-}2) & 0.851 (±1.89\text{e-}2) \\
0.002 & \textbf{0.947 (±6.56\text{e-}3)} & 0.887 (±2.10\text{e-}2) & 0.720 (±1.27\text{e-}2) & 0.877 (±1.54\text{e-}2) \\
\bottomrule
\end{tabular}
\end{table}

\begin{table}[htbp]
\captionof{table}{Precision}
\begin{tabular}{ccccc}
\toprule
Dark State Probability & CRJMCMC & Factorial HMM & Monotonic MAP & Sequential MAP \\
\midrule
0.0002 & \textbf{0.993 (±2.17\text{e-}3)} & 0.961 (±1.59\text{e-}2) & 0.906 (±1.29\text{e-}2) & 0.942 (±1.81\text{e-}2) \\
0.0004 & \textbf{0.989 (±3.88\text{e-}3)} & 0.963 (±1.14\text{e-}2) & 0.855 (±1.59\text{e-}2) & 0.919 (±2.25\text{e-}2) \\
0.0006 & \textbf{0.986 (±3.64\text{e-}3)} & 0.962 (±1.04\text{e-}2) & 0.804 (±1.73\text{e-}2) & 0.906 (±2.50\text{e-}2) \\
0.0008 & \textbf{0.980 (±5.21\text{e-}3)} & 0.940 (±1.89\text{e-}2) & 0.789 (±1.83\text{e-}2) & 0.900 (±2.62\text{e-}2) \\
0.001 & \textbf{0.970 (±6.91\text{e-}3)} & \textbf{0.947 (±1.86\text{e-}2)} & 0.759 (±1.89\text{e-}2) & 0.891 (±2.58\text{e-}2) \\
0.0012 & \textbf{0.959 (±8.21\text{e-}3)} & \textbf{0.939 (±1.47\text{e-}2)} & 0.727 (±2.02\text{e-}2) & 0.906 (±2.37\text{e-}2) \\
0.0014 & \textbf{0.948 (±9.00\text{e-}3)} & \textbf{0.940 (±1.79\text{e-}2)} & 0.715 (±1.97\text{e-}2) & 0.863 (±2.96\text{e-}2) \\
0.0016 & \textbf{0.949 (±8.06\text{e-}3)} & \textbf{0.929 (±1.97\text{e-}2)} & 0.712 (±2.05\text{e-}2) & 0.872 (±2.73\text{e-}2) \\
0.0018 & \textbf{0.936 (±9.90\text{e-}3)} & \textbf{0.918 (±2.36\text{e-}2)} & 0.683 (±2.08\text{e-}2) & 0.846 (±3.18\text{e-}2) \\
0.002 & \textbf{0.930 (±9.77\text{e-}3)} & \textbf{0.928 (±1.78\text{e-}2)} & 0.690 (±2.05\text{e-}2) & 0.859 (±2.91\text{e-}2) \\
\bottomrule
\end{tabular}
\end{table}

\begin{table}[htbp]
\captionof{table}{Sensitivity}
\begin{tabular}{ccccc}
\toprule
Dark State Probability & CRJMCMC & Factorial HMM & Monotonic MAP & Sequential MAP \\
\midrule
0.0002 & \textbf{0.983 (±4.79\text{e-}3)} & 0.780 (±4.55\text{e-}2) & 0.944 (±1.31\text{e-}2) & 0.919 (±2.39\text{e-}2) \\
0.0004 & \textbf{0.982 (±4.73\text{e-}3)} & 0.820 (±4.12\text{e-}2) & 0.895 (±1.85\text{e-}2) & 0.856 (±3.18\text{e-}2) \\
0.0006 & \textbf{0.982 (±4.10\text{e-}3)} & 0.793 (±4.57\text{e-}2) & 0.843 (±2.16\text{e-}2) & 0.838 (±3.38\text{e-}2) \\
0.0008 & \textbf{0.981 (±3.96\text{e-}3)} & 0.801 (±4.54\text{e-}2) & 0.797 (±2.39\text{e-}2) & 0.809 (±3.66\text{e-}2) \\
0.001 & \textbf{0.975 (±4.41\text{e-}3)} & 0.820 (±3.89\text{e-}2) & 0.765 (±2.47\text{e-}2) & 0.794 (±3.70\text{e-}2) \\
0.0012 & \textbf{0.969 (±5.29\text{e-}3)} & 0.804 (±4.53\text{e-}2) & 0.732 (±2.57\text{e-}2) & 0.781 (±3.78\text{e-}2) \\
0.0014 & \textbf{0.956 (±7.07\text{e-}3)} & 0.747 (±5.43\text{e-}2) & 0.708 (±2.70\text{e-}2) & 0.764 (±3.83\text{e-}2) \\
0.0016 & \textbf{0.956 (±6.64\text{e-}3)} & 0.756 (±5.32\text{e-}2) & 0.675 (±2.77\text{e-}2) & 0.767 (±3.78\text{e-}2) \\
0.0018 & \textbf{0.943 (±7.14\text{e-}3)} & 0.807 (±4.70\text{e-}2) & 0.626 (±2.77\text{e-}2) & 0.706 (±4.12\text{e-}2) \\
0.002 & \textbf{0.931 (±9.71\text{e-}3)} & 0.778 (±4.93\text{e-}2) & 0.589 (±2.81\text{e-}2) & 0.734 (±3.85\text{e-}2) \\
\bottomrule
\end{tabular}
\end{table}

\begin{table}[htbp]
\captionof{table}{Specificity}
\begin{tabular}{ccccc}
\toprule
Dark State Probability & CRJMCMC & Factorial HMM & Monotonic MAP & Sequential MAP \\
\midrule
0.0002 & \textbf{0.980 (±4.07\text{e-}3)} & 0.922 (±2.22\text{e-}2) & 0.750 (±2.91\text{e-}2) & 0.922 (±1.40\text{e-}2) \\
0.0004 & \textbf{0.979 (±4.72\text{e-}3)} & 0.930 (±1.85\text{e-}2) & 0.689 (±3.17\text{e-}2) & 0.929 (±1.32\text{e-}2) \\
0.0006 & \textbf{0.975 (±5.19\text{e-}3)} & 0.932 (±2.09\text{e-}2) & 0.639 (±3.23\text{e-}2) & 0.940 (±1.08\text{e-}2) \\
0.0008 & \textbf{0.969 (±7.76\text{e-}3)} & 0.935 (±2.03\text{e-}2) & 0.662 (±3.09\text{e-}2) & 0.946 (±1.02\text{e-}2) \\
0.001 & \textbf{0.963 (±8.08\text{e-}3)} & \textbf{0.942 (±1.76\text{e-}2)} & 0.656 (±3.05\text{e-}2) & 0.936 (±1.25\text{e-}2) \\
0.0012 & \textbf{0.956 (±8.90\text{e-}3)} & \textbf{0.942 (±1.40\text{e-}2)} & 0.653 (±3.07\text{e-}2) & \textbf{0.957 (±8.06\text{e-}3)} \\
0.0014 & \textbf{0.952 (±8.51\text{e-}3)} & \textbf{0.953 (±1.50\text{e-}2)} & 0.670 (±2.94\text{e-}2) & \textbf{0.955 (±8.27\text{e-}3)} \\
0.0016 & \textbf{0.952 (±8.20\text{e-}3)} & \textbf{0.950 (±1.56\text{e-}2)} & 0.703 (±2.80\text{e-}2) & \textbf{0.956 (±6.98\text{e-}3)} \\
0.0018 & 0.944 (±8.81\text{e-}3) & 0.945 (±1.43\text{e-}2) & 0.692 (±2.81\text{e-}2) & \textbf{0.966 (±5.86\text{e-}3)} \\
0.002 & \textbf{0.952 (±7.18\text{e-}3)} & \textbf{0.957 (±1.19\text{e-}2)} & 0.740 (±2.59\text{e-}2) & \textbf{0.963 (±6.86\text{e-}3)} \\
\bottomrule
\end{tabular}
\end{table}

\begin{table}[htbp]
\captionof{table}{Cohen's Kappa}
\begin{tabular}{ccccc}
\toprule
Dark State Probability & CRJMCMC & Factorial HMM & Monotonic MAP & Sequential MAP \\
\midrule
0.0002 & \textbf{0.970 (±6.80\text{e-}3)} & 0.741 (±4.54\text{e-}2) & 0.807 (±2.12\text{e-}2) & 0.872 (±2.45\text{e-}2) \\
0.0004 & \textbf{0.968 (±6.70\text{e-}3)} & 0.782 (±4.03\text{e-}2) & 0.709 (±2.54\text{e-}2) & 0.812 (±3.13\text{e-}2) \\
0.0006 & \textbf{0.966 (±6.29\text{e-}3)} & 0.765 (±4.35\text{e-}2) & 0.601 (±2.59\text{e-}2) & 0.803 (±3.25\text{e-}2) \\
0.0008 & \textbf{0.961 (±7.11\text{e-}3)} & 0.768 (±4.38\text{e-}2) & 0.564 (±2.63\text{e-}2) & 0.781 (±3.44\text{e-}2) \\
0.001 & \textbf{0.949 (±8.27\text{e-}3)} & 0.788 (±3.73\text{e-}2) & 0.527 (±2.60\text{e-}2) & 0.754 (±3.52\text{e-}2) \\
0.0012 & \textbf{0.937 (±9.71\text{e-}3)} & 0.781 (±4.10\text{e-}2) & 0.485 (±2.57\text{e-}2) & 0.758 (±3.52\text{e-}2) \\
0.0014 & \textbf{0.920 (±1.12\text{e-}2)} & 0.732 (±4.95\text{e-}2) & 0.460 (±2.48\text{e-}2) & 0.738 (±3.59\text{e-}2) \\
0.0016 & \textbf{0.921 (±1.01\text{e-}2)} & 0.733 (±4.98\text{e-}2) & 0.458 (±2.57\text{e-}2) & 0.743 (±3.49\text{e-}2) \\
0.0018 & \textbf{0.902 (±1.17\text{e-}2)} & 0.774 (±4.39\text{e-}2) & 0.393 (±2.26\text{e-}2) & 0.690 (±3.82\text{e-}2) \\
0.002 & \textbf{0.895 (±1.27\text{e-}2)} & 0.763 (±4.45\text{e-}2) & 0.399 (±2.17\text{e-}2) & 0.725 (±3.49\text{e-}2) \\
\bottomrule
\end{tabular}
\end{table}

\begin{table}[htbp]
\captionof{table}{RMSE (Intensity)}
\begin{tabular}{ccccc}
\toprule
Dark State Probability & CRJMCMC & Factorial HMM & Monotonic MAP & Sequential MAP \\
\midrule
0.0002 & \textbf{92.948 (±9.31)} & 199.437 (±26.74) & 321.784 (±26.82) & 208.747 (±16.87) \\
0.0004 & \textbf{101.704 (±11.00)} & 186.822 (±25.14) & 401.017 (±32.32) & 226.243 (±17.41) \\
0.0006 & \textbf{109.522 (±10.78)} & 183.822 (±24.07) & 486.953 (±35.75) & 241.987 (±18.96) \\
0.0008 & \textbf{118.616 (±13.28)} & 179.914 (±22.33) & 495.207 (±35.21) & 244.268 (±19.43) \\
0.001 & \textbf{140.264 (±15.57)} & 191.509 (±24.20) & 520.888 (±35.26) & 255.998 (±18.51) \\
0.0012 & \textbf{155.684 (±16.02)} & \textbf{187.835 (±22.62)} & 542.455 (±36.38) & 257.035 (±18.00) \\
0.0014 & \textbf{176.002 (±17.91)} & \textbf{178.826 (±21.89)} & 554.943 (±35.94) & 275.841 (±19.41) \\
0.0016 & \textbf{180.220 (±17.80)} & \textbf{179.908 (±21.81)} & 563.542 (±38.19) & 286.804 (±22.41) \\
0.0018 & \textbf{204.166 (±20.28)} & \textbf{192.638 (±21.82)} & 596.184 (±38.81) & 308.452 (±23.76) \\
0.002 & \textbf{201.434 (±19.65)} & \textbf{188.301 (±20.72)} & 566.396 (±35.42) & 289.854 (±21.45) \\
\bottomrule
\end{tabular}
\end{table}

\begin{table}[htbp]
\captionof{table}{Absolute Error ($\mu_f$)}
\begin{tabular}{ccccc}
\toprule
Dark State Probability & CRJMCMC & Factorial HMM & Monotonic MAP & Sequential MAP \\
\midrule
0.0002 & \textbf{34.384 (±3.53)} & 47.915 (±6.07) & 197.612 (±11.43) & 101.791 (±20.21) \\
0.0004 & \textbf{38.789 (±6.28)} & \textbf{48.217 (±6.28)} & 197.612 (±11.43) & 142.355 (±23.53) \\
0.0006 & \textbf{39.886 (±5.04)} & \textbf{48.727 (±6.18)} & 197.612 (±11.43) & 174.930 (±31.19) \\
0.0008 & \textbf{46.146 (±6.38)} & \textbf{49.260 (±6.21)} & 197.612 (±11.43) & 202.737 (±36.16) \\
0.001 & \textbf{58.741 (±9.69)} & \textbf{49.682 (±6.11)} & 197.612 (±11.43) & 225.680 (±32.84) \\
0.0012 & \textbf{63.406 (±10.65)} & \textbf{49.844 (±6.21)} & 197.612 (±11.43) & 241.031 (±35.34) \\
0.0014 & 70.250 (±10.35) & \textbf{48.440 (±6.29)} & 197.612 (±11.43) & 280.443 (±38.06) \\
0.0016 & 74.901 (±10.86) & \textbf{47.976 (±6.04)} & 197.612 (±11.43) & 284.138 (±43.39) \\
0.0018 & 81.342 (±12.16) & \textbf{51.307 (±6.13)} & 197.612 (±11.43) & 353.945 (±45.28) \\
0.002 & 80.178 (±12.19) & \textbf{50.200 (±6.03)} & 197.612 (±11.43) & 341.295 (±47.16) \\
\bottomrule
\end{tabular}
\end{table}

\begin{table}[htbp]
\captionof{table}{Absolute Error ($\mu_b$)}
\begin{tabular}{ccccc}
\toprule
Dark State Probability & CRJMCMC & Factorial HMM & Monotonic MAP & Sequential MAP \\
\midrule
0.0002 & \textbf{1.980 (±0.16)} & 90.490 (±17.65) & 125.419 (±7.82) & 48.186 (±9.22) \\
0.0004 & \textbf{2.083 (±0.19)} & 72.165 (±13.18) & 125.419 (±7.82) & 73.756 (±11.54) \\
0.0006 & \textbf{2.104 (±0.20)} & 66.756 (±12.50) & 125.419 (±7.82) & 105.108 (±18.34) \\
0.0008 & \textbf{2.285 (±0.26)} & 68.595 (±13.40) & 125.419 (±7.82) & 115.275 (±18.16) \\
0.001 & \textbf{2.265 (±0.26)} & 71.336 (±14.15) & 125.419 (±7.82) & 122.527 (±17.72) \\
0.0012 & \textbf{2.556 (±0.32)} & 64.990 (±12.80) & 125.419 (±7.82) & 136.435 (±17.64) \\
0.0014 & \textbf{2.863 (±0.40)} & 56.676 (±11.18) & 125.419 (±7.82) & 159.590 (±18.57) \\
0.0016 & \textbf{2.498 (±0.30)} & 66.887 (±12.89) & 125.419 (±7.82) & 165.590 (±21.99) \\
0.0018 & \textbf{2.988 (±0.43)} & 64.604 (±12.66) & 125.419 (±7.82) & 189.386 (±21.68) \\
0.002 & \textbf{3.501 (±0.51)} & 66.624 (±12.49) & 125.419 (±7.82) & 168.324 (±18.20) \\
\bottomrule
\end{tabular}
\end{table}

\begin{table}[htbp]
\captionof{table}{Absolute Error ($\sigma_f^2$)}
\input{supplementary/sim_tables/extreme_dark_freq_rmse_sd_fluoro_summary.tex}
\end{table}

\begin{table}[htbp]
\captionof{table}{Absolute Error ($\sigma_b^2$)}
\input{supplementary/sim_tables/extreme_dark_freq_rmse_sd_back_summary.tex}
\end{table}

\clearpage
\subsubsection{Varying Dark State Duration}

\begin{table}[htbp]
\captionof{table}{Accuracy}
\input{supplementary/sim_tables/extreme_dark_dur_accuracy_summary.tex}
\end{table}

\begin{table}[htbp]
\captionof{table}{Precision}
\input{supplementary/sim_tables/extreme_dark_dur_precision_summary.tex}
\end{table}

\begin{table}[htbp]
\captionof{table}{Sensitivity}
\input{supplementary/sim_tables/extreme_dark_dur_sensitivity_summary.tex}
\end{table}

\begin{table}[htbp]
\captionof{table}{Specificity}
\input{supplementary/sim_tables/extreme_dark_dur_specificity_summary.tex}
\end{table}

\begin{table}[htbp]
\captionof{table}{Cohen's kappa}
\input{supplementary/sim_tables/extreme_dark_dur_cohen_kappa_summary}
\end{table}

\begin{table}[htbp]
\captionof{table}{RMSE (Intensity)}
\input{supplementary/sim_tables/extreme_dark_dur_rmse_intensity_summary.tex}
\end{table}

\begin{table}[htbp]
\captionof{table}{Absolute Error ($\mu_f$)}
\input{supplementary/sim_tables/extreme_dark_dur_rmse_mean_fluoro_summary.tex}
\end{table}

\begin{table}[htbp]
\captionof{table}{Absolute Error ($\mu_b$)}
\input{supplementary/sim_tables/extreme_dark_dur_rmse_mean_back_summary.tex}
\end{table}

\begin{table}[htbp]
\captionof{table}{Absolute Error ($\sigma_f^2$)}
\input{supplementary/sim_tables/extreme_dark_dur_rmse_sd_fluoro_summary.tex}
\end{table}

\begin{table}[htbp]
\captionof{table}{Absolute Error ($\sigma_b^2$)}
\input{supplementary/sim_tables/extreme_dark_dur_rmse_sd_back_summary.tex}
\end{table}

\clearpage

\subsubsection{Varying Blink State Transition Probability}

\begin{table}[htbp]
\captionof{table}{Accuracy}
\input{supplementary/sim_tables/extreme_blink_freq_accuracy_summary.tex}
\end{table}

\begin{table}[htbp]
\captionof{table}{Precision}
\input{supplementary/sim_tables/extreme_blink_freq_precision_summary.tex}
\end{table}

\begin{table}[htbp]
\captionof{table}{Sensitivity}
\input{supplementary/sim_tables/extreme_blink_freq_sensitivity_summary.tex}
\end{table}

\begin{table}[htbp]
\captionof{table}{Specificity}
\input{supplementary/sim_tables/extreme_blink_freq_specificity_summary.tex}
\end{table}

\begin{table}[htbp]
\captionof{table}{Cohen's Kappa}
\input{supplementary/sim_tables/extreme_blink_freq_cohen_kappa_summary}
\end{table}

\begin{table}
\captionof{table}{RMSE (Intensity)}
\input{supplementary/sim_tables/extreme_blink_freq_rmse_intensity_summary.tex}
\end{table}

\begin{table}[htbp]
\captionof{table}{RMSE ($\mu_f$)}
\input{supplementary/sim_tables/extreme_blink_freq_rmse_mean_fluoro_summary.tex}
\end{table}

\begin{table}[htbp]
\captionof{table}{RMSE ($\mu_b$)}
\input{supplementary/sim_tables/extreme_blink_freq_rmse_mean_back_summary.tex}
\end{table}

\begin{table}
\captionof{table}{RMSE ($\sigma_f^2$)}
\input{supplementary/sim_tables/extreme_blink_freq_rmse_sd_fluoro_summary.tex}
\end{table}

\begin{table}
\captionof{table}{RMSE ($\sigma_b^2$)}
\input{supplementary/sim_tables/extreme_blink_freq_rmse_sd_back_summary.tex}
\end{table}

\clearpage
\subsubsection{Varying Blink State Duration}

\begin{table}[htbp]
\captionof{table}{Accuracy}
\input{supplementary/sim_tables/extreme_blink_dur_accuracy_summary.tex}
\end{table}

\begin{table}[htbp] \centering
\captionof{table}{Precision}
\input{supplementary/sim_tables/extreme_blink_dur_precision_summary.tex}
\end{table}

\begin{table}[htbp] \centering
\centering
\captionof{table}{Sensitivity}
\input{supplementary/sim_tables/extreme_blink_dur_sensitivity_summary.tex}
\end{table}

\begin{table}[htbp] \centering
\captionof{table}{Specificity}
\input{supplementary/sim_tables/extreme_blink_dur_specificity_summary.tex}
\end{table}

\begin{table}[htbp] \centering
\captionof{table}{Cohen's Kappa}
\input{supplementary/sim_tables/extreme_blink_dur_cohen_kappa_summary}
\end{table}

\begin{table}
\captionof{table}{RMSE (Intensity)}
\input{supplementary/sim_tables/extreme_blink_dur_rmse_intensity_summary.tex}
\end{table}

\begin{table}[htbp] \centering
\captionof{table}{Absolute Error ($\mu_f$)}
\input{supplementary/sim_tables/extreme_blink_dur_rmse_mean_fluoro_summary}
\end{table}

\begin{table}[htbp] \centering
\captionof{table}{Absolute Error ($\mu_b$)}
\input{supplementary/sim_tables/extreme_blink_dur_rmse_mean_back_summary.tex}
\end{table}

\begin{table}[htbp] \centering
\captionof{table}{Absolute Error ($\sigma_f^2$)}
\input{supplementary/sim_tables/extreme_blink_dur_rmse_sd_fluoro_summary.tex}
\end{table}

\begin{table}[htbp] \centering
\captionof{table}{Absolute Error ($\sigma_b^2$)}
\input{supplementary/sim_tables/extreme_blink_dur_rmse_sd_back_summary.tex}
\end{table}

\clearpage

\end{appendices}

\end{document}